\documentclass[%
amsmath,pra, amssymb,superscriptaddress,twocolumn,notitlepage,nofootinbib
]{revtex4-2}
\usepackage[utf8]{inputenc}
\usepackage{amsmath, amssymb, amsthm, physics}
\usepackage{mathtools}
\usepackage[countmax]{subfloat}
\usepackage{booktabs}
\usepackage{bbm}
\usepackage{graphicx}
\usepackage{lineno}
\usepackage[makeroom]{cancel}
\usepackage{comment}
\usepackage{color}
\usepackage{hyperref}
\usepackage{epstopdf}
\usepackage{dsfont}
\usepackage{multirow}
\usepackage[english]{babel}
\usepackage{graphicx}
\usepackage{float}
\usepackage{lipsum}
\usepackage{curve2e}
\usepackage{hyperref}
\usepackage{xcolor}
\hypersetup{colorlinks=true, linkbordercolor=white, linkcolor=blue, pdfborder={/S/U/W 1}}
\usepackage{braket}
\usepackage{bm}
\usepackage{modiagram}

\newcommand{\x}{\dag}
\usepackage{soul}

\graphicspath{{Images/}}
\begin{document}

\title{%Passive 
Teleportation of a genuine single-rail vacuum–one-photon qubit generated via a quantum dot source}

\author{Beatrice Polacchi}
\affiliation{Dipartimento di Fisica - Sapienza Universit\`{a} di Roma, P.le Aldo Moro 5, I-00185 Roma, Italy}

\author{Francesco Hoch}
\affiliation{Dipartimento di Fisica - Sapienza Universit\`{a} di Roma, P.le Aldo Moro 5, I-00185 Roma, Italy}

\author{Giovanni Rodari}
\affiliation{Dipartimento di Fisica - Sapienza Universit\`{a} di Roma, P.le Aldo Moro 5, I-00185 Roma, Italy}

\author{Stefano Savo}
\affiliation{Dipartimento di Fisica - Sapienza Universit\`{a} di Roma, P.le Aldo Moro 5, I-00185 Roma, Italy}

\author{Gonzalo Carvacho}
\affiliation{Dipartimento di Fisica - Sapienza Universit\`{a} di Roma, P.le Aldo Moro 5, I-00185 Roma, Italy}

\author{Nicol\`o Spagnolo}
\affiliation{Dipartimento di Fisica - Sapienza Universit\`{a} di Roma, P.le Aldo Moro 5, I-00185 Roma, Italy}

\author{Taira Giordani}
\email[Corresponding author: ]{taira.giordani@uniroma1.it}
\affiliation{Dipartimento di Fisica - Sapienza Universit\`{a} di Roma, P.le Aldo Moro 5, I-00185 Roma, Italy}

\author{Fabio Sciarrino}
\affiliation{Dipartimento di Fisica - Sapienza Universit\`{a} di Roma, P.le Aldo Moro 5, I-00185 Roma, Italy}

\begin{abstract}
\textbf{Quantum state teleportation represents a pillar of quantum information and a milestone on the roadmap towards quantum networks with a large number of nodes. Successful photonic demonstrations of this protocol have been carried out employing different qubit encodings. However, demonstrations in the Fock basis encoding are challenging, due to the impossibility of generating a coherent superposition of vacuum-one photon states on a single mode with linear optics. Indeed, previous realizations only allowed the teleportation of dual-rail entangled states, by exploiting ancillary electromagnetic modes. Here, instead, we enable the quantum teleportation of pure vacuum–one-photon qubits encoded in a single spatial mode, by exploiting coherent control of a resonantly excited semiconductor quantum dot in a micro-cavity. Within our setup, we can both teleport genuine single-rail vacuum–one-photon qubits and perform entanglement swapping. Our results may disclose new quantum information processing potentialities for this encoding, whose manipulation is achievable via quantum dot single-photon sources.}
\end{abstract}

\maketitle

\section*{Introduction}

Quantum teleportation \cite{bennett1993teleporting, exptele, boschi1998experimental} is one of the most intriguing processes arising from the theory of quantum mechanics, being at the core of quantum technologies as well as to the development of many concepts in quantum information theory. This protocol is enabled by a shared quantum entangled resource. A quantum state is jointly measured with one-half of the entangled resource in a given place and, as a result of such an operation, it is transferred to the other half in a remote location.
Such a protocol, together with entanglement swapping \cite{swapping, PhysRevLett.80.3891}, is at the core of several quantum computation \cite{Gottesman1999, Knill2001} and quantum communication schemes ranging from quantum repeaters \cite{quantumrepeater1, Repeaters_swapping}, quantum gate teleportation \cite{PhysRevLett.90.117901, Chou2018, Wan2019}, measurement-based quantum computing \cite{PhysRevA.68.022312, PhysRevLett.98.220503,  Briegel2009} as well as port-based quantum teleportation \cite{Studziński2017, PhysRevA.102.012414}. 
Over the years, several experiments have successfully shown the teleportation of unknown quantum states using different experimental setups and degrees of freedom \cite{boschi1998experimental,exptele,pirandola2015advances,hu2023progress,marcikic2003long,lombardi2002teleportation,furusawa1998unconditional,giacomini2002active,PhysRevLett.86.1370,fattal2004quantum, Pirandola2015}. In particular, the first platforms were based on photonic systems \cite{exptele, boschi1998experimental}, since photons represent natural information carriers, due to their low interaction with the environment and to the capability of their efficient manipulation through linear optics. Quantum teleportation has been implemented over hundreds of kilometers using free-space channels \cite{Ma2012, Yin2012}, across metropolitan networks \cite{Valivarthi2016,Grosshans2016, Shen2023}, using satellites stations \cite{Ren2017} and near-deterministic single-photon sources \cite{BassoBasset_tele}.
%{ \cite{Gottesman1999, Knill2001,grice2011arbitrarily}. 
%}
In this context, a major challenge consists in teleporting qubits encoded in the photon-number basis. {We identify two main reasons. Firstly, this encoding is affected by photon losses.
Secondly, generating photon-number superposition states on a single electromagnetic mode is unfeasible only by means of linear optical elements.
Nowadays, these two major experimental challenges can be addressed by  near-deterministic single-photon sources that proved to {reach high level of efficiency \cite{Wang2019, ding2023highefficiency} and to} generate genuine single-rail vacuum-one-photon states \cite{loredo2019generation,wein2022photon}. Furthermore, the recent progress in the development of high-efficiency single-photon detectors \cite{natarajan2012superconducting} and low-loss optical platforms \cite{Chanana2022} open new perspectives for the investigation of information encoding based on the number of photons.}

{Previous quantum teleportation implementations} in the photon-number basis \cite{lombardi2002teleportation, fattal2004quantum} considered the teleportation of a vacuum-one-photon qubit produced by making a photon impinging on a beam splitter and encoding the qubit into one of the two output modes. {In detail, the output state is entangled over the two output modes of the beam splitter. One mode is used to encode the qubit to be teleported while the other mode encodes an ancillary qubit used for the final verification of the protocol.}
Consequently, as also stated by the authors, the teleported state is, in fact, not a \textit{genuine} vacuum-one-photon qubit, but a subsystem of an entangled state, making this scheme {corresponding precisely} to entanglement swapping. %{This causes such schemes to be only able to teleport states obtainable as reduced density matrices of entangled states, while not being suitable for arbitrary ones.}

\begin{figure}[ht]
    \centering
    \includegraphics[width = \columnwidth]{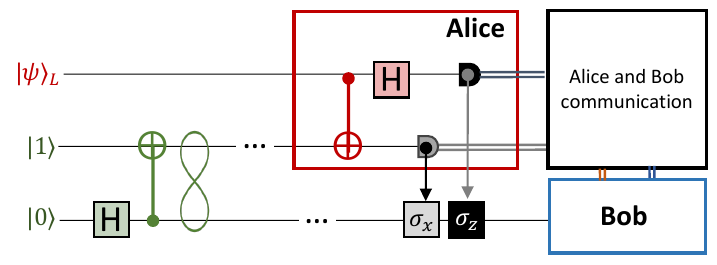}
    \caption{{\textbf{Circuit of the quantum teleportation protocol.}} The Bell state generated by the Hadamard (H) and CNOT gates (in green) is shared between Alice and Bob. Alice performs a Bell measurement (in red) between the qubit $\ket{\psi}$ to teleport and one of the two qubits in the Bell state. Bob retrieves the qubit $\ket{\psi}$ by applying $\sigma_x$ and/or $\sigma_z$ according to Alice's measurement outcomes.}
    \label{fig:circuit_scheme}
\end{figure}

Here, we propose a way to overcome such limitations, based on the nonlinear features enabled by the resonant excitation of a quantum dot single-photon source \cite{Senellart2017,Heindel:23}. 
{In detail, the main novelty of our work lies in the first quantum teleportation of arbitrary pure vacuum--one-photon qubits genuinely encoded on a single electromagnetic mode.}
Recent advances in the quantum dot sources technology have enabled a significant step forward in the generation of photonic states for quantum information protocols \cite{10.1063/5.0010193, somaschi2016near, Lu2021, BassoBasset_qkd, cao2023photonic, maring2023generalpurpose, chen2023heralded}.
Our approach relies on the generation of single photons through resonant optical excitation of a quantum dot in a microcavity \cite{somaschi2016near, Wang2019}, which was recently shown to be able to produce coherent superposition states of vacuum and one photon \cite{loredo2019generation}, as well as photon-number entangled states \cite{wein2022photon}. This is due to its capability of transferring the coherence of the excited-ground atomic state to the generated photons \cite{somaschi2016near,loredo2019generation}. By harnessing this property, we were able to precisely control the vacuum-one-photon state to be teleported, as shown by the achieved results.

Indeed, we {first} design a scheme tailored to the teleportation of genuine single-rail vacuum--one-photon qubits.
Then, we {compare our teleportation scheme with the entanglement swapping one, by implementing also the latter} in the spirit of the previous demonstrations of teleportation using the vacuum-one-photon paradigm \cite{lombardi2002teleportation, fattal2004quantum}. {These results demonstrate a step forward with respect to previous works where only entangled states could be teleported.}
The results are highly compatible with the expectations, demonstrating that our platform allows for the teleportation of both entangled and single-qubit states in the Fock basis encoding.

Our experiment represents a step forward in the field of quantum communication, being the first example of genuine quantum teleportation of a pure state in the Fock basis, and stimulating further investigations in protocols involving photon-number encoded states \cite{wein2022photon} for quantum information tasks. Furthermore, our results may offer exciting prospects for the development of quantum dot-based advanced quantum technologies.

\section*{Results}

{\subsection{Quantum state teleportation}}
{This section is divided into four paragraphs. In the first one, we recall the standard circuital description of the quantum teleportation protocol. In the second one, we translate the circuital paradigm into our photonic protocol proposal where logical qubits are encoded in the Fock basis. In the third paragraph, we describe our experimental platform for the implementation of the proposed protocol. Finally, in the fourth paragraph, we discuss our experimental results.}

{\subsubsection{Theoretical Background}}

Before going into the details of the experiment, we briefly review the quantum teleportation protocol through {its} circuital representation, shown in Fig.~\ref{fig:circuit_scheme}.

{In the quantum teleportation protocol, Alice aims at sending a generic {qubit} of the form:
\begin{equation}
{\ket{\psi}} = \alpha \ket{0}_L + \beta \ket{1}_L,
\label{eq:generic_qubit}
\end{equation}
to Bob, in absence of a direct quantum channel, by only exploiting a classical channel and a shared Bell state. Here, the index $L$ is used to indicate the logical qubit. From a logical point of view, reported in Fig. \ref{fig:circuit_scheme}a, the Bell state is produced by applying the following transformation to an initial two-qubit state:
{
\begin{equation}
\begin{split}
    \ket{\psi^+} &= \text{CNOT} (\text{H} \otimes \text{I}) \ket{01}_L \\
    &= \frac{\ket{01}_L + \ket{10}_L}{\sqrt{2}}
\end{split}
\label{eq:logical_bell_state}
\end{equation}}
To teleport her state, Alice performs a Bell-state measurement (BSM) between the two qubits on her stage, i.e. {$\ket{\psi}$} and one half of the Bell state, obtaining a two-bit outcome. Depending on the result of such an operation, the second half of the Bell state, sent to Bob's stage, is left in one of the possible states: $\alpha \ket{0}_L + \beta \ket{1}_L$, $\alpha \ket{1}_L + \beta \ket{0}_L$, $\alpha \ket{0}_L - \beta \ket{1}_L$, or $\alpha \ket{1}_L - \beta \ket{0}_L$. Therefore, to retrieve the original state {$\ket{\psi}$}, Bob may need to apply a unitary transformation to his state according to Alice's two-bit outcome.}

\begin{figure*}[t]
    \centering 
    \includegraphics[width = 0.9\textwidth]{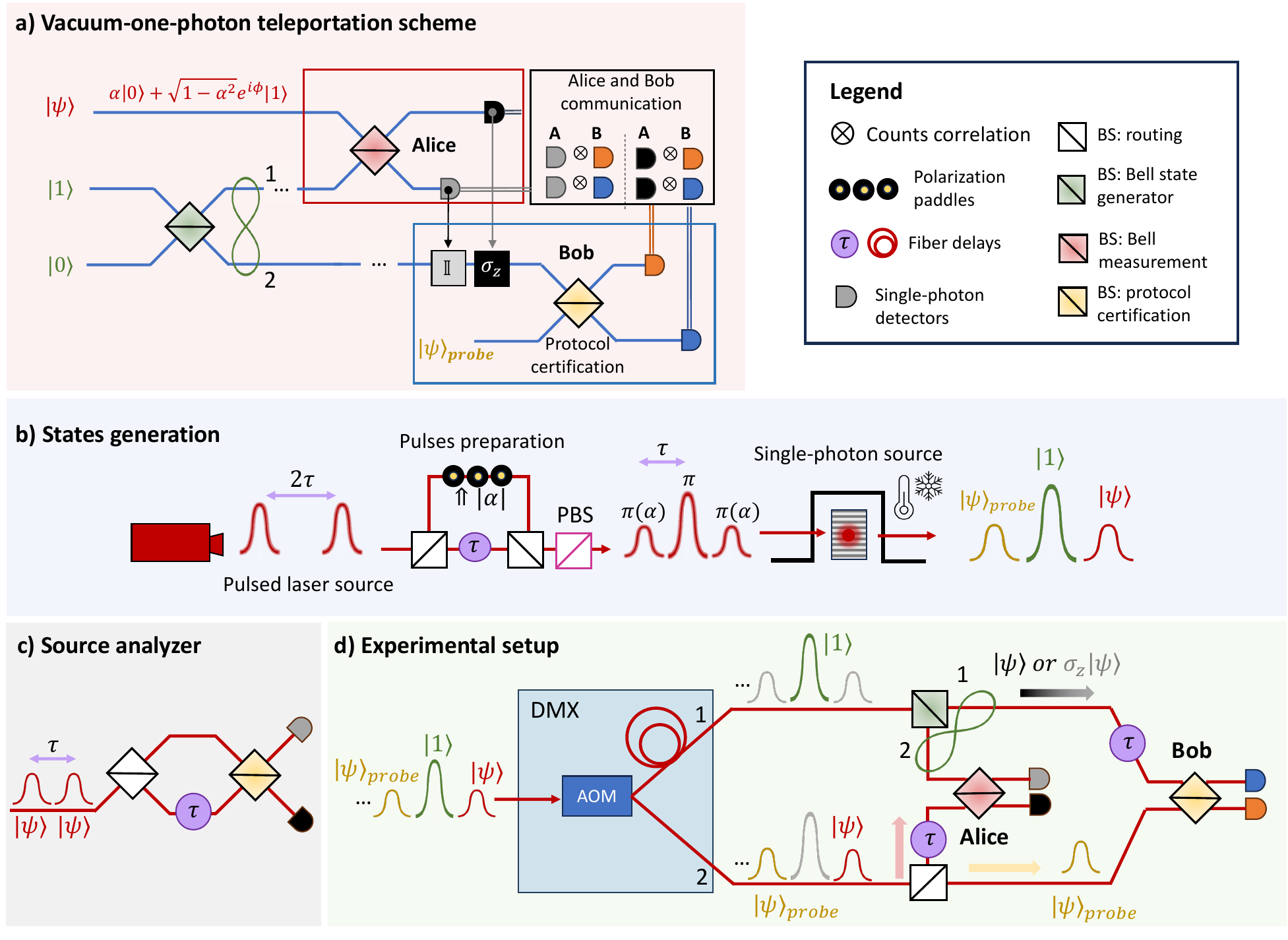}
    \caption{{\textbf{ Quantum teleportation protocol in the vacuum--one-photon encoding. a)} Circuit realization of a probabilistic quantum teleportation protocol in the vacuum-one-photon qubit encoding. The Bell state generation (green) and a probabilistic Bell measurement (red) are realized through the interference of single photons in a beamsplitter (BS). Alice communicates the outcomes of the measurement to Bob. Bob applies accordingly the identity or $\sigma_z$ and certifies the teleportation by a self-homodyne detection with a reference state performed by the last yellow BS.}
    \textbf{b)} In the first step {of our experimental setup}, we double the repetition rate of the pump pulsed laser, from $\sim 80$ Mhz to $\sim 160$ Mhz, through an in-fiber Mach-Zehnder interferometer (MZI). Polarization paddles control the pump power of the pulses after passing through a polarizing beam-splitter (PBS). Such a modulation of the pulses power allows to resonantly excite our QD to generate sequence of states like $\ket{\psi}^{t=0}$, $\ket{1}^{t=\tau}$ and $\ket{\psi}^{t=2\tau}_{probe}$, where $\ket{\psi}$ is the generic qubit of Eq.~\eqref{eq:qubit_s}. \textbf{c)} The MZI employed to characterize the vacuum-one-photon qubits generated by the source, i.e. the estimate of the $\alpha$ value from the visibility of the interference fringes. \textbf{d)} The train of states is distributed and synchronized in two different channels by a time-to-space de-multiplexer (DMX) based on an acoustic-optical modulation (AOM). In the first channel ($ch_1$), the $\ket{1}^{t=\tau}$ state is sent through the green BS to generate a Bell state, and, afterward, half of it interferes at time ${t=\tau}$ in Alice's station with $\ket{\psi}$ coming from the second channel ($ch_2$). The second half of the Bell state, i.e. the teleported state, is sent to Bob, who characterizes it through interference at time ${t=2\tau}$ with $\ket{\psi}_{\text{probe}}$ that is a copy of the original qubit $\ket{\psi}$. The red, yellow, black arrows represent the case in which the protocol succeeds due to the probabilistic routing of $\ket{\psi}$ and $\ket{\psi}_{\text{probe}}$, as all pulses take the correct delay lines. We prepared a train of $\ket{1}$ states and generated Bell states at times $t=0$ and $t=\tau$ on the first BS. Alice performs a BSM by making half of the Bell state generated at $t=0$ interfere with half of the next entangled state generated at $t=\tau$. Bob and Charlie perform the same measurement to verify the success of the entanglement swapping protocol after the communication of Alice's station outcomes.}
    \label{fig:expscheme_teleportation}
\end{figure*}

{\subsubsection{Quantum state teleportation in the vacuum-one-photon encoding}}

In Fig.~\ref{fig:expscheme_teleportation}a, we show the conceptual scheme of our {quantum teleportation} experiment, that implements the teleportation of a vacuum-one photon state, i.e. where the logical {qubits} $\ket{0}_L$ {and} $\ket{1}_L$ {are} encoded in the {physical Fock states with zero and one photon, respectively, i.e., $\ket{0}_L \coloneqq \ket{0}$ and $\ket{1}_L \coloneqq \ket{1}$}.
In detail, we initially generate a genuine vacuum-one photon {qubit} in the state:
\begin{equation}
    \ket{\psi} = \alpha \ket{0} + \sqrt{1- \alpha^2}e^{i\phi} \ket{1}
    \label{eq:qubit_s}
\end{equation}
that is equivalent to the one in Eq.~\eqref{eq:generic_qubit}. 
{We suppose that the coefficient $\alpha$ is a real number. Moreover,} we take the phase $\phi$ as a reference and, therefore, for simplicity, we can assume that it amounts to zero.
The {one-photon} Fock state {$\ket{1}$} is {used to produce {the} Bell state {shared between Alice and Bob as} required by the teleportation protocol, by sending it onto} a symmetric BS with modes labeled as $1$ and $2$. 
This optical element transforms the creation operators according to $a_1^{\x} \longrightarrow (a_{1}^{\x} + a_{2}^{\x})/\sqrt{2}$, so that the resulting state reads:
\begin{equation}
    \ket{1_10_2} \longrightarrow \frac{ \ket{1_{1}0_{2}} + \ket{0_{1}1_{2}} }{\sqrt{2}}
\label{eq:bell_state}
\end{equation}
{In the vacuum-one photon encoding previously defined, the state above is a maximally entangled state equivalent to the one in Eq.~\eqref{eq:logical_bell_state}}.
{The subsystem in mode 1 is sent to Alice while the subsystem in mode 2 is sent to Bob.}
{In her station, Alice performs} a partial BSM between the qubit $\ket{\psi}$ and {her} half of the Bell state by means of {the red} symmetric BS.
{After Alice's {BSM},} the half of the Bell state {held by Bob in mode 2}  is left with a success probability $p_{\pm}=1/4$ either in the state {$\ket{\psi}$} or {$\sigma_z \ket{\psi}$}, depending on Alice's measurement outcome.
Finally, Bob validates the protocol by performing a suitable measurement between the received qubit and a reference state $\ket{\psi}_{\text{probe}}$, {which is only used as a probe to certify the protocol success}. This measurement aims at witnessing the coherence between the two logical states, i.e. the vacuum and the one-photon components.
As $\ket{\psi}_{\text{probe}}$, we employ a copy of the original qubit $\ket{\psi}$ {of Eq.~\eqref{eq:qubit_s}}. Such a method, known as {self-homodyne} detection, allows measuring coherence by observing interference fringes in the single-photon counts at the output of the BS, as a function of the relative phase between the optical paths. The presence of such interference fringes is a signature of the coherence in the input states, since it cannot be obtained by mixed states. Additionally, its observation requires the preparation of indistinguishable photons to enable quantum interference. Furthermore, the visibility of such fringes depends on the vacuum population of the two qubits \cite{loredo2019generation}. Such a method can be used to characterize the output of the teleportation protocol, thus verifying that it has been carried out successfully. Further descriptions of the self-homodyne detection in our protocol can be found in Supplementary Notes~{I-IV}.

\begin{figure*}[t]
    \centering
    \includegraphics[width=\textwidth]{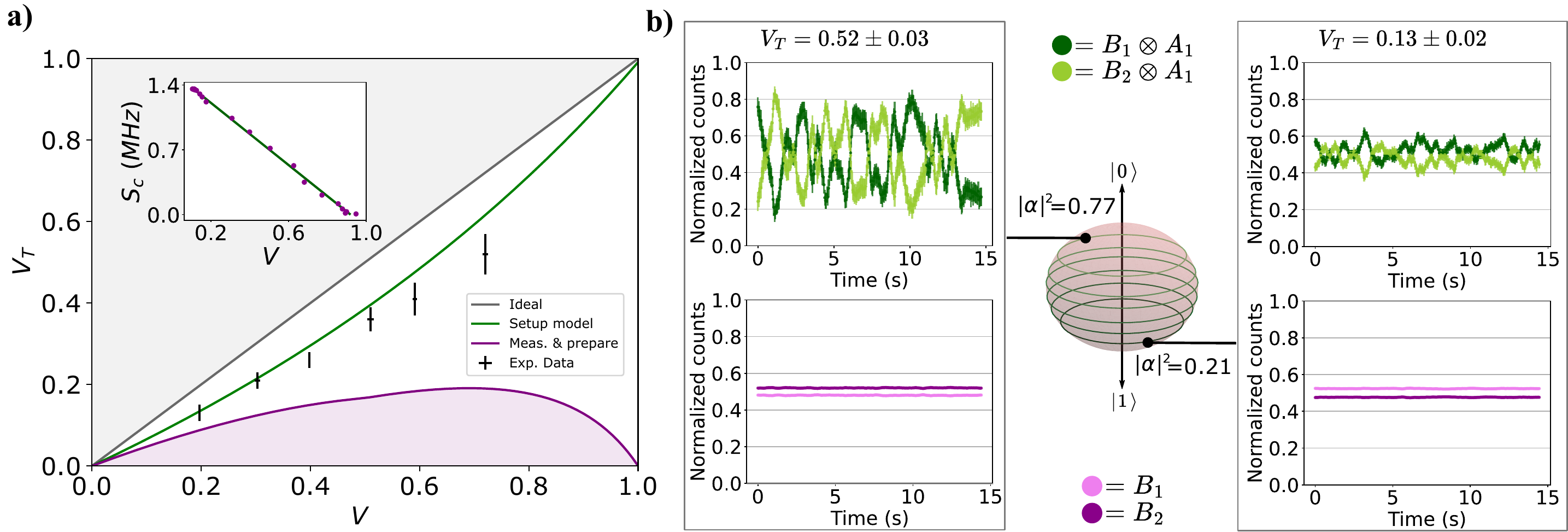}
    \caption{\textbf{Theoretical expectations and experimental results for the teleportation protocol. {a)}} In grey, we show the expected visibility $V_T$ of the teleported state as a function of the probe state visibility $V$ within an ideal platform, with deterministic routing in every MZI, photon-number resolving detectors, the same self-homodyne measurement station, and perfect photon indistinguishability. In green, we show how such expectations change when considering that, in our platform, the separation of the probe and the qubit to be teleported is not deterministic and threshold detectors are employed. Furthermore, this model is formulated in the limit of high losses and considering imperfections such as the state {conditional} purity and partial photon distinguishability. In purple, instead, we report the same quantities when using the {fully classical \textit{measure-and-prepare} protocol, which allows for the highest fidelity achievable with only classical resources} \cite{Classical_limit_1995} in the same lossy regime. In the inset, we report the single count rate ($S_c$) in one of two outputs of the self-homodyne station (see Fig. \ref{fig:expscheme_teleportation}c) as a function of the visibility, for power pulse areas below or equal to $\pi$. From this plot, we estimate a state {conditional} purity $\lambda \approx 0.98$. \textbf{b)} Two examples of the free evolution of the single-photon counts observed on Bob's detectors after the teleportation protocol and corresponding to two different states. The green plots show events in Bob's station, conditioned on the presence of one photon in Alice's detector number 1. In pink, instead, we show the unfiltered single count traces detected on Bob's side.
    Each pair of traces is normalized to the sum of the two single-photon signals. In this figure, we also illustrate that the states we are able to teleport correspond to parallel planes on the Bloch sphere, since no phase information can be retrieved with our measurements. All the shown uncertainties amount to one standard deviation and were computed by assuming Poissonian events.}
    \label{fig:alpha_data}
\end{figure*}

\subsubsection{Experimental apparatus}

Our experimental setup is depicted in {Fig.~\ref{fig:expscheme_teleportation}b-d}. As a single-photon source, we use a commercially available (Quandela \emph{e-Delight}) semiconductor quantum dot (QD). Such source consists of an InGaAs matrix placed in a nanoscale electrically controlled micropillar cavity \cite{somaschi2016near} kept at cryogenic temperature (around 4K) by an \emph{Attocube-Attodry800} He-closed cycle cryostat. The QD is optically excited by a pulsed laser resonant with the cavity characteristic wavelength ($928.05$~nm), in the {resonance fluorescence (RF) regime}  \cite{somaschi2016near, Wang2019}.
We achieve single-photon collection by means of a single-mode fiber (SMF) located inside the cryostat {(collection efficiency $\approx 10\%$)} and detect photons with avalanche photodiodes (APDs) {with efficiency amounting to $\approx 30 \%$}. Photons are separated from the residual pumping laser in a cross-polarization scheme. The nominal repetition rate of the laser amounts to $79$~MHz, but we double it by introducing a delay between two consecutive pulses and then recombining them, as shown in the ``Pulses preparation" section of Fig.~\ref{fig:expscheme_teleportation}b.
{The procedure of doubling the laser repetition rate is key to our protocol. Indeed,} such a preparation stage allows us to control the polarization of either pulse independently and, consequently, their power, as the laser light passes through a polarizing beam splitter (PBS) before optically exciting the QD. 
{This is needed because}, in the RF excitation scheme, an excitation power lower than the $\pi$-pulse translates into the generation of coherent vacuum-one photon states of the same form of Eq.~\eqref{eq:qubit_s}, where the population of the state $\ket{1}$, i.e. {$1- \alpha ^2$}, grows with the excitation power \cite{loredo2019generation}, achieving the value $1$ at the $\pi$-pulse.
{Therefore, as depicted in Fig.~\ref{fig:expscheme_teleportation}b, by independently manipulating the power of the two pulse trains, we make the source generate two different kinds of states, namely the qubit to be teleported $\ket{\psi}$, which is arbitrarily chosen, its probe copy $\ket{\psi}_{probe}$, and the one-photon Fock state $\ket{1}$, used to share entanglement between Alice and Bob, as explained in the previous section.}

In the Methods section, we report the typical performances of the source measured with the Mach-Zehnder interferometer (MZI) of Fig.~\ref{fig:expscheme_teleportation}c for what concerns the count rate, the second-order auto-correlation function $g^{(2)}$, the Hong-Ou-Mandel visibility and the conditional purity of the generated vacuum-one photon qubits, {which is the purity conditioned on the success of the protocol (for further information see Methods and Supplementary Note I)}.\\

{In our protocol, we have two characterization stages. Before the teleportation protocol, we characterize the qubit $\ket{\psi}$.
Then, after the teleportation protocol, Bob characterizes the teleported qubit by means of a probe copy of qubit $\ket{\psi}$, denoted as $\ket{\psi}_{probe}$. The details of such characterizations are explained in the following.}

{Characterization of the {qubit $\ket{\psi}$} is performed independently by using a time-unbalanced MZI, as the one shown in Fig.~\ref{fig:expscheme_teleportation}c.
More specifically, a train of states prepared in the target state, characterized by amplitude $\alpha$ for the vacuum component, is prepared in the same spatial mode and injected in one input port of a symmetric BS. A delay $\tau$, equal to the time separation between two input pulses, is then applied in one of the output modes of the BS. Then, the output modes interfere in a second symmetric BS, and the output fringes are measured from the single-photon counts in each of the output modes, as a function of the relative free evolving phase in the interferometer. The visibility $V$ of such a pattern provides a direct estimate of $\alpha^2$ (see Methods and Supplementary Note I).}

For the teleportation protocol, we set the two laser pulse trains at two different powers to produce alternatively the states $\ket{\psi}$ and $\ket{1}$, respectively at times $t=n\tau$ and $t=(n+1)\tau$, where $n=0,1,...$ and $\tau \approx 6$~ns. We trigger each protocol run in a way such to take as timing reference the following sequence: $\ket{\psi}_{\text{probe}}^{t=2\tau}$, $\ket{1}^{t=\tau}$, and $\ket{\psi}^{t=0}$. The generated states are sent through a Quandela commercially available temporal-to-spatial demultiplexer (DMX) which actively separates the incoming pulses into two synchronized but spatially different $\approx 150~$ns-long trains (see Fig.~\ref{fig:expscheme_teleportation}d).
The $\ket{1}^{t=\tau}$ state brought by the DMX channel 1 is sent on a BS to produce the Bell state in Eq.~\eqref{eq:bell_state}. 
{Details about the generation of the experimental Bell state and its characterization are discussed in Supplementary Note~{II}.}
One half of such Bell state is sent to Alice, who performs a partial BSM between it and the $\ket{\psi}^{t=0}$ state brought by the DMX channel 2. The second half of the Bell state is sent to Bob, who characterizes it through interference with the  $\ket{\psi}_{\text{probe}}^{t=2\tau}$ state, also brought by the DMX channel 2. More specifically, we record Bob's single counts traces conditioned on the presence of one photon in Alice's station $\approx 6$~ns before, by taking a two-fold coincidence window equal to $1.5~$ns. Note that the routing of $\ket{\psi}$ and $\ket{\psi}_{\text{probe}}$ is not deterministic, thus lowering the success probability to $p_{\pm}=1/16$.

\subsubsection{Teleportation results}
Here, we report and discuss our experimental results for  the quantum teleportation {protocol.}

To demonstrate that the teleportation protocol was successful, we analyzed the interference fringes which are obtained by the self-homodyning technique \cite{loredo2019generation} described above. More specifically, the main concept is to have a quantum state, in principle unknown and that needs to be characterized, interfere on a beam splitter with a reference probe prepared in a precise state such as the qubit in Eq.~\eqref{eq:generic_qubit}.
It can be shown (as discussed in the Methods and in Supplementary Note~I) that the visibility of interference fringes leads to a direct estimation of the vacuum population {and the conditional purity} of the unknown state. 
Moreover, the presence of interference fringes is a genuine signature of the coherence between the vacuum $\vert 0 \rangle$ and single-photon term $\vert 1 \rangle$ in the state, given that an input mixed state results in no interference in the output pattern as a function of the phase between the modes. 
The standard figure of merit to assess the quality of a quantum teleportation protocol is the average fidelity between teleported and target states. However, as investigated in \cite{cavalcanti2017all,carvacho2018experimental}, the average fidelity can be lower than the classical bound even if genuine quantum resources were used to implement teleportation and alternative witnesses of nonclassicality can be found {\cite{bussandri2024challenges}}. Therefore, to demonstrate the nonclassicality of our protocol, we opted for {comparing the fringe visibility of the teleported state $V_T$ with the target's one $V$}. Such a choice is motivated by observing that {the relation between $V_T$ and $V$ has different behaviors in the quantum and in the classical case. In particular, as we demonstrate in Supplementary Note~III, the visibilities that would be achieved with the fully classical \textit{measure-and-prepare} strategy, which allows for the highest fidelity achievable with only classical resources \cite{Classical_limit_1995}, define a bounded region, highlighted in purple in Fig.~3a, which has no intersection with the visibility achievable by using quantum resources. Moreover, our experimental points are compatible with the theoretical expectations within at least two standard deviation in the worst case. In contrast, the distance from the measure-and-prepare model is always larger than two standard deviation even in the region where the quantum and the classical curves are expected to be close.} 
{In summary, the teleported visibility achievable with quantum resources violates the classical limits defined by the region accessible through the \textit{measure-and-prepare} strategy.} Therefore, {it represents a quantitative figure of merit that} can be used to demonstrate genuine quantum teleportation as soon as the experimental visibilities {lie outside such a region.}

Our results are shown in Fig.~\ref{fig:alpha_data}.
{In the inset of Fig. \ref{fig:alpha_data}a, {we report the data used to extrapolate the {conditional} purity of the generated states, which amounts to $\approx 0.98$, as explained in the Methods section.}
{In detail,} we show the relationship between $V$ and the single-count rate recorded by one of the two detectors at the output of the last BS, when varying the vacuum population $\alpha^2$, i.e the power of the pump pulses. In the case of lossy apparatuses and  identical pulses, such {to} generate trains of qubits with the same $\alpha$, it can be shown that both the $V$ and single-photon counts are linearly proportional to $\alpha^2$ (see Methods and Supplementary Note~I for the full derivation).} 
In {the main plot of} Fig.~\ref{fig:alpha_data}a, {instead, we report the experimental results of the teleportation protocol, represented as black dots}. We compare {our} observations with three different theoretical models for the teleported visibility $V_T$ of the fringes in Bob's counts conditioned by Alice's measurements as a function of the probe state visibility $V$.
In the first two models (grey and green curves) we assume correct teleportation, i.e the teleported state is an exact copy of $\ket{\psi}$. 
In grey, we show the trend $V_T = V$ that would be observed if our setup was equipped with fully deterministic routing of $\ket{\psi}$ and $\ket{\psi}_{\text{probe}}$, {as in the scheme in Fig. \ref{fig:expscheme_teleportation}a,} and photon-number resolving detectors.
In the green plot, instead, we show a model taking into account the features of our platform, where deterministic routing is only applied to separate the central pulse from the two copies of the $\ket{\psi}$ state, and threshold detectors are employed.
Furthermore, we also consider the imperfections of the source for what concerns the qubits {conditional} purity and the partial photon distinguishability (see Methods and Supplementary Note~{IV} for the full derivation). Finally, the purple region corresponds to what would be observed in the case of the {\textit{measure-and-prepare} protocol}  ~\cite{Classical_limit_1995}, as we {demonstrate} in Supplementary Note~{III}.
In this case, the classically teleported (CT) state would be the following statistical mixture:
$\rho_{CT} = \frac{1}{3} \ket{\psi}\bra{\psi} + \frac{1}{3}  \ketbra{0}{0}+ \frac{1}{3}  \ketbra{1}{1}
$ that corresponds to the {closest-to-target state that can be achieved classically, having a} fidelity $F = 2/3$.
Our experimental observations are compatible with the green model within two standard deviations and {violate the \textit{measure-and-prepare} bound} thus showing that our protocol outperforms its best classical counterpart.

{Moreover, to demonstrate another pivotal element of quantum teleportation, i.e., the classical communication between Alice and Bob,} in Fig.~\ref{fig:alpha_data}b, we report in green (violet) our protocol performances when (not) allowing classical communication between Alice and Bob, for two different teleported coefficients $\alpha^2$. In the green plots, where Bob's single-photon counts are conditioned on the presence of one photon on one of Alice's detectors, we observe free-evolving fringes corresponding to successful teleportation. In the violet plots, instead, where Bob's single-photon counts are unfiltered, we cannot observe any fringes, due to lack of information about Alice's measurement outcome.
The same plot illustrates that we teleport quantum states lying on a plane of the Bloch sphere, as no phase information can be retrieved with our measurements.\\

{In the next section, we show the implementation of an entanglement swapping protocol within the apparatus. The results we are going to present have a two-fold goal. On one hand, they highlight the difference between the platform needed for teleportation and the one needed for entanglement swapping in the vacuum--one-photon encoding, which share the same quantum channel. On the other hand, they allow us to obtain a further certification of the teleportation protocol since, from the swapping results, we retrieve additional information about the quality of the quantum channel that was employed to implement both the genuine teleportation and the swapping protocols.}

\subsection{Entanglement swapping in the vacuum--one-photon encoding}

{This section is divided into three paragraphs. In the first one, we recall the circuital description of the entanglement swapping protocol and its photonic implementation. In the second one, we describe our experimental setup for the implementation of the protocol. In the third section, we discuss our experimental results.}

{\subsection{Theoretical Background}}

\begin{figure}[t]
    \centering
    \includegraphics[width = \columnwidth]{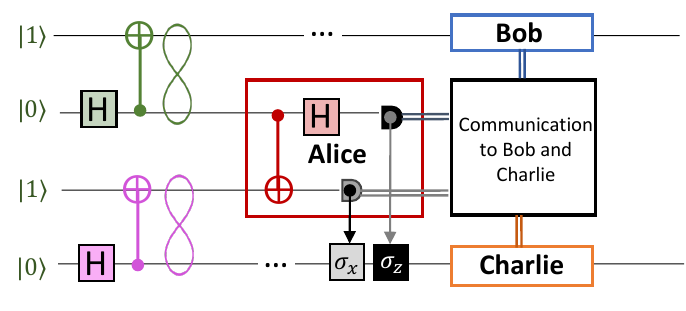}
    \caption{\textbf{{Circuital scheme of the entanglement swapping protocol}}.{ Alice performs a Bell measurement on two qubits that belong to two independent Bell states. Bob and Charlie will share a Bell state after the communication of Alice's measurement outcomes and the consequent applications of $\sigma_x$ or/and $\sigma_z$ operators on Charlie's qubit.}}
    \label{fig:circuit_scheme2}
\end{figure}

{In the entanglement swapping protocol, whose scheme is shown in Fig.~\ref{fig:circuit_scheme2}, entanglement is transferred from a given pair of particles to another at distant nodes. In detail, two independent Bell pairs are initially generated. One-half of both pairs is sent to a central observer, Alice, who performs a BSM. The remaining particles, each belonging to initially distinct pairs, are sent to two arbitrarily distant parties, Bob and Charlie. The interaction of the two subsystems in Alice's station results in the state shared by Bob and Charlie ending up in a maximally entangled state. As in the quantum teleportation protocol, Bob and Charlie may need to apply a unitary transformation according to Alice's outcome to retrieve the initial Bell state.}

\begin{figure*}[t]
    \centering
    \includegraphics[width = \textwidth]{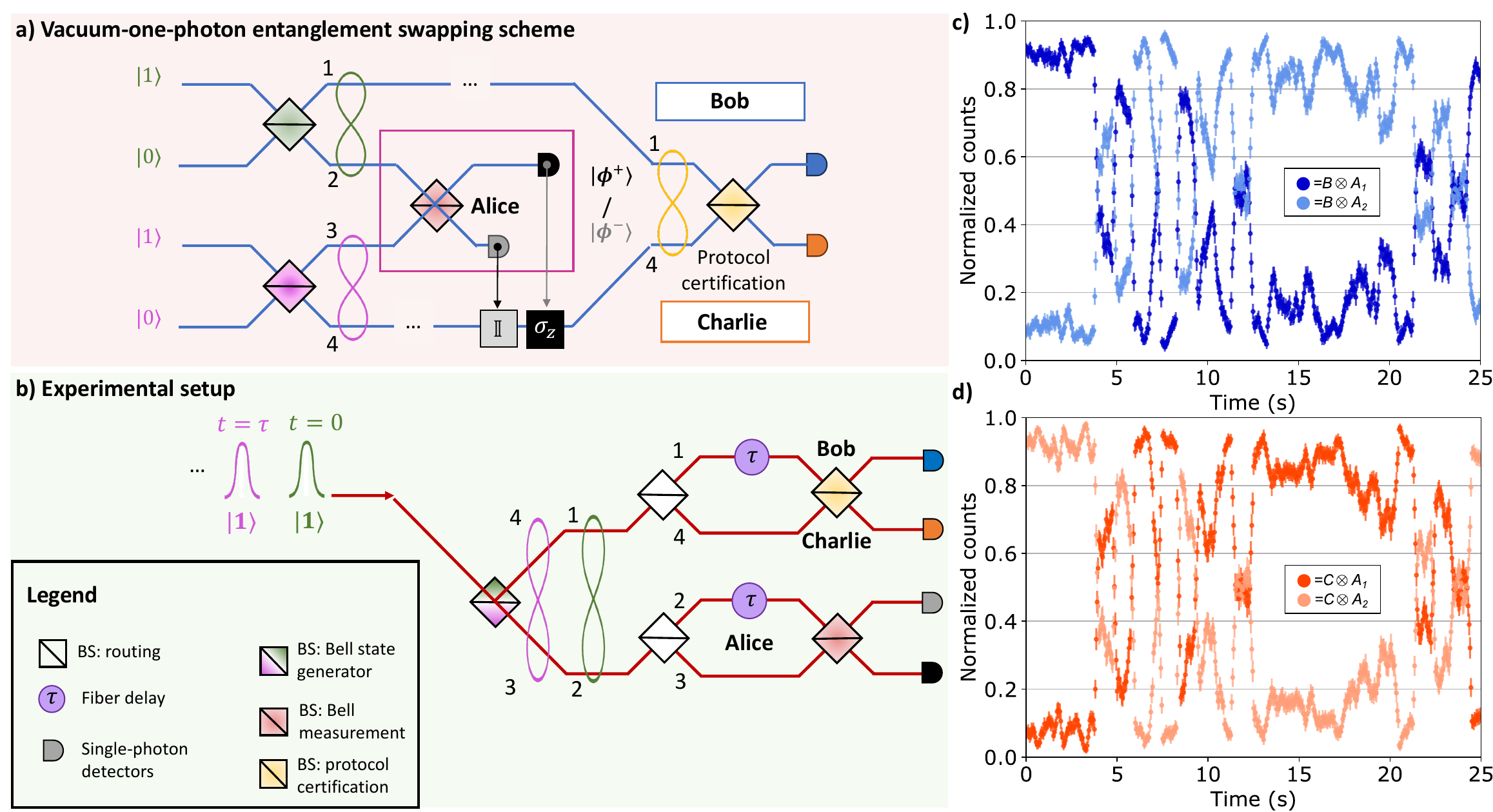}
    \caption{{\textbf{Entanglement swapping in the vacuum-one-photon encoding. a)} Entanglement swapping in the vacuum-one-photon encoding. Two single photons enter in two different BSs (in green) that generate two Bell states. Alice performs the probabilistic Bell measurement (red BS) and communicates to Charlie and Bob which detector clicks. Then, Charlie applies the identity or $\sigma_z$. Bob and Charlie perform a further probabilistic Bell measurement to identify the entangled state they share. \textbf{b)} We prepared a train of $\ket{1}$ states and generated Bell states at times $t=0$ and $t=\tau$ on the first BS. Alice performs a BSM by making interfere half of the Bell state generated at $t=0$ with half of the next entangled state generated at $t=\tau$. Bob and Charlie perform the same measurement to verify the success of the entanglement swapping protocol after the communication of Alice's station outcomes.} \textbf{c)}  Freely evolved fringes recorded by Bob's detector (blue and light blue data) and \textbf{d)} by Charlie's one (red and light red traces), triggered by the two Alice's BSM outcomes. Each pair of traces is normalized to the sum of the two single-photon signals. The uncertainties derive from the Poissonian statistics of single-photon counts.}
    \label{fig:swapping}
\end{figure*}

{In what follows, we present and demonstrate a probabilistic version of the entanglement swapping protocol} {encoded in the photon-number basis, which is depicted in Fig.~\ref{fig:swapping}a}. 
The two Bell states (Eq.~\eqref{eq:bell_state}) are generated by two single-photon states, that are injected in two separate BSs. Then, Alice performs a partial BSM through a symmetric BS, analogously to the teleportation protocol described above. Bob and Charlie then retrieve their maximal entangled states $\ket{\psi^+}$ or $\ket{\psi^-}$ after Alice communicates the measurement outcome obtained at her stage, i.e. which of the two detectors counts. The success probability of the protocol is again $p_{\pm}$ = 1/4. Certification of the output state after the entanglement swapping is then obtained via a further BSM between the modes at Bob's and Charlie's stages.

\subsection{Experimental apparatus}

{To perform the {photonic} entanglement swapping protocol {in Fig.~\ref{fig:swapping}a}, {we employ the setup} shown in Fig.~\ref{fig:swapping}b. Unlike the teleportation protocol, we {set the excitation laser power such to} solely generate $\pi$-pulses and consider the initial state $\ket{1}^{t=\tau}\ket{1}^{t=0}$.
Then, a symmetric BS generates two Bell states of the form given in Eq.~\eqref{eq:bell_state} {and highlighted in green and pink respectively in Fig.~\ref{fig:swapping}a-b, to stress their correspondence with the Bell pairs of the circuital scheme in Fig.~\ref{fig:circuit_scheme}d}.
{The first Bell pair, arriving at time $t=0$, is distributed along modes 1 and 2 yielding the following dual-rail state:
\begin{equation}
    \ket{\psi^+}^{t=0} = \frac{\ket{1}_{1} \ket{0}_{2} + \ket{0}_{1} \ket{1}_{2}}{\sqrt{2}}
\label{eq:first_bell_state_swap}
\end{equation}
The second pair arrives at time $t = \tau$ and is distributed along modes 3 and 4 analogously.
}
{Subsystems 2 and 3 are} {} sent to Alice's Mach-Zehnder interferometer (MZI) to perform a partial BSM.
{Subsystems 1 and 4, instead, are respectively }sent to Bob { and} Charlie {, who can be located, in principle, at an arbitrarily large distance.} {At this point, the spatial modes employed are four, as in Fig.~4a. The remainder of the protocol is thus equivalent to a probabilistic entanglement swapping.  Indeed, Alice performs a partial BSM between subsystems 2 and 3 through a delay $t=\tau$ on mode 2 that synchronizes the two pulses and the interference on the red BS in Fig.~4b. According to Alice's outcome, Bob and Charlie will end up with {the following} maximally entangled state {both in the time bin and the path degrees of freedom}:
    \begin{equation*}
        {\ket{\psi^{\pm}}_{swap} = \frac{ \ket{1}_{1}^{t=0} \ket{0}_{4}^{t=\tau} \pm \ket{0}_{1}^{t=0} \ket{1}_{4}^{t=\tau}}{\sqrt{2}}}
    \end{equation*}
    {The above state has}} the same form as the {one in Eq.~\eqref{eq:first_bell_state_swap} up to a unitary transformation, and is characterized} through interference on a BS
    and a delay $t=\tau$ applied to mode 1. Then, we measure the visibility of the fringes recorded in the single-photon counts of Bob and Charlie station conditioned to the Alice's measurement outcomes. As for the teleportation protocol, we consider a two-fold coincidence window equal to $1.5~$ns.  The probabilistic routing of the states lowers the success probability of the protocol to $p_{\pm}=1/16$.}

\subsection{Results}

{In {Fig.~\ref{fig:swapping}c-d}, we show the single-count traces observed when the output modes at Bob's and Charlie's stages interfere in a BS. Indeed, similarly to the teleportation protocol, the interference fringes measured at the output can be employed as a means to verify that the entanglement swapping procedure was successful. 
Respectively, the measured visibilities for the four possible two-fold events are:
\begin{equation}
    \begin{cases}
        V_{A_1,C} = 0.942 \pm 0.002\\
        V_{A_1,B} = 0.862 \pm 0.002\\
        V_{A_2,C} = 0.879 \pm 0.002\\
        V_{A_2,B} = 0.903 \pm 0.002\\
    \end{cases}
    V_{\text{ave}} = 0.896 \pm 0.001.
\end{equation}
As demonstrated in Supplementary Note~{V}, in the ideal case the visibility of the state shared between Charlie and Bob at the output of the BS is 1.
However, taking into account the partial distinguishability of photons, the expected visibility amounts to $V^{\text{theo}} = 0.902 $ and, therefore, it is in good agreement with our experimental results.
Small deviations of the individual visibilities from this ideal value are mainly due to non-ideal reflectivities of the BSs and differences in the coupling and detection efficiencies.
We estimated the experimental visibility through the same method used in the previous protocol and illustrated in the Methods. {The corresponding fidelity of the teleported entangled state can be computed with the formula reported in \cite{lombardi2002teleportation,fattal2004quantum} and it amounts to $F_{\text{ave}} = \frac{1+V_{\text{ave}}}{2} = 0.9480 \pm 0.0005$. The high swapping fidelity with the Bell state $\ket{\psi^+}$ mirrors the quality of the entangled resource and of the quantum channel used for both the entanglement swapping and quantum teleportation protocols reported above in this work. Therefore, this result further supports the conclusion that our teleportation protocol employs genuine quantum resources.}

\section*{Discussions}
In this work, we addressed a fundamental open problem, corresponding to the quantum teleportation of a general qubit encoded in the Fock basis. Indeed, due to technological constraints, such a problem was addressed until now by using linear optics, and, in particular, BSM operations. Vacuum-one photon states were indeed generated by letting a photon impinge on a BS and taking one of the output modes as the target mode. Such a procedure results in the quantum teleportation of one subsystem of an entangled state, while not permitting the quantum teleportation of genuine qubit states in such an encoding without the use of ancillary modes. 
To address this issue, we exploited the nonlinear properties of a semiconductor quantum dot optically excited through resonance fluorescence. Indeed, it was recently demonstrated that such a procedure yields the production of coherent superposition states of vacuum and photon-number states. Within such a platform, we were able to quantum teleport six different pure genuine vacuum-one photon states without employing any ancillary mode, and achieving results in good agreement with expectations. Moreover, within this setup, we were able to successfully perform entanglement swapping thus further extending the potentialities of our scheme to more complex scenarios.

We believe that our findings may represent an important step toward the development of large-scale quantum networks based on photon-number basis encoding, an approach that is widely investigated in quantum computation \cite{Renema_BS_superp, Flamini_rev} and communication tasks \cite{wein2022photon,Caspar2020,Monteiro2017,erkilicc2023surpassing, karli2023controlling}. 
{Further improvements of our protocol regard the implementation of full BSMs, which are still very challenging to realize in every photonic setup and encoding. We foresee as a future perspective using nonlinearities to realize such a kind of measurement, considering the recent advances in the field  \cite{PhysRevLett.86.1370, Yang_sorting_2022}.}
Our results may encourage new applications of quantum dot-based single-photon sources for several quantum information tasks.

\section*{Methods}
\subsection*{Characterization of the single-photon source}
We use a Hanbury-Brown-Twiss setup to measure the values of the second-order auto-correlation function $g_2(0)$ of our single-photon source and a MZI to measure the Hong-Ou-Mandel (HOM) visibility $V_{\text{HOM}}$, for each measurement station. For the quantum teleportation experiment, we only have two participants, Alice and Bob. For the entanglement swapping experiment, instead, we also have Charlie. However, for both experiments, we call more generally Bob's HOM visibility, the one characterizing the second MZI.

Typical values throughout the whole experiment are (see also the Supplementary Note~{VI}): 
\begin{gather*}
\begin{split}
    g_2^{\text{Alice}}(0) &= 0.0146 \pm 0.0006, \\
    g_2^{\text{Bob}}(0) &=0.0192 \pm 0.0007, \\
    V_{\text{HOM}}^{\text{Alice}} &= 0.9055 \pm 0.0015, \\
    V_{\text{HOM}}^{\text{Bob}} &= 0.8987 \pm 0.0012.
\end{split}
\end{gather*}

Additionally, we characterized the {conditional} purity of the state generated by our single-photon source by measuring the fringe visibility for different values of the excitation power, as also suggested in \cite{loredo2019generation}. 
Indeed, it is necessary to consider the generation of a mixed state of the form:

\begin{equation*}
    \rho = \lambda \rho_{\text{pure}} + (1-\lambda) \rho_{\text{mixed}}
\end{equation*}

Here $\rho_{\text{pure}} = \ket{\psi}\bra{\psi}$ is a pure state, analogous to the one in Eq.~\eqref{eq:generic_qubit} and encoded in the photon-number basis, while 
$ \rho_{\text{mixed}} = \text{diag} \{ |\alpha|^2, |\beta|^2  \} $ 
is a mixture of the vacuum and one-photon populations.
As demonstrated in \cite{loredo2019generation}, one can extract the value of $\lambda$ by observing the dependence of the fringe visibility $V$ from the vacuum population in a self-homodyne detection. {For a more detailed discussion about the estimation of the purity $\lambda$ conditioned on detection events, see Supplementary Note~I.}
In our apparatus, $\lambda$ is retrieved from the slope of the linear fit performed on the data in the inset of Fig. \ref{fig:alpha_data}a. It amounts to $\lambda \approx 0.98$.

\subsection*{Characterization of the probe state} 
We characterize the probe state by using an independent procedure with respect to the protocol implementation. In detail, we observe the interference of two same probe states at the output of a standard MZI (see Fig.~\ref{fig:expscheme_teleportation}c) in Alice's station. The $\alpha^2$ coefficient is then computed from the measured fringe visibility by inverting the following formula:
\begin{equation*}
    V =  \lambda^2 \sqrt{V_{\text{HOM}}^{\text{Alice}}} \alpha^2,
\end{equation*}
where $\lambda$ is the state {conditional} purity. We report its full formal derivation in Supplementary Note~I.

\subsection*{Characterization of the teleported state and experimental imperfections} 
We characterize the teleported state by observing the fringe visibility at the output of Bob's last BS. In the hypothesis of correct teleportation, which means that the state $\ket{\psi}_{\text{probe}}$ and the teleported state have the same vacuum population $\alpha^2$, we expect that $V_T = V$. The imperfections of the experimental apparatus, such as the use of threshold detectors and of probabilistic routing of the states $\ket{\psi}$  and $\ket{\psi}_{\text{probe}}$, limit the value of the measured $V_T$, such that $V_T < V$. By taking into account such limitations and considering the other imperfections summarized by $\lambda$, $V_{\text{HOM}}^{\text{Alice}}$ and $V_{\text{HOM}}^{\text{Bob}}$, we derive the following formula for the visibility of the teleported state as a function of $V$:
\begin{equation*}
    V_T = \frac{2 \lambda^2 \sqrt{V_{\text{HOM}}^{\text{Alice}}V_{\text{HOM}}^{\text{Bob}}} V }{3\lambda^2 \sqrt{V_{\text{HOM}}^{\text{Alice}}} - V}
\end{equation*}
The full derivation of such an equation is reported in Supplementary Note~{IV}. The calculation has been carried out in the regime of high losses as for the derivation of $V$.

\subsection*{Data analysis}
We derived a procedure that improves significantly the accuracy of the measured visibility, even in the case of low event statistics, that we briefly discuss below. For details on the derivation we refer to Supplementary Note~{VII}. We indicate with $n(\phi(t))$ a single-photon count trace, where $\phi(t)$ is the freely evolving phase in a MZI. The visibility can be evaluated as:
\begin{equation*}
    V = \sqrt{2 \frac{\langle n^2\rangle _{t} - \langle n \rangle ^2_{t} - \langle n \rangle_{t}}{{\langle n \rangle_{t}^2}}}
\end{equation*}
where $\braket{\cdot}_t$ is the time average, and for brevity we have omitted in the notation the dependence of $n$ on $\phi(t)$. The measured single-photon count traces for the teleportation (Fig. \ref{fig:alpha_data}b) and entanglement swapping (Fig. \ref{fig:swapping}) were recorded for a time of $\sim 10^5$ bins of $50$ ms that is around 1 hour. The average number $\braket{n}$ of photons in each time bin ranges from 5 to 100 depending on the vacuum population.

\section*{Acknowledgements}

We thank \textit{Quandela} 
for all the support provided. We also acknowledge support from the ERC Advanced Grant QU-BOSS (QUantum advantage via nonlinear BOSon Sampling, grant agreement no. 884676) and from PNRR MUR project
PE0000023-NQSTI (Spoke 4). \\

\section*{Author contributions}

B.P., F.H., G.R., T..G., and F.S. conceived the experiment. B.P., G.R., F.H.,  S.S., T..G.  carried out the experiment. F.H., B.P., G.R., S.S., G.C., N.S., T.G. and F.S.  performed the data analysis. All the authors discussed the results and contributed to the writing of the paper.

\section*{Competing Interests}
The authors declare no competing interests.

\section*{Data availability}
The data that support the findings of this study are available from the corresponding author upon reasonable request.

%\bibliography{biblio.bib}

\begin{thebibliography}{59}%
	\makeatletter
	\providecommand \@ifxundefined [1]{%
		\@ifx{#1\undefined}
	}%
	\providecommand \@ifnum [1]{%
		\ifnum #1\expandafter \@firstoftwo
		\else \expandafter \@secondoftwo
		\fi
	}%
	\providecommand \@ifx [1]{%
		\ifx #1\expandafter \@firstoftwo
		\else \expandafter \@secondoftwo
		\fi
	}%
	\providecommand \natexlab [1]{#1}%
	\providecommand \enquote  [1]{``#1''}%
	\providecommand \bibnamefont  [1]{#1}%
	\providecommand \bibfnamefont [1]{#1}%
	\providecommand \citenamefont [1]{#1}%
	\providecommand \href@noop [0]{\@secondoftwo}%
	\providecommand \href [0]{\begingroup \@sanitize@url \@href}%
	\providecommand \@href[1]{\@@startlink{#1}\@@href}%
	\providecommand \@@href[1]{\endgroup#1\@@endlink}%
	\providecommand \@sanitize@url [0]{\catcode `\\12\catcode `\$12\catcode
		`\&12\catcode `\#12\catcode `\^12\catcode `\_12\catcode `\%12\relax}%
	\providecommand \@@startlink[1]{}%
	\providecommand \@@endlink[0]{}%
	\providecommand \url  [0]{\begingroup\@sanitize@url \@url }%
	\providecommand \@url [1]{\endgroup\@href {#1}{\urlprefix }}%
	\providecommand \urlprefix  [0]{URL }%
	\providecommand \Eprint [0]{\href }%
	\providecommand \doibase [0]{https://doi.org/}%
	\providecommand \selectlanguage [0]{\@gobble}%
	\providecommand \bibinfo  [0]{\@secondoftwo}%
	\providecommand \bibfield  [0]{\@secondoftwo}%
	\providecommand \translation [1]{[#1]}%
	\providecommand \BibitemOpen [0]{}%
	\providecommand \bibitemStop [0]{}%
	\providecommand \bibitemNoStop [0]{.\EOS\space}%
	\providecommand \EOS [0]{\spacefactor3000\relax}%
	\providecommand \BibitemShut  [1]{\csname bibitem#1\endcsname}%
	\let\auto@bib@innerbib\@empty
	%</preamble>
	\bibitem [{\citenamefont {Bennett}\ \emph {et~al.}(1993)\citenamefont
		{Bennett}, \citenamefont {Brassard}, \citenamefont {Cr{\'e}peau},
		\citenamefont {Jozsa}, \citenamefont {Peres},\ and\ \citenamefont
		{Wootters}}]{bennett1993teleporting}%
	\BibitemOpen
	\bibfield  {author} {\bibinfo {author} {\bibfnamefont {C.~H.}\ \bibnamefont
			{Bennett}}, \bibinfo {author} {\bibfnamefont {G.}~\bibnamefont {Brassard}},
		\bibinfo {author} {\bibfnamefont {C.}~\bibnamefont {Cr{\'e}peau}}, \bibinfo
		{author} {\bibfnamefont {R.}~\bibnamefont {Jozsa}}, \bibinfo {author}
		{\bibfnamefont {A.}~\bibnamefont {Peres}},\ and\ \bibinfo {author}
		{\bibfnamefont {W.~K.}\ \bibnamefont {Wootters}},\ }\bibfield  {title}
	{\bibinfo {title} {Teleporting an unknown quantum state via dual classical
			and einstein-podolsky-rosen channels},\ }\href
	{https://doi.org/10.1103/PhysRevLett.70.1895} {\bibfield  {journal} {\bibinfo
			{journal} {Phys. Rev. Lett.}\ }\textbf {\bibinfo {volume} {70}},\ \bibinfo
		{pages} {1895} (\bibinfo {year} {1993})}\BibitemShut {NoStop}%
	\bibitem [{\citenamefont {Bouwmeester}\ \emph {et~al.}(1997)\citenamefont
		{Bouwmeester}, \citenamefont {Pan}, \citenamefont {Mattle}, \citenamefont
		{Eibl}, \citenamefont {Weinfurter},\ and\ \citenamefont
		{Zeilinger}}]{exptele}%
	\BibitemOpen
	\bibfield  {author} {\bibinfo {author} {\bibfnamefont {D.}~\bibnamefont
			{Bouwmeester}}, \bibinfo {author} {\bibfnamefont {J.-W.}\ \bibnamefont
			{Pan}}, \bibinfo {author} {\bibfnamefont {K.}~\bibnamefont {Mattle}},
		\bibinfo {author} {\bibfnamefont {M.}~\bibnamefont {Eibl}}, \bibinfo {author}
		{\bibfnamefont {H.}~\bibnamefont {Weinfurter}},\ and\ \bibinfo {author}
		{\bibfnamefont {A.}~\bibnamefont {Zeilinger}},\ }\bibfield  {title} {\bibinfo
		{title} {Experimental quantum teleportation},\ }\href
	{https://doi.org/10.1038/37539} {\bibfield  {journal} {\bibinfo  {journal}
			{Nature}\ }\textbf {\bibinfo {volume} {390}},\ \bibinfo {pages} {575}
		(\bibinfo {year} {1997})}\BibitemShut {NoStop}%
	\bibitem [{\citenamefont {Boschi}\ \emph {et~al.}(1998)\citenamefont {Boschi},
		\citenamefont {Branca}, \citenamefont {De~Martini}, \citenamefont {Hardy},\
		and\ \citenamefont {Popescu}}]{boschi1998experimental}%
	\BibitemOpen
	\bibfield  {author} {\bibinfo {author} {\bibfnamefont {D.}~\bibnamefont
			{Boschi}}, \bibinfo {author} {\bibfnamefont {S.}~\bibnamefont {Branca}},
		\bibinfo {author} {\bibfnamefont {F.}~\bibnamefont {De~Martini}}, \bibinfo
		{author} {\bibfnamefont {L.}~\bibnamefont {Hardy}},\ and\ \bibinfo {author}
		{\bibfnamefont {S.}~\bibnamefont {Popescu}},\ }\bibfield  {title} {\bibinfo
		{title} {Experimental realization of teleporting an unknown pure quantum
			state via dual classical and einstein-podolsky-rosen channels},\ }\href
	{https://doi.org/10.1103/PhysRevLett.80.1121} {\bibfield  {journal} {\bibinfo
			{journal} {Phys. Rev. Lett.}\ }\textbf {\bibinfo {volume} {80}},\ \bibinfo
		{pages} {1121} (\bibinfo {year} {1998})}\BibitemShut {NoStop}%
	\bibitem [{\citenamefont {\ifmmode~\dot{Z}\else \.{Z}\fi{}ukowski}\ \emph
		{et~al.}(1993)\citenamefont {\ifmmode~\dot{Z}\else \.{Z}\fi{}ukowski},
		\citenamefont {Zeilinger}, \citenamefont {Horne},\ and\ \citenamefont
		{Ekert}}]{swapping}%
	\BibitemOpen
	\bibfield  {author} {\bibinfo {author} {\bibfnamefont {M.}~\bibnamefont
			{\ifmmode~\dot{Z}\else \.{Z}\fi{}ukowski}}, \bibinfo {author} {\bibfnamefont
			{A.}~\bibnamefont {Zeilinger}}, \bibinfo {author} {\bibfnamefont {M.~A.}\
			\bibnamefont {Horne}},\ and\ \bibinfo {author} {\bibfnamefont {A.~K.}\
			\bibnamefont {Ekert}},\ }\bibfield  {title} {\bibinfo {title}
		{``event-ready-detectors'' bell experiment via entanglement swapping},\
	}\href {https://doi.org/10.1103/PhysRevLett.71.4287} {\bibfield  {journal}
		{\bibinfo  {journal} {Phys. Rev. Lett.}\ }\textbf {\bibinfo {volume} {71}},\
		\bibinfo {pages} {4287} (\bibinfo {year} {1993})}\BibitemShut {NoStop}%
	\bibitem [{\citenamefont {Pan}\ \emph {et~al.}(1998)\citenamefont {Pan},
		\citenamefont {Bouwmeester}, \citenamefont {Weinfurter},\ and\ \citenamefont
		{Zeilinger}}]{PhysRevLett.80.3891}%
	\BibitemOpen
	\bibfield  {author} {\bibinfo {author} {\bibfnamefont {J.-W.}\ \bibnamefont
			{Pan}}, \bibinfo {author} {\bibfnamefont {D.}~\bibnamefont {Bouwmeester}},
		\bibinfo {author} {\bibfnamefont {H.}~\bibnamefont {Weinfurter}},\ and\
		\bibinfo {author} {\bibfnamefont {A.}~\bibnamefont {Zeilinger}},\ }\bibfield
	{title} {\bibinfo {title} {Experimental entanglement swapping: Entangling
			photons that never interacted},\ }\href
	{https://doi.org/10.1103/PhysRevLett.80.3891} {\bibfield  {journal} {\bibinfo
			{journal} {Phys. Rev. Lett.}\ }\textbf {\bibinfo {volume} {80}},\ \bibinfo
		{pages} {3891} (\bibinfo {year} {1998})}\BibitemShut {NoStop}%
	\bibitem [{\citenamefont {Gottesman}\ and\ \citenamefont
		{Chuang}(1999)}]{Gottesman1999}%
	\BibitemOpen
	\bibfield  {author} {\bibinfo {author} {\bibfnamefont {D.}~\bibnamefont
			{Gottesman}}\ and\ \bibinfo {author} {\bibfnamefont {I.~L.}\ \bibnamefont
			{Chuang}},\ }\bibfield  {title} {\bibinfo {title} {Demonstrating the
			viability of universal quantum computation using teleportation and
			single-qubit operations},\ }\href {https://doi.org/10.1038/46503} {\bibfield
		{journal} {\bibinfo  {journal} {Nature}\ }\textbf {\bibinfo {volume} {402}},\
		\bibinfo {pages} {390} (\bibinfo {year} {1999})}\BibitemShut {NoStop}%
	\bibitem [{\citenamefont {Knill}\ \emph {et~al.}(2001)\citenamefont {Knill},
		\citenamefont {Laflamme},\ and\ \citenamefont {Milburn}}]{Knill2001}%
	\BibitemOpen
	\bibfield  {author} {\bibinfo {author} {\bibfnamefont {E.}~\bibnamefont
			{Knill}}, \bibinfo {author} {\bibfnamefont {R.}~\bibnamefont {Laflamme}},\
		and\ \bibinfo {author} {\bibfnamefont {G.~J.}\ \bibnamefont {Milburn}},\
	}\bibfield  {title} {\bibinfo {title} {A scheme for efficient quantum
			computation with linear optics},\ }\href {https://doi.org/10.1038/35051009}
	{\bibfield  {journal} {\bibinfo  {journal} {Nature}\ }\textbf {\bibinfo
			{volume} {409}},\ \bibinfo {pages} {46} (\bibinfo {year} {2001})}\BibitemShut
	{NoStop}%
	\bibitem [{\citenamefont {Briegel}\ \emph {et~al.}(1998)\citenamefont
		{Briegel}, \citenamefont {D\"ur}, \citenamefont {Cirac},\ and\ \citenamefont
		{Zoller}}]{quantumrepeater1}%
	\BibitemOpen
	\bibfield  {author} {\bibinfo {author} {\bibfnamefont {H.-J.}\ \bibnamefont
			{Briegel}}, \bibinfo {author} {\bibfnamefont {W.}~\bibnamefont {D\"ur}},
		\bibinfo {author} {\bibfnamefont {J.~I.}\ \bibnamefont {Cirac}},\ and\
		\bibinfo {author} {\bibfnamefont {P.}~\bibnamefont {Zoller}},\ }\bibfield
	{title} {\bibinfo {title} {Quantum repeaters: The role of imperfect local
			operations in quantum communication},\ }\href
	{https://doi.org/10.1103/PhysRevLett.81.5932} {\bibfield  {journal} {\bibinfo
			{journal} {Phys. Rev. Lett.}\ }\textbf {\bibinfo {volume} {81}},\ \bibinfo
		{pages} {5932} (\bibinfo {year} {1998})}\BibitemShut {NoStop}%
	\bibitem [{\citenamefont {D\"ur}\ \emph {et~al.}(1999)\citenamefont {D\"ur},
		\citenamefont {Briegel}, \citenamefont {Cirac},\ and\ \citenamefont
		{Zoller}}]{Repeaters_swapping}%
	\BibitemOpen
	\bibfield  {author} {\bibinfo {author} {\bibfnamefont {W.}~\bibnamefont
			{D\"ur}}, \bibinfo {author} {\bibfnamefont {H.-J.}\ \bibnamefont {Briegel}},
		\bibinfo {author} {\bibfnamefont {J.~I.}\ \bibnamefont {Cirac}},\ and\
		\bibinfo {author} {\bibfnamefont {P.}~\bibnamefont {Zoller}},\ }\bibfield
	{title} {\bibinfo {title} {Quantum repeaters based on entanglement
			purification},\ }\href {https://doi.org/10.1103/PhysRevA.59.169} {\bibfield
		{journal} {\bibinfo  {journal} {Phys. Rev. A}\ }\textbf {\bibinfo {volume}
			{59}},\ \bibinfo {pages} {169} (\bibinfo {year} {1999})}\BibitemShut
	{NoStop}%
	\bibitem [{\citenamefont {Bartlett}\ and\ \citenamefont
		{Munro}(2003)}]{PhysRevLett.90.117901}%
	\BibitemOpen
	\bibfield  {author} {\bibinfo {author} {\bibfnamefont {S.~D.}\ \bibnamefont
			{Bartlett}}\ and\ \bibinfo {author} {\bibfnamefont {W.~J.}\ \bibnamefont
			{Munro}},\ }\bibfield  {title} {\bibinfo {title} {Quantum teleportation of
			optical quantum gates},\ }\href
	{https://doi.org/10.1103/PhysRevLett.90.117901} {\bibfield  {journal}
		{\bibinfo  {journal} {Phys. Rev. Lett.}\ }\textbf {\bibinfo {volume} {90}},\
		\bibinfo {pages} {117901} (\bibinfo {year} {2003})}\BibitemShut {NoStop}%
	\bibitem [{\citenamefont {Chou}\ \emph {et~al.}(2018)\citenamefont {Chou},
		\citenamefont {Blumoff}, \citenamefont {Wang}, \citenamefont {Reinhold},
		\citenamefont {Axline}, \citenamefont {Gao}, \citenamefont {Frunzio},
		\citenamefont {Devoret}, \citenamefont {Jiang},\ and\ \citenamefont
		{Schoelkopf}}]{Chou2018}%
	\BibitemOpen
	\bibfield  {author} {\bibinfo {author} {\bibfnamefont {K.~S.}\ \bibnamefont
			{Chou}}, \bibinfo {author} {\bibfnamefont {J.~Z.}\ \bibnamefont {Blumoff}},
		\bibinfo {author} {\bibfnamefont {C.~S.}\ \bibnamefont {Wang}}, \bibinfo
		{author} {\bibfnamefont {P.~C.}\ \bibnamefont {Reinhold}}, \bibinfo {author}
		{\bibfnamefont {C.~J.}\ \bibnamefont {Axline}}, \bibinfo {author}
		{\bibfnamefont {Y.~Y.}\ \bibnamefont {Gao}}, \bibinfo {author} {\bibfnamefont
			{L.}~\bibnamefont {Frunzio}}, \bibinfo {author} {\bibfnamefont {M.~H.}\
			\bibnamefont {Devoret}}, \bibinfo {author} {\bibfnamefont {L.}~\bibnamefont
			{Jiang}},\ and\ \bibinfo {author} {\bibfnamefont {R.~J.}\ \bibnamefont
			{Schoelkopf}},\ }\bibfield  {title} {\bibinfo {title} {Deterministic
			teleportation of a quantum gate between two logical qubits},\ }\href
	{https://doi.org/10.1038/s41586-018-0470-y} {\bibfield  {journal} {\bibinfo
			{journal} {Nature}\ }\textbf {\bibinfo {volume} {561}},\ \bibinfo {pages}
		{368} (\bibinfo {year} {2018})}\BibitemShut {NoStop}%
	\bibitem [{\citenamefont {Wan}\ \emph {et~al.}(2019)\citenamefont {Wan},
		\citenamefont {Kienzler}, \citenamefont {Erickson}, \citenamefont {Mayer},
		\citenamefont {Tan}, \citenamefont {Wu}, \citenamefont {Vasconcelos},
		\citenamefont {Glancy}, \citenamefont {Knill}, \citenamefont {Wineland},
		\citenamefont {Wilson},\ and\ \citenamefont {Leibfried}}]{Wan2019}%
	\BibitemOpen
	\bibfield  {author} {\bibinfo {author} {\bibfnamefont {Y.}~\bibnamefont
			{Wan}}, \bibinfo {author} {\bibfnamefont {D.}~\bibnamefont {Kienzler}},
		\bibinfo {author} {\bibfnamefont {S.~D.}\ \bibnamefont {Erickson}}, \bibinfo
		{author} {\bibfnamefont {K.~H.}\ \bibnamefont {Mayer}}, \bibinfo {author}
		{\bibfnamefont {T.~R.}\ \bibnamefont {Tan}}, \bibinfo {author} {\bibfnamefont
			{J.~J.}\ \bibnamefont {Wu}}, \bibinfo {author} {\bibfnamefont {H.~M.}\
			\bibnamefont {Vasconcelos}}, \bibinfo {author} {\bibfnamefont
			{S.}~\bibnamefont {Glancy}}, \bibinfo {author} {\bibfnamefont
			{E.}~\bibnamefont {Knill}}, \bibinfo {author} {\bibfnamefont {D.~J.}\
			\bibnamefont {Wineland}}, \bibinfo {author} {\bibfnamefont {A.~C.}\
			\bibnamefont {Wilson}},\ and\ \bibinfo {author} {\bibfnamefont
			{D.}~\bibnamefont {Leibfried}},\ }\bibfield  {title} {\bibinfo {title}
		{Quantum gate teleportation between separated qubits in a trapped-ion
			processor},\ }\href {https://doi.org/10.1126/science.aaw9415} {\bibfield
		{journal} {\bibinfo  {journal} {Science}\ }\textbf {\bibinfo {volume}
			{364}},\ \bibinfo {pages} {875} (\bibinfo {year} {2019})}\BibitemShut
	{NoStop}%
	\bibitem [{\citenamefont {Raussendorf}\ \emph {et~al.}(2003)\citenamefont
		{Raussendorf}, \citenamefont {Browne},\ and\ \citenamefont
		{Briegel}}]{PhysRevA.68.022312}%
	\BibitemOpen
	\bibfield  {author} {\bibinfo {author} {\bibfnamefont {R.}~\bibnamefont
			{Raussendorf}}, \bibinfo {author} {\bibfnamefont {D.~E.}\ \bibnamefont
			{Browne}},\ and\ \bibinfo {author} {\bibfnamefont {H.~J.}\ \bibnamefont
			{Briegel}},\ }\bibfield  {title} {\bibinfo {title} {Measurement-based quantum
			computation on cluster states},\ }\href
	{https://doi.org/10.1103/PhysRevA.68.022312} {\bibfield  {journal} {\bibinfo
			{journal} {Phys. Rev. A}\ }\textbf {\bibinfo {volume} {68}},\ \bibinfo
		{pages} {022312} (\bibinfo {year} {2003})}\BibitemShut {NoStop}%
	\bibitem [{\citenamefont {Gross}\ and\ \citenamefont
		{Eisert}(2007)}]{PhysRevLett.98.220503}%
	\BibitemOpen
	\bibfield  {author} {\bibinfo {author} {\bibfnamefont {D.}~\bibnamefont
			{Gross}}\ and\ \bibinfo {author} {\bibfnamefont {J.}~\bibnamefont {Eisert}},\
	}\bibfield  {title} {\bibinfo {title} {Novel schemes for measurement-based
			quantum computation},\ }\href {https://doi.org/10.1103/PhysRevLett.98.220503}
	{\bibfield  {journal} {\bibinfo  {journal} {Phys. Rev. Lett.}\ }\textbf
		{\bibinfo {volume} {98}},\ \bibinfo {pages} {220503} (\bibinfo {year}
		{2007})}\BibitemShut {NoStop}%
	\bibitem [{\citenamefont {Briegel}\ \emph {et~al.}(2009)\citenamefont
		{Briegel}, \citenamefont {Browne}, \citenamefont {D{\"u}r}, \citenamefont
		{Raussendorf},\ and\ \citenamefont {Van~den Nest}}]{Briegel2009}%
	\BibitemOpen
	\bibfield  {author} {\bibinfo {author} {\bibfnamefont {H.~J.}\ \bibnamefont
			{Briegel}}, \bibinfo {author} {\bibfnamefont {D.~E.}\ \bibnamefont {Browne}},
		\bibinfo {author} {\bibfnamefont {W.}~\bibnamefont {D{\"u}r}}, \bibinfo
		{author} {\bibfnamefont {R.}~\bibnamefont {Raussendorf}},\ and\ \bibinfo
		{author} {\bibfnamefont {M.}~\bibnamefont {Van~den Nest}},\ }\bibfield
	{title} {\bibinfo {title} {Measurement-based quantum computation},\ }\href
	{https://doi.org/10.1038/nphys1157} {\bibfield  {journal} {\bibinfo
			{journal} {Nature Physics}\ }\textbf {\bibinfo {volume} {5}},\ \bibinfo
		{pages} {19} (\bibinfo {year} {2009})}\BibitemShut {NoStop}%
	\bibitem [{\citenamefont {Studzi{\'{n}}ski}\ \emph {et~al.}(2017)\citenamefont
		{Studzi{\'{n}}ski}, \citenamefont {Strelchuk}, \citenamefont {Mozrzymas},\
		and\ \citenamefont {Horodecki}}]{Studziński2017}%
	\BibitemOpen
	\bibfield  {author} {\bibinfo {author} {\bibfnamefont {M.}~\bibnamefont
			{Studzi{\'{n}}ski}}, \bibinfo {author} {\bibfnamefont {S.}~\bibnamefont
			{Strelchuk}}, \bibinfo {author} {\bibfnamefont {M.}~\bibnamefont
			{Mozrzymas}},\ and\ \bibinfo {author} {\bibfnamefont {M.}~\bibnamefont
			{Horodecki}},\ }\bibfield  {title} {\bibinfo {title} {Port-based
			teleportation in arbitrary dimension},\ }\href
	{https://doi.org/10.1038/s41598-017-10051-4} {\bibfield  {journal} {\bibinfo
			{journal} {Scientific Reports}\ }\textbf {\bibinfo {volume} {7}},\ \bibinfo
		{pages} {10871} (\bibinfo {year} {2017})}\BibitemShut {NoStop}%
	\bibitem [{\citenamefont {Jeong}\ \emph {et~al.}(2020)\citenamefont {Jeong},
		\citenamefont {Kim},\ and\ \citenamefont {Lee}}]{PhysRevA.102.012414}%
	\BibitemOpen
	\bibfield  {author} {\bibinfo {author} {\bibfnamefont {K.}~\bibnamefont
			{Jeong}}, \bibinfo {author} {\bibfnamefont {J.}~\bibnamefont {Kim}},\ and\
		\bibinfo {author} {\bibfnamefont {S.}~\bibnamefont {Lee}},\ }\bibfield
	{title} {\bibinfo {title} {Generalization of port-based teleportation and
			controlled teleportation capability},\ }\href
	{https://doi.org/10.1103/PhysRevA.102.012414} {\bibfield  {journal} {\bibinfo
			{journal} {Phys. Rev. A}\ }\textbf {\bibinfo {volume} {102}},\ \bibinfo
		{pages} {012414} (\bibinfo {year} {2020})}\BibitemShut {NoStop}%
	\bibitem [{\citenamefont {Pirandola}\ \emph
		{et~al.}(2015{\natexlab{a}})\citenamefont {Pirandola}, \citenamefont
		{Eisert}, \citenamefont {Weedbrook}, \citenamefont {Furusawa},\ and\
		\citenamefont {Braunstein}}]{pirandola2015advances}%
	\BibitemOpen
	\bibfield  {author} {\bibinfo {author} {\bibfnamefont {S.}~\bibnamefont
			{Pirandola}}, \bibinfo {author} {\bibfnamefont {J.}~\bibnamefont {Eisert}},
		\bibinfo {author} {\bibfnamefont {C.}~\bibnamefont {Weedbrook}}, \bibinfo
		{author} {\bibfnamefont {A.}~\bibnamefont {Furusawa}},\ and\ \bibinfo
		{author} {\bibfnamefont {S.~L.}\ \bibnamefont {Braunstein}},\ }\bibfield
	{title} {\bibinfo {title} {Advances in quantum teleportation},\ }\href
	{https://doi.org/10.1038/nphoton.2015.154} {\bibfield  {journal} {\bibinfo
			{journal} {Nature Photonics}\ }\textbf {\bibinfo {volume} {9}},\ \bibinfo
		{pages} {641} (\bibinfo {year} {2015}{\natexlab{a}})}\BibitemShut {NoStop}%
	\bibitem [{\citenamefont {Hu}\ \emph {et~al.}(2023)\citenamefont {Hu},
		\citenamefont {Guo}, \citenamefont {Liu}, \citenamefont {Li},\ and\
		\citenamefont {Guo}}]{hu2023progress}%
	\BibitemOpen
	\bibfield  {author} {\bibinfo {author} {\bibfnamefont {X.-M.}\ \bibnamefont
			{Hu}}, \bibinfo {author} {\bibfnamefont {Y.}~\bibnamefont {Guo}}, \bibinfo
		{author} {\bibfnamefont {B.-H.}\ \bibnamefont {Liu}}, \bibinfo {author}
		{\bibfnamefont {C.-F.}\ \bibnamefont {Li}},\ and\ \bibinfo {author}
		{\bibfnamefont {G.-C.}\ \bibnamefont {Guo}},\ }\bibfield  {title} {\bibinfo
		{title} {Progress in quantum teleportation},\ }\href
	{https://doi.org/10.1038/s42254-023-00588-x} {\bibfield  {journal} {\bibinfo
			{journal} {Nature Reviews Physics}\ }\textbf {\bibinfo {volume} {5}},\
		\bibinfo {pages} {339} (\bibinfo {year} {2023})}\BibitemShut {NoStop}%
	\bibitem [{\citenamefont {Marcikic}\ \emph {et~al.}(2003)\citenamefont
		{Marcikic}, \citenamefont {De~Riedmatten}, \citenamefont {Tittel},
		\citenamefont {Zbinden},\ and\ \citenamefont {Gisin}}]{marcikic2003long}%
	\BibitemOpen
	\bibfield  {author} {\bibinfo {author} {\bibfnamefont {I.}~\bibnamefont
			{Marcikic}}, \bibinfo {author} {\bibfnamefont {H.}~\bibnamefont
			{De~Riedmatten}}, \bibinfo {author} {\bibfnamefont {W.}~\bibnamefont
			{Tittel}}, \bibinfo {author} {\bibfnamefont {H.}~\bibnamefont {Zbinden}},\
		and\ \bibinfo {author} {\bibfnamefont {N.}~\bibnamefont {Gisin}},\ }\bibfield
	{title} {\bibinfo {title} {Long-distance teleportation of qubits at
			telecommunication wavelengths},\ }\href {https://doi.org/10.1038/nature01376}
	{\bibfield  {journal} {\bibinfo  {journal} {Nature}\ }\textbf {\bibinfo
			{volume} {421}},\ \bibinfo {pages} {509} (\bibinfo {year}
		{2003})}\BibitemShut {NoStop}%
	\bibitem [{\citenamefont {Lombardi}\ \emph {et~al.}(2002)\citenamefont
		{Lombardi}, \citenamefont {Sciarrino}, \citenamefont {Popescu},\ and\
		\citenamefont {De~Martini}}]{lombardi2002teleportation}%
	\BibitemOpen
	\bibfield  {author} {\bibinfo {author} {\bibfnamefont {E.}~\bibnamefont
			{Lombardi}}, \bibinfo {author} {\bibfnamefont {F.}~\bibnamefont {Sciarrino}},
		\bibinfo {author} {\bibfnamefont {S.}~\bibnamefont {Popescu}},\ and\ \bibinfo
		{author} {\bibfnamefont {F.}~\bibnamefont {De~Martini}},\ }\bibfield  {title}
	{\bibinfo {title} {Teleportation of a vacuum--one-photon qubit},\ }\href
	{https://doi.org/10.1103/PhysRevLett.88.070402} {\bibfield  {journal}
		{\bibinfo  {journal} {Phys. Rev. Lett.}\ }\textbf {\bibinfo {volume} {88}},\
		\bibinfo {pages} {070402} (\bibinfo {year} {2002})}\BibitemShut {NoStop}%
	\bibitem [{\citenamefont {Furusawa}\ \emph {et~al.}(1998)\citenamefont
		{Furusawa}, \citenamefont {S{\o}rensen}, \citenamefont {Braunstein},
		\citenamefont {Fuchs}, \citenamefont {Kimble},\ and\ \citenamefont
		{Polzik}}]{furusawa1998unconditional}%
	\BibitemOpen
	\bibfield  {author} {\bibinfo {author} {\bibfnamefont {A.}~\bibnamefont
			{Furusawa}}, \bibinfo {author} {\bibfnamefont {J.~L.}\ \bibnamefont
			{S{\o}rensen}}, \bibinfo {author} {\bibfnamefont {S.~L.}\ \bibnamefont
			{Braunstein}}, \bibinfo {author} {\bibfnamefont {C.~A.}\ \bibnamefont
			{Fuchs}}, \bibinfo {author} {\bibfnamefont {H.~J.}\ \bibnamefont {Kimble}},\
		and\ \bibinfo {author} {\bibfnamefont {E.~S.}\ \bibnamefont {Polzik}},\
	}\bibfield  {title} {\bibinfo {title} {Unconditional quantum teleportation},\
	}\href {https://doi.org/10.1126/science.282.5389.706} {\bibfield  {journal}
		{\bibinfo  {journal} {Science}\ }\textbf {\bibinfo {volume} {282}},\ \bibinfo
		{pages} {706} (\bibinfo {year} {1998})}\BibitemShut {NoStop}%
	\bibitem [{\citenamefont {Giacomini}\ \emph {et~al.}(2002)\citenamefont
		{Giacomini}, \citenamefont {Sciarrino}, \citenamefont {Lombardi},\ and\
		\citenamefont {De~Martini}}]{giacomini2002active}%
	\BibitemOpen
	\bibfield  {author} {\bibinfo {author} {\bibfnamefont {S.}~\bibnamefont
			{Giacomini}}, \bibinfo {author} {\bibfnamefont {F.}~\bibnamefont
			{Sciarrino}}, \bibinfo {author} {\bibfnamefont {E.}~\bibnamefont
			{Lombardi}},\ and\ \bibinfo {author} {\bibfnamefont {F.}~\bibnamefont
			{De~Martini}},\ }\bibfield  {title} {\bibinfo {title} {Active teleportation
			of a quantum bit},\ }\href {https://doi.org/10.1103/PhysRevA.66.030302}
	{\bibfield  {journal} {\bibinfo  {journal} {Phys. Rev. A}\ }\textbf {\bibinfo
			{volume} {66}},\ \bibinfo {pages} {030302} (\bibinfo {year}
		{2002})}\BibitemShut {NoStop}%
	\bibitem [{\citenamefont {Kim}\ \emph {et~al.}(2001)\citenamefont {Kim},
		\citenamefont {Kulik},\ and\ \citenamefont {Shih}}]{PhysRevLett.86.1370}%
	\BibitemOpen
	\bibfield  {author} {\bibinfo {author} {\bibfnamefont {Y.-H.}\ \bibnamefont
			{Kim}}, \bibinfo {author} {\bibfnamefont {S.~P.}\ \bibnamefont {Kulik}},\
		and\ \bibinfo {author} {\bibfnamefont {Y.}~\bibnamefont {Shih}},\ }\bibfield
	{title} {\bibinfo {title} {Quantum teleportation of a polarization state with
			a complete bell state measurement},\ }\href
	{https://doi.org/10.1103/PhysRevLett.86.1370} {\bibfield  {journal} {\bibinfo
			{journal} {Phys. Rev. Lett.}\ }\textbf {\bibinfo {volume} {86}},\ \bibinfo
		{pages} {1370} (\bibinfo {year} {2001})}\BibitemShut {NoStop}%
	\bibitem [{\citenamefont {Fattal}\ \emph {et~al.}(2004)\citenamefont {Fattal},
		\citenamefont {Diamanti}, \citenamefont {Inoue},\ and\ \citenamefont
		{Yamamoto}}]{fattal2004quantum}%
	\BibitemOpen
	\bibfield  {author} {\bibinfo {author} {\bibfnamefont {D.}~\bibnamefont
			{Fattal}}, \bibinfo {author} {\bibfnamefont {E.}~\bibnamefont {Diamanti}},
		\bibinfo {author} {\bibfnamefont {K.}~\bibnamefont {Inoue}},\ and\ \bibinfo
		{author} {\bibfnamefont {Y.}~\bibnamefont {Yamamoto}},\ }\bibfield  {title}
	{\bibinfo {title} {Quantum teleportation with a quantum dot single photon
			source},\ }\href {https://doi.org/10.1103/PhysRevLett.92.037904} {\bibfield
		{journal} {\bibinfo  {journal} {Phys. Rev. Lett.}\ }\textbf {\bibinfo
			{volume} {92}},\ \bibinfo {pages} {037904} (\bibinfo {year}
		{2004})}\BibitemShut {NoStop}%
	\bibitem [{\citenamefont {Pirandola}\ \emph
		{et~al.}(2015{\natexlab{b}})\citenamefont {Pirandola}, \citenamefont
		{Eisert}, \citenamefont {Weedbrook}, \citenamefont {Furusawa},\ and\
		\citenamefont {Braunstein}}]{Pirandola2015}%
	\BibitemOpen
	\bibfield  {author} {\bibinfo {author} {\bibfnamefont {S.}~\bibnamefont
			{Pirandola}}, \bibinfo {author} {\bibfnamefont {J.}~\bibnamefont {Eisert}},
		\bibinfo {author} {\bibfnamefont {C.}~\bibnamefont {Weedbrook}}, \bibinfo
		{author} {\bibfnamefont {A.}~\bibnamefont {Furusawa}},\ and\ \bibinfo
		{author} {\bibfnamefont {S.~L.}\ \bibnamefont {Braunstein}},\ }\bibfield
	{title} {\bibinfo {title} {Advances in quantum teleportation},\ }\href
	{https://doi.org/10.1038/nphoton.2015.154} {\bibfield  {journal} {\bibinfo
			{journal} {Nature Photonics}\ }\textbf {\bibinfo {volume} {9}},\ \bibinfo
		{pages} {641} (\bibinfo {year} {2015}{\natexlab{b}})}\BibitemShut {NoStop}%
	\bibitem [{\citenamefont {Ma}\ \emph {et~al.}(2012)\citenamefont {Ma},
		\citenamefont {Herbst}, \citenamefont {Scheidl}, \citenamefont {Wang},
		\citenamefont {Kropatschek}, \citenamefont {Naylor}, \citenamefont
		{Wittmann}, \citenamefont {Mech}, \citenamefont {Kofler}, \citenamefont
		{Anisimova}, \citenamefont {Makarov}, \citenamefont {Jennewein},
		\citenamefont {Ursin},\ and\ \citenamefont {Zeilinger}}]{Ma2012}%
	\BibitemOpen
	\bibfield  {author} {\bibinfo {author} {\bibfnamefont {X.-S.}\ \bibnamefont
			{Ma}}, \bibinfo {author} {\bibfnamefont {T.}~\bibnamefont {Herbst}}, \bibinfo
		{author} {\bibfnamefont {T.}~\bibnamefont {Scheidl}}, \bibinfo {author}
		{\bibfnamefont {D.}~\bibnamefont {Wang}}, \bibinfo {author} {\bibfnamefont
			{S.}~\bibnamefont {Kropatschek}}, \bibinfo {author} {\bibfnamefont
			{W.}~\bibnamefont {Naylor}}, \bibinfo {author} {\bibfnamefont
			{B.}~\bibnamefont {Wittmann}}, \bibinfo {author} {\bibfnamefont
			{A.}~\bibnamefont {Mech}}, \bibinfo {author} {\bibfnamefont {J.}~\bibnamefont
			{Kofler}}, \bibinfo {author} {\bibfnamefont {E.}~\bibnamefont {Anisimova}},
		\bibinfo {author} {\bibfnamefont {V.}~\bibnamefont {Makarov}}, \bibinfo
		{author} {\bibfnamefont {T.}~\bibnamefont {Jennewein}}, \bibinfo {author}
		{\bibfnamefont {R.}~\bibnamefont {Ursin}},\ and\ \bibinfo {author}
		{\bibfnamefont {A.}~\bibnamefont {Zeilinger}},\ }\bibfield  {title} {\bibinfo
		{title} {Quantum teleportation over 143 kilometres using active
			feed-forward},\ }\href {https://doi.org/10.1038/nature11472} {\bibfield
		{journal} {\bibinfo  {journal} {Nature}\ }\textbf {\bibinfo {volume} {489}},\
		\bibinfo {pages} {269} (\bibinfo {year} {2012})}\BibitemShut {NoStop}%
	\bibitem [{\citenamefont {Yin}\ \emph {et~al.}(2012)\citenamefont {Yin},
		\citenamefont {Ren}, \citenamefont {Lu}, \citenamefont {Cao}, \citenamefont
		{Yong}, \citenamefont {Wu}, \citenamefont {Liu}, \citenamefont {Liao},
		\citenamefont {Zhou}, \citenamefont {Jiang}, \citenamefont {Cai},
		\citenamefont {Xu}, \citenamefont {Pan}, \citenamefont {Jia}, \citenamefont
		{Huang}, \citenamefont {Yin}, \citenamefont {Wang}, \citenamefont {Chen},
		\citenamefont {Peng},\ and\ \citenamefont {Pan}}]{Yin2012}%
	\BibitemOpen
	\bibfield  {author} {\bibinfo {author} {\bibfnamefont {J.}~\bibnamefont
			{Yin}}, \bibinfo {author} {\bibfnamefont {J.-G.}\ \bibnamefont {Ren}},
		\bibinfo {author} {\bibfnamefont {H.}~\bibnamefont {Lu}}, \bibinfo {author}
		{\bibfnamefont {Y.}~\bibnamefont {Cao}}, \bibinfo {author} {\bibfnamefont
			{H.-L.}\ \bibnamefont {Yong}}, \bibinfo {author} {\bibfnamefont {Y.-P.}\
			\bibnamefont {Wu}}, \bibinfo {author} {\bibfnamefont {C.}~\bibnamefont
			{Liu}}, \bibinfo {author} {\bibfnamefont {S.-K.}\ \bibnamefont {Liao}},
		\bibinfo {author} {\bibfnamefont {F.}~\bibnamefont {Zhou}}, \bibinfo {author}
		{\bibfnamefont {Y.}~\bibnamefont {Jiang}}, \bibinfo {author} {\bibfnamefont
			{X.-D.}\ \bibnamefont {Cai}}, \bibinfo {author} {\bibfnamefont
			{P.}~\bibnamefont {Xu}}, \bibinfo {author} {\bibfnamefont {G.-S.}\
			\bibnamefont {Pan}}, \bibinfo {author} {\bibfnamefont {J.-J.}\ \bibnamefont
			{Jia}}, \bibinfo {author} {\bibfnamefont {Y.-M.}\ \bibnamefont {Huang}},
		\bibinfo {author} {\bibfnamefont {H.}~\bibnamefont {Yin}}, \bibinfo {author}
		{\bibfnamefont {J.-Y.}\ \bibnamefont {Wang}}, \bibinfo {author}
		{\bibfnamefont {Y.-A.}\ \bibnamefont {Chen}}, \bibinfo {author}
		{\bibfnamefont {C.-Z.}\ \bibnamefont {Peng}},\ and\ \bibinfo {author}
		{\bibfnamefont {J.-W.}\ \bibnamefont {Pan}},\ }\bibfield  {title} {\bibinfo
		{title} {Quantum teleportation and entanglement distribution over
			100-kilometre free-space channels},\ }\href
	{https://doi.org/10.1038/nature11332} {\bibfield  {journal} {\bibinfo
			{journal} {Nature}\ }\textbf {\bibinfo {volume} {488}},\ \bibinfo {pages}
		{185} (\bibinfo {year} {2012})}\BibitemShut {NoStop}%
	\bibitem [{\citenamefont {Valivarthi}\ \emph {et~al.}(2016)\citenamefont
		{Valivarthi}, \citenamefont {Puigibert}, \citenamefont {Zhou}, \citenamefont
		{Aguilar}, \citenamefont {Verma}, \citenamefont {Marsili}, \citenamefont
		{Shaw}, \citenamefont {Nam}, \citenamefont {Oblak},\ and\ \citenamefont
		{Tittel}}]{Valivarthi2016}%
	\BibitemOpen
	\bibfield  {author} {\bibinfo {author} {\bibfnamefont {R.}~\bibnamefont
			{Valivarthi}}, \bibinfo {author} {\bibfnamefont {M.~G.}\ \bibnamefont
			{Puigibert}}, \bibinfo {author} {\bibfnamefont {Q.}~\bibnamefont {Zhou}},
		\bibinfo {author} {\bibfnamefont {G.~H.}\ \bibnamefont {Aguilar}}, \bibinfo
		{author} {\bibfnamefont {V.~B.}\ \bibnamefont {Verma}}, \bibinfo {author}
		{\bibfnamefont {F.}~\bibnamefont {Marsili}}, \bibinfo {author} {\bibfnamefont
			{M.~D.}\ \bibnamefont {Shaw}}, \bibinfo {author} {\bibfnamefont {S.~W.}\
			\bibnamefont {Nam}}, \bibinfo {author} {\bibfnamefont {D.}~\bibnamefont
			{Oblak}},\ and\ \bibinfo {author} {\bibfnamefont {W.}~\bibnamefont
			{Tittel}},\ }\bibfield  {title} {\bibinfo {title} {Quantum teleportation
			across a metropolitan fibre network},\ }\href
	{https://doi.org/10.1038/nphoton.2016.180} {\bibfield  {journal} {\bibinfo
			{journal} {Nature Photonics}\ }\textbf {\bibinfo {volume} {10}},\ \bibinfo
		{pages} {676} (\bibinfo {year} {2016})}\BibitemShut {NoStop}%
	\bibitem [{\citenamefont {Grosshans}(2016)}]{Grosshans2016}%
	\BibitemOpen
	\bibfield  {author} {\bibinfo {author} {\bibfnamefont {F.}~\bibnamefont
			{Grosshans}},\ }\bibfield  {title} {\bibinfo {title} {Teleportation becomes
			streetwise},\ }\href {https://doi.org/10.1038/nphoton.2016.190} {\bibfield
		{journal} {\bibinfo  {journal} {Nature Photonics}\ }\textbf {\bibinfo
			{volume} {10}},\ \bibinfo {pages} {623} (\bibinfo {year} {2016})}\BibitemShut
	{NoStop}%
	\bibitem [{\citenamefont {Shen}\ \emph {et~al.}(2023)\citenamefont {Shen},
		\citenamefont {Yuan}, \citenamefont {Zhang}, \citenamefont {Yu},
		\citenamefont {Zhang}, \citenamefont {Yang}, \citenamefont {Li},
		\citenamefont {Wang}, \citenamefont {Wang}, \citenamefont {Deng},
		\citenamefont {Song}, \citenamefont {You}, \citenamefont {Fan}, \citenamefont
		{Guo},\ and\ \citenamefont {Zhou}}]{Shen2023}%
	\BibitemOpen
	\bibfield  {author} {\bibinfo {author} {\bibfnamefont {S.}~\bibnamefont
			{Shen}}, \bibinfo {author} {\bibfnamefont {C.}~\bibnamefont {Yuan}}, \bibinfo
		{author} {\bibfnamefont {Z.}~\bibnamefont {Zhang}}, \bibinfo {author}
		{\bibfnamefont {H.}~\bibnamefont {Yu}}, \bibinfo {author} {\bibfnamefont
			{R.}~\bibnamefont {Zhang}}, \bibinfo {author} {\bibfnamefont
			{C.}~\bibnamefont {Yang}}, \bibinfo {author} {\bibfnamefont {H.}~\bibnamefont
			{Li}}, \bibinfo {author} {\bibfnamefont {Z.}~\bibnamefont {Wang}}, \bibinfo
		{author} {\bibfnamefont {Y.}~\bibnamefont {Wang}}, \bibinfo {author}
		{\bibfnamefont {G.}~\bibnamefont {Deng}}, \bibinfo {author} {\bibfnamefont
			{H.}~\bibnamefont {Song}}, \bibinfo {author} {\bibfnamefont {L.}~\bibnamefont
			{You}}, \bibinfo {author} {\bibfnamefont {Y.}~\bibnamefont {Fan}}, \bibinfo
		{author} {\bibfnamefont {G.}~\bibnamefont {Guo}},\ and\ \bibinfo {author}
		{\bibfnamefont {Q.}~\bibnamefont {Zhou}},\ }\bibfield  {title} {\bibinfo
		{title} {Hertz-rate metropolitan quantum teleportation},\ }\href
	{https://doi.org/10.1038/s41377-023-01158-7} {\bibfield  {journal} {\bibinfo
			{journal} {Light: Science {\&} Applications}\ }\textbf {\bibinfo {volume}
			{12}},\ \bibinfo {pages} {115} (\bibinfo {year} {2023})}\BibitemShut
	{NoStop}%
	\bibitem [{\citenamefont {Ren}\ \emph {et~al.}(2017)\citenamefont {Ren},
		\citenamefont {Xu}, \citenamefont {Yong}, \citenamefont {Zhang},
		\citenamefont {Liao}, \citenamefont {Yin}, \citenamefont {Liu}, \citenamefont
		{Cai}, \citenamefont {Yang}, \citenamefont {Li}, \citenamefont {Yang},
		\citenamefont {Han}, \citenamefont {Yao}, \citenamefont {Li}, \citenamefont
		{Wu}, \citenamefont {Wan}, \citenamefont {Liu}, \citenamefont {Liu},
		\citenamefont {Kuang}, \citenamefont {He}, \citenamefont {Shang},
		\citenamefont {Guo}, \citenamefont {Zheng}, \citenamefont {Tian},
		\citenamefont {Zhu}, \citenamefont {Liu}, \citenamefont {Lu}, \citenamefont
		{Shu}, \citenamefont {Chen}, \citenamefont {Peng}, \citenamefont {Wang},\
		and\ \citenamefont {Pan}}]{Ren2017}%
	\BibitemOpen
	\bibfield  {author} {\bibinfo {author} {\bibfnamefont {J.-G.}\ \bibnamefont
			{Ren}}, \bibinfo {author} {\bibfnamefont {P.}~\bibnamefont {Xu}}, \bibinfo
		{author} {\bibfnamefont {H.-L.}\ \bibnamefont {Yong}}, \bibinfo {author}
		{\bibfnamefont {L.}~\bibnamefont {Zhang}}, \bibinfo {author} {\bibfnamefont
			{S.-K.}\ \bibnamefont {Liao}}, \bibinfo {author} {\bibfnamefont
			{J.}~\bibnamefont {Yin}}, \bibinfo {author} {\bibfnamefont {W.-Y.}\
			\bibnamefont {Liu}}, \bibinfo {author} {\bibfnamefont {W.-Q.}\ \bibnamefont
			{Cai}}, \bibinfo {author} {\bibfnamefont {M.}~\bibnamefont {Yang}}, \bibinfo
		{author} {\bibfnamefont {L.}~\bibnamefont {Li}}, \bibinfo {author}
		{\bibfnamefont {K.-X.}\ \bibnamefont {Yang}}, \bibinfo {author}
		{\bibfnamefont {X.}~\bibnamefont {Han}}, \bibinfo {author} {\bibfnamefont
			{Y.-Q.}\ \bibnamefont {Yao}}, \bibinfo {author} {\bibfnamefont
			{J.}~\bibnamefont {Li}}, \bibinfo {author} {\bibfnamefont {H.-Y.}\
			\bibnamefont {Wu}}, \bibinfo {author} {\bibfnamefont {S.}~\bibnamefont
			{Wan}}, \bibinfo {author} {\bibfnamefont {L.}~\bibnamefont {Liu}}, \bibinfo
		{author} {\bibfnamefont {D.-Q.}\ \bibnamefont {Liu}}, \bibinfo {author}
		{\bibfnamefont {Y.-W.}\ \bibnamefont {Kuang}}, \bibinfo {author}
		{\bibfnamefont {Z.-P.}\ \bibnamefont {He}}, \bibinfo {author} {\bibfnamefont
			{P.}~\bibnamefont {Shang}}, \bibinfo {author} {\bibfnamefont
			{C.}~\bibnamefont {Guo}}, \bibinfo {author} {\bibfnamefont {R.-H.}\
			\bibnamefont {Zheng}}, \bibinfo {author} {\bibfnamefont {K.}~\bibnamefont
			{Tian}}, \bibinfo {author} {\bibfnamefont {Z.-C.}\ \bibnamefont {Zhu}},
		\bibinfo {author} {\bibfnamefont {N.-L.}\ \bibnamefont {Liu}}, \bibinfo
		{author} {\bibfnamefont {C.-Y.}\ \bibnamefont {Lu}}, \bibinfo {author}
		{\bibfnamefont {R.}~\bibnamefont {Shu}}, \bibinfo {author} {\bibfnamefont
			{Y.-A.}\ \bibnamefont {Chen}}, \bibinfo {author} {\bibfnamefont {C.-Z.}\
			\bibnamefont {Peng}}, \bibinfo {author} {\bibfnamefont {J.-Y.}\ \bibnamefont
			{Wang}},\ and\ \bibinfo {author} {\bibfnamefont {J.-W.}\ \bibnamefont
			{Pan}},\ }\bibfield  {title} {\bibinfo {title} {Ground-to-satellite quantum
			teleportation},\ }\href {https://doi.org/10.1038/nature23675} {\bibfield
		{journal} {\bibinfo  {journal} {Nature}\ }\textbf {\bibinfo {volume} {549}},\
		\bibinfo {pages} {70} (\bibinfo {year} {2017})}\BibitemShut {NoStop}%
	\bibitem [{\citenamefont {Basso~Basset}\ \emph {et~al.}(2019)\citenamefont
		{Basso~Basset}, \citenamefont {Rota}, \citenamefont {Schimpf}, \citenamefont
		{Tedeschi}, \citenamefont {Zeuner}, \citenamefont {Covre~da Silva},
		\citenamefont {Reindl}, \citenamefont {Zwiller}, \citenamefont {J\"ons},
		\citenamefont {Rastelli},\ and\ \citenamefont {Trotta}}]{BassoBasset_tele}%
	\BibitemOpen
	\bibfield  {author} {\bibinfo {author} {\bibfnamefont {F.}~\bibnamefont
			{Basso~Basset}}, \bibinfo {author} {\bibfnamefont {M.~B.}\ \bibnamefont
			{Rota}}, \bibinfo {author} {\bibfnamefont {C.}~\bibnamefont {Schimpf}},
		\bibinfo {author} {\bibfnamefont {D.}~\bibnamefont {Tedeschi}}, \bibinfo
		{author} {\bibfnamefont {K.~D.}\ \bibnamefont {Zeuner}}, \bibinfo {author}
		{\bibfnamefont {S.~F.}\ \bibnamefont {Covre~da Silva}}, \bibinfo {author}
		{\bibfnamefont {M.}~\bibnamefont {Reindl}}, \bibinfo {author} {\bibfnamefont
			{V.}~\bibnamefont {Zwiller}}, \bibinfo {author} {\bibfnamefont {K.~D.}\
			\bibnamefont {J\"ons}}, \bibinfo {author} {\bibfnamefont {A.}~\bibnamefont
			{Rastelli}},\ and\ \bibinfo {author} {\bibfnamefont {R.}~\bibnamefont
			{Trotta}},\ }\bibfield  {title} {\bibinfo {title} {Entanglement swapping with
			photons generated on demand by a quantum dot},\ }\href
	{https://doi.org/10.1103/PhysRevLett.123.160501} {\bibfield  {journal}
		{\bibinfo  {journal} {Phys. Rev. Lett.}\ }\textbf {\bibinfo {volume} {123}},\
		\bibinfo {pages} {160501} (\bibinfo {year} {2019})}\BibitemShut {NoStop}%
	\bibitem [{\citenamefont {Wang}\ \emph {et~al.}(2019)\citenamefont {Wang},
		\citenamefont {He}, \citenamefont {Chung}, \citenamefont {Hu}, \citenamefont
		{Yu}, \citenamefont {Chen}, \citenamefont {Ding}, \citenamefont {Chen},
		\citenamefont {Qin}, \citenamefont {Yang}, \citenamefont {Liu}, \citenamefont
		{Duan}, \citenamefont {Li}, \citenamefont {Gerhardt}, \citenamefont
		{Winkler}, \citenamefont {Jurkat}, \citenamefont {Wang}, \citenamefont
		{Gregersen}, \citenamefont {Huo}, \citenamefont {Dai}, \citenamefont {Yu},
		\citenamefont {H{\"o}fling}, \citenamefont {Lu},\ and\ \citenamefont
		{Pan}}]{Wang2019}%
	\BibitemOpen
	\bibfield  {author} {\bibinfo {author} {\bibfnamefont {H.}~\bibnamefont
			{Wang}}, \bibinfo {author} {\bibfnamefont {Y.-M.}\ \bibnamefont {He}},
		\bibinfo {author} {\bibfnamefont {T.-H.}\ \bibnamefont {Chung}}, \bibinfo
		{author} {\bibfnamefont {H.}~\bibnamefont {Hu}}, \bibinfo {author}
		{\bibfnamefont {Y.}~\bibnamefont {Yu}}, \bibinfo {author} {\bibfnamefont
			{S.}~\bibnamefont {Chen}}, \bibinfo {author} {\bibfnamefont {X.}~\bibnamefont
			{Ding}}, \bibinfo {author} {\bibfnamefont {M.-C.}\ \bibnamefont {Chen}},
		\bibinfo {author} {\bibfnamefont {J.}~\bibnamefont {Qin}}, \bibinfo {author}
		{\bibfnamefont {X.}~\bibnamefont {Yang}}, \bibinfo {author} {\bibfnamefont
			{R.-Z.}\ \bibnamefont {Liu}}, \bibinfo {author} {\bibfnamefont {Z.-C.}\
			\bibnamefont {Duan}}, \bibinfo {author} {\bibfnamefont {J.-P.}\ \bibnamefont
			{Li}}, \bibinfo {author} {\bibfnamefont {S.}~\bibnamefont {Gerhardt}},
		\bibinfo {author} {\bibfnamefont {K.}~\bibnamefont {Winkler}}, \bibinfo
		{author} {\bibfnamefont {J.}~\bibnamefont {Jurkat}}, \bibinfo {author}
		{\bibfnamefont {L.-J.}\ \bibnamefont {Wang}}, \bibinfo {author}
		{\bibfnamefont {N.}~\bibnamefont {Gregersen}}, \bibinfo {author}
		{\bibfnamefont {Y.-H.}\ \bibnamefont {Huo}}, \bibinfo {author} {\bibfnamefont
			{Q.}~\bibnamefont {Dai}}, \bibinfo {author} {\bibfnamefont {S.}~\bibnamefont
			{Yu}}, \bibinfo {author} {\bibfnamefont {S.}~\bibnamefont {H{\"o}fling}},
		\bibinfo {author} {\bibfnamefont {C.-Y.}\ \bibnamefont {Lu}},\ and\ \bibinfo
		{author} {\bibfnamefont {J.-W.}\ \bibnamefont {Pan}},\ }\bibfield  {title}
	{\bibinfo {title} {Towards optimal single-photon sources from polarized
			microcavities},\ }\href {https://doi.org/10.1038/s41566-019-0494-3}
	{\bibfield  {journal} {\bibinfo  {journal} {Nature Photonics}\ }\textbf
		{\bibinfo {volume} {13}},\ \bibinfo {pages} {770} (\bibinfo {year}
		{2019})}\BibitemShut {NoStop}%
	\bibitem [{\citenamefont {Ding}\ \emph {et~al.}(2023)\citenamefont {Ding},
		\citenamefont {Guo}, \citenamefont {Xu}, \citenamefont {Liu}, \citenamefont
		{Zou}, \citenamefont {Zhao}, \citenamefont {Ge}, \citenamefont {Zhang},
		\citenamefont {Liu}, \citenamefont {Chen} \emph
		{et~al.}}]{ding2023highefficiency}%
	\BibitemOpen
	\bibfield  {author} {\bibinfo {author} {\bibfnamefont {X.}~\bibnamefont
			{Ding}}, \bibinfo {author} {\bibfnamefont {Y.-P.}\ \bibnamefont {Guo}},
		\bibinfo {author} {\bibfnamefont {M.-C.}\ \bibnamefont {Xu}}, \bibinfo
		{author} {\bibfnamefont {R.-Z.}\ \bibnamefont {Liu}}, \bibinfo {author}
		{\bibfnamefont {G.-Y.}\ \bibnamefont {Zou}}, \bibinfo {author} {\bibfnamefont
			{J.-Y.}\ \bibnamefont {Zhao}}, \bibinfo {author} {\bibfnamefont {Z.-X.}\
			\bibnamefont {Ge}}, \bibinfo {author} {\bibfnamefont {Q.-H.}\ \bibnamefont
			{Zhang}}, \bibinfo {author} {\bibfnamefont {H.-L.}\ \bibnamefont {Liu}},
		\bibinfo {author} {\bibfnamefont {M.-C.}\ \bibnamefont {Chen}}, \emph
		{et~al.},\ }\bibfield  {title} {\bibinfo {title} {High-efficiency
			single-photon source above the loss-tolerant threshold for efficient linear
			optical quantum computing},\ }\href@noop {} {\bibfield  {journal} {\bibinfo
			{journal} {arXiv preprint arXiv:2311.08347}\ } (\bibinfo {year}
		{2023})}\BibitemShut {NoStop}%
	\bibitem [{\citenamefont {Loredo}\ \emph {et~al.}(2019)\citenamefont {Loredo},
		\citenamefont {Ant{\'o}n}, \citenamefont {Reznychenko}, \citenamefont
		{Hilaire}, \citenamefont {Harouri}, \citenamefont {Millet}, \citenamefont
		{Ollivier}, \citenamefont {Somaschi}, \citenamefont {De~Santis},
		\citenamefont {Lema{\^\i}tre}, \citenamefont {Sagnes}, \citenamefont {Lanco},
		\citenamefont {Auff{\`e}ves}, \citenamefont {Krebs},\ and\ \citenamefont
		{Senellart}}]{loredo2019generation}%
	\BibitemOpen
	\bibfield  {author} {\bibinfo {author} {\bibfnamefont {J.}~\bibnamefont
			{Loredo}}, \bibinfo {author} {\bibfnamefont {C.}~\bibnamefont {Ant{\'o}n}},
		\bibinfo {author} {\bibfnamefont {B.}~\bibnamefont {Reznychenko}}, \bibinfo
		{author} {\bibfnamefont {P.}~\bibnamefont {Hilaire}}, \bibinfo {author}
		{\bibfnamefont {A.}~\bibnamefont {Harouri}}, \bibinfo {author} {\bibfnamefont
			{C.}~\bibnamefont {Millet}}, \bibinfo {author} {\bibfnamefont
			{H.}~\bibnamefont {Ollivier}}, \bibinfo {author} {\bibfnamefont
			{N.}~\bibnamefont {Somaschi}}, \bibinfo {author} {\bibfnamefont
			{L.}~\bibnamefont {De~Santis}}, \bibinfo {author} {\bibfnamefont
			{A.}~\bibnamefont {Lema{\^\i}tre}}, \bibinfo {author} {\bibfnamefont
			{I.}~\bibnamefont {Sagnes}}, \bibinfo {author} {\bibfnamefont
			{L.}~\bibnamefont {Lanco}}, \bibinfo {author} {\bibfnamefont
			{A.}~\bibnamefont {Auff{\`e}ves}}, \bibinfo {author} {\bibfnamefont
			{O.}~\bibnamefont {Krebs}},\ and\ \bibinfo {author} {\bibfnamefont
			{P.}~\bibnamefont {Senellart}},\ }\bibfield  {title} {\bibinfo {title}
		{Generation of non-classical light in a photon-number superposition},\ }\href
	{https://doi.org/Generation of non-classical light in a photon-number
		superposition} {\bibfield  {journal} {\bibinfo  {journal} {Nature Photonics}\
		}\textbf {\bibinfo {volume} {13}},\ \bibinfo {pages} {803} (\bibinfo {year}
		{2019})}\BibitemShut {NoStop}%
	\bibitem [{\citenamefont {Wein}\ \emph {et~al.}(2022)\citenamefont {Wein},
		\citenamefont {Loredo}, \citenamefont {Maffei}, \citenamefont {Hilaire},
		\citenamefont {Harouri}, \citenamefont {Somaschi}, \citenamefont
		{Lema{\^\i}tre}, \citenamefont {Sagnes}, \citenamefont {Lanco}, \citenamefont
		{Krebs}, \citenamefont {Auff{\`e}ves}, \citenamefont {Simon}, \citenamefont
		{Senellart},\ and\ \citenamefont {Ant{\'o}n-Solanas}}]{wein2022photon}%
	\BibitemOpen
	\bibfield  {author} {\bibinfo {author} {\bibfnamefont {S.~C.}\ \bibnamefont
			{Wein}}, \bibinfo {author} {\bibfnamefont {J.~C.}\ \bibnamefont {Loredo}},
		\bibinfo {author} {\bibfnamefont {M.}~\bibnamefont {Maffei}}, \bibinfo
		{author} {\bibfnamefont {P.}~\bibnamefont {Hilaire}}, \bibinfo {author}
		{\bibfnamefont {A.}~\bibnamefont {Harouri}}, \bibinfo {author} {\bibfnamefont
			{N.}~\bibnamefont {Somaschi}}, \bibinfo {author} {\bibfnamefont
			{A.}~\bibnamefont {Lema{\^\i}tre}}, \bibinfo {author} {\bibfnamefont
			{I.}~\bibnamefont {Sagnes}}, \bibinfo {author} {\bibfnamefont
			{L.}~\bibnamefont {Lanco}}, \bibinfo {author} {\bibfnamefont
			{O.}~\bibnamefont {Krebs}}, \bibinfo {author} {\bibfnamefont
			{A.}~\bibnamefont {Auff{\`e}ves}}, \bibinfo {author} {\bibfnamefont
			{C.}~\bibnamefont {Simon}}, \bibinfo {author} {\bibfnamefont
			{P.}~\bibnamefont {Senellart}},\ and\ \bibinfo {author} {\bibfnamefont
			{C.}~\bibnamefont {Ant{\'o}n-Solanas}},\ }\bibfield  {title} {\bibinfo
		{title} {Photon-number entanglement generated by sequential excitation of a
			two-level atom},\ }\href {https://doi.org/10.1038/s41566-022-00979-z}
	{\bibfield  {journal} {\bibinfo  {journal} {Nature Photonics}\ }\textbf
		{\bibinfo {volume} {16}},\ \bibinfo {pages} {374} (\bibinfo {year}
		{2022})}\BibitemShut {NoStop}%
	\bibitem [{\citenamefont {Natarajan}\ \emph {et~al.}(2012)\citenamefont
		{Natarajan}, \citenamefont {Tanner},\ and\ \citenamefont
		{Hadfield}}]{natarajan2012superconducting}%
	\BibitemOpen
	\bibfield  {author} {\bibinfo {author} {\bibfnamefont {C.~M.}\ \bibnamefont
			{Natarajan}}, \bibinfo {author} {\bibfnamefont {M.~G.}\ \bibnamefont
			{Tanner}},\ and\ \bibinfo {author} {\bibfnamefont {R.~H.}\ \bibnamefont
			{Hadfield}},\ }\bibfield  {title} {\bibinfo {title} {Superconducting nanowire
			single-photon detectors: physics and applications},\ }\href@noop {}
	{\bibfield  {journal} {\bibinfo  {journal} {Superconductor science and
				technology}\ }\textbf {\bibinfo {volume} {25}},\ \bibinfo {pages} {063001}
		(\bibinfo {year} {2012})}\BibitemShut {NoStop}%
	\bibitem [{\citenamefont {Chanana}\ \emph {et~al.}(2022)\citenamefont
		{Chanana}, \citenamefont {Larocque}, \citenamefont {Moreira}, \citenamefont
		{Carolan}, \citenamefont {Guha}, \citenamefont {Melo}, \citenamefont {Anant},
		\citenamefont {Song}, \citenamefont {Englund}, \citenamefont {Blumenthal},
		\citenamefont {Srinivasan},\ and\ \citenamefont {Davanco}}]{Chanana2022}%
	\BibitemOpen
	\bibfield  {author} {\bibinfo {author} {\bibfnamefont {A.}~\bibnamefont
			{Chanana}}, \bibinfo {author} {\bibfnamefont {H.}~\bibnamefont {Larocque}},
		\bibinfo {author} {\bibfnamefont {R.}~\bibnamefont {Moreira}}, \bibinfo
		{author} {\bibfnamefont {J.}~\bibnamefont {Carolan}}, \bibinfo {author}
		{\bibfnamefont {B.}~\bibnamefont {Guha}}, \bibinfo {author} {\bibfnamefont
			{E.~G.}\ \bibnamefont {Melo}}, \bibinfo {author} {\bibfnamefont
			{V.}~\bibnamefont {Anant}}, \bibinfo {author} {\bibfnamefont
			{J.}~\bibnamefont {Song}}, \bibinfo {author} {\bibfnamefont {D.}~\bibnamefont
			{Englund}}, \bibinfo {author} {\bibfnamefont {D.~J.}\ \bibnamefont
			{Blumenthal}}, \bibinfo {author} {\bibfnamefont {K.}~\bibnamefont
			{Srinivasan}},\ and\ \bibinfo {author} {\bibfnamefont {M.}~\bibnamefont
			{Davanco}},\ }\bibfield  {title} {\bibinfo {title} {Ultra-low loss quantum
			photonic circuits integrated with single quantum emitters},\ }\bibfield
	{journal} {\bibinfo  {journal} {Nature Communications}\ }\textbf {\bibinfo
		{volume} {13}},\ \href {https://doi.org/10.1038/s41467-022-35332-z}
	{10.1038/s41467-022-35332-z} (\bibinfo {year} {2022})\BibitemShut {NoStop}%
	\bibitem [{\citenamefont {Senellart}\ \emph {et~al.}(2017)\citenamefont
		{Senellart}, \citenamefont {Solomon},\ and\ \citenamefont
		{White}}]{Senellart2017}%
	\BibitemOpen
	\bibfield  {author} {\bibinfo {author} {\bibfnamefont {P.}~\bibnamefont
			{Senellart}}, \bibinfo {author} {\bibfnamefont {G.}~\bibnamefont {Solomon}},\
		and\ \bibinfo {author} {\bibfnamefont {A.}~\bibnamefont {White}},\ }\bibfield
	{title} {\bibinfo {title} {High-performance semiconductor quantum-dot
			single-photon sources},\ }\href {https://doi.org/10.1038/nnano.2017.218}
	{\bibfield  {journal} {\bibinfo  {journal} {Nature Nanotechnology}\ }\textbf
		{\bibinfo {volume} {12}},\ \bibinfo {pages} {1026} (\bibinfo {year}
		{2017})}\BibitemShut {NoStop}%
	\bibitem [{\citenamefont {Heindel}\ \emph {et~al.}(2023)\citenamefont
		{Heindel}, \citenamefont {Kim}, \citenamefont {Gregersen}, \citenamefont
		{Rastelli},\ and\ \citenamefont {Reitzenstein}}]{Heindel:23}%
	\BibitemOpen
	\bibfield  {author} {\bibinfo {author} {\bibfnamefont {T.}~\bibnamefont
			{Heindel}}, \bibinfo {author} {\bibfnamefont {J.-H.}\ \bibnamefont {Kim}},
		\bibinfo {author} {\bibfnamefont {N.}~\bibnamefont {Gregersen}}, \bibinfo
		{author} {\bibfnamefont {A.}~\bibnamefont {Rastelli}},\ and\ \bibinfo
		{author} {\bibfnamefont {S.}~\bibnamefont {Reitzenstein}},\ }\bibfield
	{title} {\bibinfo {title} {Quantum dots for photonic quantum information
			technology},\ }\href {https://doi.org/10.1364/AOP.490091} {\bibfield
		{journal} {\bibinfo  {journal} {Adv. Opt. Photon.}\ }\textbf {\bibinfo
			{volume} {15}},\ \bibinfo {pages} {613} (\bibinfo {year} {2023})}\BibitemShut
	{NoStop}%
	\bibitem [{\citenamefont {Arakawa}\ and\ \citenamefont
		{Holmes}(2020)}]{10.1063/5.0010193}%
	\BibitemOpen
	\bibfield  {author} {\bibinfo {author} {\bibfnamefont {Y.}~\bibnamefont
			{Arakawa}}\ and\ \bibinfo {author} {\bibfnamefont {M.~J.}\ \bibnamefont
			{Holmes}},\ }\bibfield  {title} {\bibinfo {title} {{Progress in quantum-dot
				single photon sources for quantum information technologies: A broad spectrum
				overview}},\ }\href {https://doi.org/10.1063/5.0010193} {\bibfield  {journal}
		{\bibinfo  {journal} {Applied Physics Reviews}\ }\textbf {\bibinfo {volume}
			{7}},\ \bibinfo {pages} {021309} (\bibinfo {year} {2020})}\BibitemShut
	{NoStop}%
	\bibitem [{\citenamefont {Somaschi}\ \emph {et~al.}(2016)\citenamefont
		{Somaschi}, \citenamefont {Giesz}, \citenamefont {De~Santis}, \citenamefont
		{Loredo}, \citenamefont {Almeida}, \citenamefont {Hornecker}, \citenamefont
		{Portalupi}, \citenamefont {Grange}, \citenamefont {Ant{\'o}n}, \citenamefont
		{Demory}, \citenamefont {Gomez}, \citenamefont {Sagnes}, \citenamefont
		{Lanzillotti-Kimura}, \citenamefont {Lemaitre}, \citenamefont {Auff{\`e}ves},
		\citenamefont {White}, \citenamefont {Lanco},\ and\ \citenamefont
		{Senellart}}]{somaschi2016near}%
	\BibitemOpen
	\bibfield  {author} {\bibinfo {author} {\bibfnamefont {N.}~\bibnamefont
			{Somaschi}}, \bibinfo {author} {\bibfnamefont {V.}~\bibnamefont {Giesz}},
		\bibinfo {author} {\bibfnamefont {L.}~\bibnamefont {De~Santis}}, \bibinfo
		{author} {\bibfnamefont {J.}~\bibnamefont {Loredo}}, \bibinfo {author}
		{\bibfnamefont {M.~P.}\ \bibnamefont {Almeida}}, \bibinfo {author}
		{\bibfnamefont {G.}~\bibnamefont {Hornecker}}, \bibinfo {author}
		{\bibfnamefont {S.~L.}\ \bibnamefont {Portalupi}}, \bibinfo {author}
		{\bibfnamefont {T.}~\bibnamefont {Grange}}, \bibinfo {author} {\bibfnamefont
			{C.}~\bibnamefont {Ant{\'o}n}}, \bibinfo {author} {\bibfnamefont
			{J.}~\bibnamefont {Demory}}, \bibinfo {author} {\bibfnamefont
			{C.}~\bibnamefont {Gomez}}, \bibinfo {author} {\bibfnamefont
			{I.}~\bibnamefont {Sagnes}}, \bibinfo {author} {\bibfnamefont {N.~D.}\
			\bibnamefont {Lanzillotti-Kimura}}, \bibinfo {author} {\bibfnamefont
			{A.}~\bibnamefont {Lemaitre}}, \bibinfo {author} {\bibfnamefont
			{A.}~\bibnamefont {Auff{\`e}ves}}, \bibinfo {author} {\bibfnamefont {A.~G.}\
			\bibnamefont {White}}, \bibinfo {author} {\bibfnamefont {L.}~\bibnamefont
			{Lanco}},\ and\ \bibinfo {author} {\bibfnamefont {P.}~\bibnamefont
			{Senellart}},\ }\bibfield  {title} {\bibinfo {title} {Near-optimal
			single-photon sources in the solid state},\ }\href
	{https://doi.org/10.1038/nphoton.2016.23} {\bibfield  {journal} {\bibinfo
			{journal} {Nature Photonics}\ }\textbf {\bibinfo {volume} {10}},\ \bibinfo
		{pages} {340} (\bibinfo {year} {2016})}\BibitemShut {NoStop}%
	\bibitem [{\citenamefont {Lu}\ and\ \citenamefont {Pan}(2021)}]{Lu2021}%
	\BibitemOpen
	\bibfield  {author} {\bibinfo {author} {\bibfnamefont {C.-Y.}\ \bibnamefont
			{Lu}}\ and\ \bibinfo {author} {\bibfnamefont {J.-W.}\ \bibnamefont {Pan}},\
	}\bibfield  {title} {\bibinfo {title} {Quantum-dot single-photon sources for
			the quantum internet},\ }\href {https://doi.org/10.1038/s41565-021-01033-9}
	{\bibfield  {journal} {\bibinfo  {journal} {Nature Nanotechnology}\ }\textbf
		{\bibinfo {volume} {16}},\ \bibinfo {pages} {1294} (\bibinfo {year}
		{2021})}\BibitemShut {NoStop}%
	\bibitem [{\citenamefont {{Basso Basset}}\ \emph {et~al.}(2021)\citenamefont
		{{Basso Basset}}, \citenamefont {Valeri}, \citenamefont {Roccia},
		\citenamefont {Muredda}, \citenamefont {Poderini}, \citenamefont {Neuwirth},
		\citenamefont {Spagnolo}, \citenamefont {Rota}, \citenamefont {Carvacho},
		\citenamefont {Sciarrino},\ and\ \citenamefont {Trotta}}]{BassoBasset_qkd}%
	\BibitemOpen
	\bibfield  {author} {\bibinfo {author} {\bibfnamefont {F.}~\bibnamefont
			{{Basso Basset}}}, \bibinfo {author} {\bibfnamefont {M.}~\bibnamefont
			{Valeri}}, \bibinfo {author} {\bibfnamefont {E.}~\bibnamefont {Roccia}},
		\bibinfo {author} {\bibfnamefont {V.}~\bibnamefont {Muredda}}, \bibinfo
		{author} {\bibfnamefont {D.}~\bibnamefont {Poderini}}, \bibinfo {author}
		{\bibfnamefont {J.}~\bibnamefont {Neuwirth}}, \bibinfo {author}
		{\bibfnamefont {N.}~\bibnamefont {Spagnolo}}, \bibinfo {author}
		{\bibfnamefont {M.~B.}\ \bibnamefont {Rota}}, \bibinfo {author}
		{\bibfnamefont {G.}~\bibnamefont {Carvacho}}, \bibinfo {author}
		{\bibfnamefont {F.}~\bibnamefont {Sciarrino}},\ and\ \bibinfo {author}
		{\bibfnamefont {R.}~\bibnamefont {Trotta}},\ }\bibfield  {title} {\bibinfo
		{title} {Quantum key distribution with entangled photons generated on demand
			by a quantum dot},\ }\href {https://doi.org/10.1126/sciadv.abe6379}
	{\bibfield  {journal} {\bibinfo  {journal} {Science Advances}\ }\textbf
		{\bibinfo {volume} {7}},\ \bibinfo {pages} {eabe6379} (\bibinfo {year}
		{2021})}\BibitemShut {NoStop}%
	\bibitem [{\citenamefont {Cao}\ \emph {et~al.}(2024)\citenamefont {Cao},
		\citenamefont {Hansen}, \citenamefont {Giorgino}, \citenamefont {Carosini},
		\citenamefont {Zah\'alka}, \citenamefont {Zilk}, \citenamefont {Loredo},\
		and\ \citenamefont {Walther}}]{cao2023photonic}%
	\BibitemOpen
	\bibfield  {author} {\bibinfo {author} {\bibfnamefont {H.}~\bibnamefont
			{Cao}}, \bibinfo {author} {\bibfnamefont {L.~M.}\ \bibnamefont {Hansen}},
		\bibinfo {author} {\bibfnamefont {F.}~\bibnamefont {Giorgino}}, \bibinfo
		{author} {\bibfnamefont {L.}~\bibnamefont {Carosini}}, \bibinfo {author}
		{\bibfnamefont {P.}~\bibnamefont {Zah\'alka}}, \bibinfo {author}
		{\bibfnamefont {F.}~\bibnamefont {Zilk}}, \bibinfo {author} {\bibfnamefont
			{J.~C.}\ \bibnamefont {Loredo}},\ and\ \bibinfo {author} {\bibfnamefont
			{P.}~\bibnamefont {Walther}},\ }\bibfield  {title} {\bibinfo {title}
		{Photonic source of heralded greenberger-horne-zeilinger states},\ }\href
	{https://doi.org/10.1103/PhysRevLett.132.130604} {\bibfield  {journal}
		{\bibinfo  {journal} {Phys. Rev. Lett.}\ }\textbf {\bibinfo {volume} {132}},\
		\bibinfo {pages} {130604} (\bibinfo {year} {2024})}\BibitemShut {NoStop}%
	\bibitem [{\citenamefont {Maring}\ \emph {et~al.}(2024)\citenamefont {Maring},
		\citenamefont {Fyrillas}, \citenamefont {Pont}, \citenamefont {Ivanov},
		\citenamefont {Stepanov}, \citenamefont {Margaria}, \citenamefont {Hease},
		\citenamefont {Pishchagin}, \citenamefont {Lema{\^i}tre}, \citenamefont
		{Sagnes}, \citenamefont {Au}, \citenamefont {Boissier}, \citenamefont
		{Bertasi}, \citenamefont {Baert}, \citenamefont {Valdivia}, \citenamefont
		{Billard}, \citenamefont {Acar}, \citenamefont {Brieussel}, \citenamefont
		{Mezher}, \citenamefont {Wein}, \citenamefont {Salavrakos}, \citenamefont
		{Sinnott}, \citenamefont {Fioretto}, \citenamefont {Emeriau}, \citenamefont
		{Belabas}, \citenamefont {Mansfield}, \citenamefont {Senellart},
		\citenamefont {Senellart},\ and\ \citenamefont
		{Somaschi}}]{maring2023generalpurpose}%
	\BibitemOpen
	\bibfield  {author} {\bibinfo {author} {\bibfnamefont {N.}~\bibnamefont
			{Maring}}, \bibinfo {author} {\bibfnamefont {A.}~\bibnamefont {Fyrillas}},
		\bibinfo {author} {\bibfnamefont {M.}~\bibnamefont {Pont}}, \bibinfo {author}
		{\bibfnamefont {E.}~\bibnamefont {Ivanov}}, \bibinfo {author} {\bibfnamefont
			{P.}~\bibnamefont {Stepanov}}, \bibinfo {author} {\bibfnamefont
			{N.}~\bibnamefont {Margaria}}, \bibinfo {author} {\bibfnamefont
			{W.}~\bibnamefont {Hease}}, \bibinfo {author} {\bibfnamefont
			{A.}~\bibnamefont {Pishchagin}}, \bibinfo {author} {\bibfnamefont
			{A.}~\bibnamefont {Lema{\^i}tre}}, \bibinfo {author} {\bibfnamefont
			{I.}~\bibnamefont {Sagnes}}, \bibinfo {author} {\bibfnamefont {T.~H.}\
			\bibnamefont {Au}}, \bibinfo {author} {\bibfnamefont {S.}~\bibnamefont
			{Boissier}}, \bibinfo {author} {\bibfnamefont {E.}~\bibnamefont {Bertasi}},
		\bibinfo {author} {\bibfnamefont {A.}~\bibnamefont {Baert}}, \bibinfo
		{author} {\bibfnamefont {M.}~\bibnamefont {Valdivia}}, \bibinfo {author}
		{\bibfnamefont {M.}~\bibnamefont {Billard}}, \bibinfo {author} {\bibfnamefont
			{O.}~\bibnamefont {Acar}}, \bibinfo {author} {\bibfnamefont {A.}~\bibnamefont
			{Brieussel}}, \bibinfo {author} {\bibfnamefont {R.}~\bibnamefont {Mezher}},
		\bibinfo {author} {\bibfnamefont {S.~C.}\ \bibnamefont {Wein}}, \bibinfo
		{author} {\bibfnamefont {A.}~\bibnamefont {Salavrakos}}, \bibinfo {author}
		{\bibfnamefont {P.}~\bibnamefont {Sinnott}}, \bibinfo {author} {\bibfnamefont
			{D.~A.}\ \bibnamefont {Fioretto}}, \bibinfo {author} {\bibfnamefont {P.-E.}\
			\bibnamefont {Emeriau}}, \bibinfo {author} {\bibfnamefont {N.}~\bibnamefont
			{Belabas}}, \bibinfo {author} {\bibfnamefont {S.}~\bibnamefont {Mansfield}},
		\bibinfo {author} {\bibfnamefont {P.}~\bibnamefont {Senellart}}, \bibinfo
		{author} {\bibfnamefont {J.}~\bibnamefont {Senellart}},\ and\ \bibinfo
		{author} {\bibfnamefont {N.}~\bibnamefont {Somaschi}},\ }\bibfield  {title}
	{\bibinfo {title} {A versatile single-photon-based quantum computing
			platform},\ }\bibfield  {journal} {\bibinfo  {journal} {Nature Photonics}\
	}\href {https://doi.org/10.1038/s41566-024-01403-4}
	{10.1038/s41566-024-01403-4} (\bibinfo {year} {2024})\BibitemShut {NoStop}%
	\bibitem [{\citenamefont {Chen}\ \emph {et~al.}(2024)\citenamefont {Chen},
		\citenamefont {Peng}, \citenamefont {Guo}, \citenamefont {Gu}, \citenamefont
		{Ding}, \citenamefont {Liu}, \citenamefont {Zhao}, \citenamefont {You},
		\citenamefont {Qin}, \citenamefont {Wang}, \citenamefont {He}, \citenamefont
		{Renema}, \citenamefont {Huo}, \citenamefont {Wang}, \citenamefont {Lu},\
		and\ \citenamefont {Pan}}]{chen2023heralded}%
	\BibitemOpen
	\bibfield  {author} {\bibinfo {author} {\bibfnamefont {S.}~\bibnamefont
			{Chen}}, \bibinfo {author} {\bibfnamefont {L.-C.}\ \bibnamefont {Peng}},
		\bibinfo {author} {\bibfnamefont {Y.-P.}\ \bibnamefont {Guo}}, \bibinfo
		{author} {\bibfnamefont {X.-M.}\ \bibnamefont {Gu}}, \bibinfo {author}
		{\bibfnamefont {X.}~\bibnamefont {Ding}}, \bibinfo {author} {\bibfnamefont
			{R.-Z.}\ \bibnamefont {Liu}}, \bibinfo {author} {\bibfnamefont {J.-Y.}\
			\bibnamefont {Zhao}}, \bibinfo {author} {\bibfnamefont {X.}~\bibnamefont
			{You}}, \bibinfo {author} {\bibfnamefont {J.}~\bibnamefont {Qin}}, \bibinfo
		{author} {\bibfnamefont {Y.-F.}\ \bibnamefont {Wang}}, \bibinfo {author}
		{\bibfnamefont {Y.-M.}\ \bibnamefont {He}}, \bibinfo {author} {\bibfnamefont
			{J.~J.}\ \bibnamefont {Renema}}, \bibinfo {author} {\bibfnamefont {Y.-H.}\
			\bibnamefont {Huo}}, \bibinfo {author} {\bibfnamefont {H.}~\bibnamefont
			{Wang}}, \bibinfo {author} {\bibfnamefont {C.-Y.}\ \bibnamefont {Lu}},\ and\
		\bibinfo {author} {\bibfnamefont {J.-W.}\ \bibnamefont {Pan}},\ }\bibfield
	{title} {\bibinfo {title} {Heralded three-photon entanglement from a
			single-photon source on a photonic chip},\ }\href
	{https://doi.org/10.1103/PhysRevLett.132.130603} {\bibfield  {journal}
		{\bibinfo  {journal} {Phys. Rev. Lett.}\ }\textbf {\bibinfo {volume} {132}},\
		\bibinfo {pages} {130603} (\bibinfo {year} {2024})}\BibitemShut {NoStop}%
	\bibitem [{\citenamefont {Massar}\ and\ \citenamefont
		{Popescu}(1995)}]{Classical_limit_1995}%
	\BibitemOpen
	\bibfield  {author} {\bibinfo {author} {\bibfnamefont {S.}~\bibnamefont
			{Massar}}\ and\ \bibinfo {author} {\bibfnamefont {S.}~\bibnamefont
			{Popescu}},\ }\bibfield  {title} {\bibinfo {title} {Optimal extraction of
			information from finite quantum ensembles},\ }\href
	{https://doi.org/10.1103/PhysRevLett.74.1259} {\bibfield  {journal} {\bibinfo
			{journal} {Phys. Rev. Lett.}\ }\textbf {\bibinfo {volume} {74}},\ \bibinfo
		{pages} {1259} (\bibinfo {year} {1995})}\BibitemShut {NoStop}%
	\bibitem [{\citenamefont {Cavalcanti}\ \emph {et~al.}(2017)\citenamefont
		{Cavalcanti}, \citenamefont {Skrzypczyk},\ and\ \citenamefont
		{{\v{S}}upi{\'c}}}]{cavalcanti2017all}%
	\BibitemOpen
	\bibfield  {author} {\bibinfo {author} {\bibfnamefont {D.}~\bibnamefont
			{Cavalcanti}}, \bibinfo {author} {\bibfnamefont {P.}~\bibnamefont
			{Skrzypczyk}},\ and\ \bibinfo {author} {\bibfnamefont {I.}~\bibnamefont
			{{\v{S}}upi{\'c}}},\ }\bibfield  {title} {\bibinfo {title} {All entangled
			states can demonstrate nonclassical teleportation},\ }\href@noop {}
	{\bibfield  {journal} {\bibinfo  {journal} {Physical Review Letters}\
		}\textbf {\bibinfo {volume} {119}},\ \bibinfo {pages} {110501} (\bibinfo
		{year} {2017})}\BibitemShut {NoStop}%
	\bibitem [{\citenamefont {Carvacho}\ \emph {et~al.}(2018)\citenamefont
		{Carvacho}, \citenamefont {Andreoli}, \citenamefont {Santodonato},
		\citenamefont {Bentivegna}, \citenamefont {D’Ambrosio}, \citenamefont
		{Skrzypczyk}, \citenamefont {{\v{S}}upi{\'c}}, \citenamefont {Cavalcanti},\
		and\ \citenamefont {Sciarrino}}]{carvacho2018experimental}%
	\BibitemOpen
	\bibfield  {author} {\bibinfo {author} {\bibfnamefont {G.}~\bibnamefont
			{Carvacho}}, \bibinfo {author} {\bibfnamefont {F.}~\bibnamefont {Andreoli}},
		\bibinfo {author} {\bibfnamefont {L.}~\bibnamefont {Santodonato}}, \bibinfo
		{author} {\bibfnamefont {M.}~\bibnamefont {Bentivegna}}, \bibinfo {author}
		{\bibfnamefont {V.}~\bibnamefont {D’Ambrosio}}, \bibinfo {author}
		{\bibfnamefont {P.}~\bibnamefont {Skrzypczyk}}, \bibinfo {author}
		{\bibfnamefont {I.}~\bibnamefont {{\v{S}}upi{\'c}}}, \bibinfo {author}
		{\bibfnamefont {D.}~\bibnamefont {Cavalcanti}},\ and\ \bibinfo {author}
		{\bibfnamefont {F.}~\bibnamefont {Sciarrino}},\ }\bibfield  {title} {\bibinfo
		{title} {Experimental study of nonclassical teleportation beyond average
			fidelity},\ }\href@noop {} {\bibfield  {journal} {\bibinfo  {journal}
			{Physical Review Letters}\ }\textbf {\bibinfo {volume} {121}},\ \bibinfo
		{pages} {140501} (\bibinfo {year} {2018})}\BibitemShut {NoStop}%
	\bibitem [{\citenamefont {Bussandri}\ \emph {et~al.}(2024)\citenamefont
		{Bussandri}, \citenamefont {Bosyk},\ and\ \citenamefont
		{Toscano}}]{bussandri2024challenges}%
	\BibitemOpen
	\bibfield  {author} {\bibinfo {author} {\bibfnamefont {D.}~\bibnamefont
			{Bussandri}}, \bibinfo {author} {\bibfnamefont {G.}~\bibnamefont {Bosyk}},\
		and\ \bibinfo {author} {\bibfnamefont {F.}~\bibnamefont {Toscano}},\
	}\bibfield  {title} {\bibinfo {title} {Challenges in certifying quantum
			teleportation: Moving beyond the conventional fidelity benchmark},\
	}\href@noop {} {\bibfield  {journal} {\bibinfo  {journal} {Physical Review
				A}\ }\textbf {\bibinfo {volume} {109}},\ \bibinfo {pages} {032618} (\bibinfo
		{year} {2024})}\BibitemShut {NoStop}%
	\bibitem [{\citenamefont {Renema}(2020)}]{Renema_BS_superp}%
	\BibitemOpen
	\bibfield  {author} {\bibinfo {author} {\bibfnamefont {J.~J.}\ \bibnamefont
			{Renema}},\ }\bibfield  {title} {\bibinfo {title} {Simulability of partially
			distinguishable superposition and gaussian boson sampling},\ }\href
	{https://doi.org/10.1103/PhysRevA.101.063840} {\bibfield  {journal} {\bibinfo
			{journal} {Phys. Rev. A}\ }\textbf {\bibinfo {volume} {101}},\ \bibinfo
		{pages} {063840} (\bibinfo {year} {2020})}\BibitemShut {NoStop}%
	\bibitem [{\citenamefont {Flamini}\ \emph {et~al.}(2018)\citenamefont
		{Flamini}, \citenamefont {Spagnolo},\ and\ \citenamefont
		{Sciarrino}}]{Flamini_rev}%
	\BibitemOpen
	\bibfield  {author} {\bibinfo {author} {\bibfnamefont {F.}~\bibnamefont
			{Flamini}}, \bibinfo {author} {\bibfnamefont {N.}~\bibnamefont {Spagnolo}},\
		and\ \bibinfo {author} {\bibfnamefont {F.}~\bibnamefont {Sciarrino}},\
	}\bibfield  {title} {\bibinfo {title} {Photonic quantum information
			processing: a review},\ }\href {https://doi.org/10.1088/1361-6633/aad5b2}
	{\bibfield  {journal} {\bibinfo  {journal} {Rep. Prog. Phys.}\ }\textbf
		{\bibinfo {volume} {82}},\ \bibinfo {pages} {016001} (\bibinfo {year}
		{2018})}\BibitemShut {NoStop}%
	\bibitem [{\citenamefont {Caspar}\ \emph {et~al.}(2020)\citenamefont {Caspar},
		\citenamefont {Verbanis}, \citenamefont {Oudot}, \citenamefont {Maring},
		\citenamefont {Samara}, \citenamefont {Caloz}, \citenamefont {Perrenoud},
		\citenamefont {Sekatski}, \citenamefont {Martin}, \citenamefont {Sangouard},
		\citenamefont {Zbinden},\ and\ \citenamefont {Thew}}]{Caspar2020}%
	\BibitemOpen
	\bibfield  {author} {\bibinfo {author} {\bibfnamefont {P.}~\bibnamefont
			{Caspar}}, \bibinfo {author} {\bibfnamefont {E.}~\bibnamefont {Verbanis}},
		\bibinfo {author} {\bibfnamefont {E.}~\bibnamefont {Oudot}}, \bibinfo
		{author} {\bibfnamefont {N.}~\bibnamefont {Maring}}, \bibinfo {author}
		{\bibfnamefont {F.}~\bibnamefont {Samara}}, \bibinfo {author} {\bibfnamefont
			{M.}~\bibnamefont {Caloz}}, \bibinfo {author} {\bibfnamefont
			{M.}~\bibnamefont {Perrenoud}}, \bibinfo {author} {\bibfnamefont
			{P.}~\bibnamefont {Sekatski}}, \bibinfo {author} {\bibfnamefont
			{A.}~\bibnamefont {Martin}}, \bibinfo {author} {\bibfnamefont
			{N.}~\bibnamefont {Sangouard}}, \bibinfo {author} {\bibfnamefont
			{H.}~\bibnamefont {Zbinden}},\ and\ \bibinfo {author} {\bibfnamefont {R.~T.}\
			\bibnamefont {Thew}},\ }\bibfield  {title} {\bibinfo {title} {Heralded
			distribution of single-photon path entanglement},\ }\href
	{https://doi.org/10.1103/PhysRevLett.125.110506} {\bibfield  {journal}
		{\bibinfo  {journal} {Phys. Rev. Lett.}\ }\textbf {\bibinfo {volume} {125}},\
		\bibinfo {pages} {110506} (\bibinfo {year} {2020})}\BibitemShut {NoStop}%
	\bibitem [{\citenamefont {Monteiro}\ \emph {et~al.}(2017)\citenamefont
		{Monteiro}, \citenamefont {Verbanis}, \citenamefont {Vivoli}, \citenamefont
		{Martin}, \citenamefont {Gisin}, \citenamefont {Zbinden},\ and\ \citenamefont
		{Thew}}]{Monteiro2017}%
	\BibitemOpen
	\bibfield  {author} {\bibinfo {author} {\bibfnamefont {F.}~\bibnamefont
			{Monteiro}}, \bibinfo {author} {\bibfnamefont {E.}~\bibnamefont {Verbanis}},
		\bibinfo {author} {\bibfnamefont {V.~C.}\ \bibnamefont {Vivoli}}, \bibinfo
		{author} {\bibfnamefont {A.}~\bibnamefont {Martin}}, \bibinfo {author}
		{\bibfnamefont {N.}~\bibnamefont {Gisin}}, \bibinfo {author} {\bibfnamefont
			{H.}~\bibnamefont {Zbinden}},\ and\ \bibinfo {author} {\bibfnamefont {R.~T.}\
			\bibnamefont {Thew}},\ }\bibfield  {title} {\bibinfo {title} {Heralded
			amplification of path entangled quantum states},\ }\href
	{https://doi.org/10.1088/2058-9565/aa70ad} {\bibfield  {journal} {\bibinfo
			{journal} {Quantum Science and Technology}\ }\textbf {\bibinfo {volume}
			{2}},\ \bibinfo {pages} {024008} (\bibinfo {year} {2017})}\BibitemShut
	{NoStop}%
	\bibitem [{\citenamefont {Erk{\i}l{\i}{\c{c}}}\ \emph
		{et~al.}(2023)\citenamefont {Erk{\i}l{\i}{\c{c}}}, \citenamefont {Conlon},
		\citenamefont {Shajilal}, \citenamefont {Kish}, \citenamefont {Tserkis},
		\citenamefont {Kim}, \citenamefont {Lam},\ and\ \citenamefont
		{Assad}}]{erkilicc2023surpassing}%
	\BibitemOpen
	\bibfield  {author} {\bibinfo {author} {\bibfnamefont {{\"O}.}~\bibnamefont
			{Erk{\i}l{\i}{\c{c}}}}, \bibinfo {author} {\bibfnamefont {L.}~\bibnamefont
			{Conlon}}, \bibinfo {author} {\bibfnamefont {B.}~\bibnamefont {Shajilal}},
		\bibinfo {author} {\bibfnamefont {S.}~\bibnamefont {Kish}}, \bibinfo {author}
		{\bibfnamefont {S.}~\bibnamefont {Tserkis}}, \bibinfo {author} {\bibfnamefont
			{Y.-S.}\ \bibnamefont {Kim}}, \bibinfo {author} {\bibfnamefont {P.~K.}\
			\bibnamefont {Lam}},\ and\ \bibinfo {author} {\bibfnamefont {S.~M.}\
			\bibnamefont {Assad}},\ }\bibfield  {title} {\bibinfo {title} {Surpassing the
			repeaterless bound with a photon-number encoded
			measurement-device-independent quantum key distribution protocol},\ }\href
	{https://doi.org/10.1038/s41534-023-00698-5} {\bibfield  {journal} {\bibinfo
			{journal} {npj Quantum Information}\ }\textbf {\bibinfo {volume} {9}},\
		\bibinfo {pages} {29} (\bibinfo {year} {2023})}\BibitemShut {NoStop}%
	\bibitem [{\citenamefont {Karli}\ \emph {et~al.}(2024)\citenamefont {Karli},
		\citenamefont {Vajner}, \citenamefont {Kappe}, \citenamefont {Hagen},
		\citenamefont {Hansen}, \citenamefont {Schwarz}, \citenamefont {Bracht},
		\citenamefont {Schimpf}, \citenamefont {Covre~da Silva}, \citenamefont
		{Walther}, \citenamefont {Rastelli}, \citenamefont {Axt}, \citenamefont
		{Loredo}, \citenamefont {Remesh}, \citenamefont {Heindel}, \citenamefont
		{Reiter},\ and\ \citenamefont {Weihs}}]{karli2023controlling}%
	\BibitemOpen
	\bibfield  {author} {\bibinfo {author} {\bibfnamefont {Y.}~\bibnamefont
			{Karli}}, \bibinfo {author} {\bibfnamefont {D.~A.}\ \bibnamefont {Vajner}},
		\bibinfo {author} {\bibfnamefont {F.}~\bibnamefont {Kappe}}, \bibinfo
		{author} {\bibfnamefont {P.~C.~A.}\ \bibnamefont {Hagen}}, \bibinfo {author}
		{\bibfnamefont {L.~M.}\ \bibnamefont {Hansen}}, \bibinfo {author}
		{\bibfnamefont {R.}~\bibnamefont {Schwarz}}, \bibinfo {author} {\bibfnamefont
			{T.~K.}\ \bibnamefont {Bracht}}, \bibinfo {author} {\bibfnamefont
			{C.}~\bibnamefont {Schimpf}}, \bibinfo {author} {\bibfnamefont {S.~F.}\
			\bibnamefont {Covre~da Silva}}, \bibinfo {author} {\bibfnamefont
			{P.}~\bibnamefont {Walther}}, \bibinfo {author} {\bibfnamefont
			{A.}~\bibnamefont {Rastelli}}, \bibinfo {author} {\bibfnamefont {V.~M.}\
			\bibnamefont {Axt}}, \bibinfo {author} {\bibfnamefont {J.~C.}\ \bibnamefont
			{Loredo}}, \bibinfo {author} {\bibfnamefont {V.}~\bibnamefont {Remesh}},
		\bibinfo {author} {\bibfnamefont {T.}~\bibnamefont {Heindel}}, \bibinfo
		{author} {\bibfnamefont {D.~E.}\ \bibnamefont {Reiter}},\ and\ \bibinfo
		{author} {\bibfnamefont {G.}~\bibnamefont {Weihs}},\ }\bibfield  {title}
	{\bibinfo {title} {Controlling the photon number coherence of solid-state
			quantum light sources for quantum cryptography},\ }\href
	{https://doi.org/10.1038/s41534-024-00811-2} {\bibfield  {journal} {\bibinfo
			{journal} {npj Quantum Information}\ }\textbf {\bibinfo {volume} {10}},\
		\bibinfo {pages} {17} (\bibinfo {year} {2024})}\BibitemShut {NoStop}%
	\bibitem [{\citenamefont {Yang}\ \emph {et~al.}(2022)\citenamefont {Yang},
		\citenamefont {Lund}, \citenamefont {Pohl}, \citenamefont {Lodahl},\ and\
		\citenamefont {M\o{}lmer}}]{Yang_sorting_2022}%
	\BibitemOpen
	\bibfield  {author} {\bibinfo {author} {\bibfnamefont {F.}~\bibnamefont
			{Yang}}, \bibinfo {author} {\bibfnamefont {M.~M.}\ \bibnamefont {Lund}},
		\bibinfo {author} {\bibfnamefont {T.}~\bibnamefont {Pohl}}, \bibinfo {author}
		{\bibfnamefont {P.}~\bibnamefont {Lodahl}},\ and\ \bibinfo {author}
		{\bibfnamefont {K.}~\bibnamefont {M\o{}lmer}},\ }\bibfield  {title} {\bibinfo
		{title} {Deterministic photon sorting in waveguide qed systems},\ }\href
	{https://doi.org/10.1103/PhysRevLett.128.213603} {\bibfield  {journal}
		{\bibinfo  {journal} {Phys. Rev. Lett.}\ }\textbf {\bibinfo {volume} {128}},\
		\bibinfo {pages} {213603} (\bibinfo {year} {2022})}\BibitemShut {NoStop}%
\end{thebibliography}


\end{document}

\title{Supplementary Information: Teleportation of a genuine single-rail vacuum–one-photon qubit generated via a quantum dot source}

%-----AUTHORS BLOCK-----
%\author{Author List}
%\affiliation{Dipartimento di Fisica, Sapienza Universit\`{a} di Roma, Piazzale Aldo Moro 5, I-00185 Roma, Italy}

\author{Beatrice Polacchi}
\affiliation{Dipartimento di Fisica - Sapienza Universit\`{a} di Roma, P.le Aldo Moro 5, I-00185 Roma, Italy}

\author{Francesco Hoch}
\affiliation{Dipartimento di Fisica - Sapienza Universit\`{a} di Roma, P.le Aldo Moro 5, I-00185 Roma, Italy}

\author{Giovanni Rodari}
\affiliation{Dipartimento di Fisica - Sapienza Universit\`{a} di Roma, P.le Aldo Moro 5, I-00185 Roma, Italy}

\author{Stefano Savo}
\affiliation{Dipartimento di Fisica - Sapienza Universit\`{a} di Roma, P.le Aldo Moro 5, I-00185 Roma, Italy}

\author{Gonzalo Carvacho}
\affiliation{Dipartimento di Fisica - Sapienza Universit\`{a} di Roma, P.le Aldo Moro 5, I-00185 Roma, Italy}

\author{Nicol\`o Spagnolo}
\affiliation{Dipartimento di Fisica - Sapienza Universit\`{a} di Roma, P.le Aldo Moro 5, I-00185 Roma, Italy}

\author{Taira Giordani}
\email{taira.giordani@uniroma1.it}
\affiliation{Dipartimento di Fisica - Sapienza Universit\`{a} di Roma, P.le Aldo Moro 5, I-00185 Roma, Italy}

\author{Fabio Sciarrino}
\affiliation{Dipartimento di Fisica - Sapienza Universit\`{a} di Roma, P.le Aldo Moro 5, I-00185 Roma, Italy}

\maketitle

\section{Characterization of the probe state via self-homodyning in a Mach-Zehnder interferometer}
In this Section, we describe how the vacuum population of our target states is formally linked to their measured fringe visibility when measured in a time-unbalanced MZI. These results will allow us to extrapolate the purity of the states generated by our source, as well as to find the relationship between the target state visibility $V$ and the visibility of the teleported state $V_T$, as reported in the next section.
%Performing a complete analysis of the Mach-Zehnder interferometer is quite challenging, however, certain considerations make it possible to simplify the calculations.

We start by considering the quantum state entering the MZI, which is composed, in principle, by a train of $n$ photon states.  However, due to the losses in our setup, mainly caused by a low source extraction ($\approx 10\%$) and detection efficiency ($\approx 30\%$), we map the interferometer of Fig. 2c in the main text as if the two states $\ket{\psi} = \alpha \ket{0} + \beta \ket{1}$ and $\ket{\psi}_{\phi} = \alpha \ket{0} + \beta e^{i\phi} \ket{1}$, respectively, were impinging on a beam splitter, with high losses at the output. Under these assumptions, the input state is:
\begin{equation}
    \ket{\psi_{\mathrm{in}}} = \ket{\psi} \ket{\psi}_{\phi} =  \left(\alpha +\beta a_1^\x\right)\left(\alpha+\beta e^{i\phi} a_2^\x\right) \ket{0}
\end{equation}
After the beam splitter, the state evolves into:
\begin{equation}
    \ket{\psi_{\mathrm{out}}} = \left(\alpha +\beta \frac{a_1^\x+a_2^\x}{\sqrt{2}} \right)\left(\alpha+\beta e^{i\phi} \frac{a_1^\x-a_2^\x}{\sqrt{2}}\right) \ket{0} = \left[\alpha^2+ \frac{\alpha \beta (1+ e^{i\phi})}{\sqrt{2}} a_1^\x+\frac{\alpha \beta (1- e^{i\phi})}{\sqrt{2}} a_2^\x+ \frac{\beta^2e^{i\phi}}{2}(a_1^{\x 2}-a_2^{\x 2})\right] \ket{0}.
\end{equation}
Considering that, for threshold detectors, a single photon has probability $\eta$ to be detected, while two-photon inputs trigger the detector with probability $1 - (1 - \eta)^2$, we find that the single-photon detection  probability is:
\begin{equation}
\label{eq:prob_det_1}
    P(\phi)_{\ket{\psi}} = \eta \frac{\abs{\alpha}^2\abs{\beta}^2}{2}(2+2\cos \phi)+\frac{\abs{\beta}^4}{4}2 [\eta^2+2\eta(1-\eta)] = \frac{\eta\beta^2}{2}(2-\abs{\beta}^2\eta + 2 \abs{\alpha}^2\cos \phi),
\end{equation}
being $\eta$ the overall loss parameter. By defining the visibility of the fringe pattern as:
\begin{equation}
V = \frac{\mathrm{max}_{\phi}P(\phi)_{\ket{\psi}}  - \mathrm{min}_{\phi}P(\phi)_{\ket{\psi}} }{\mathrm{max}_{\phi}P(\phi)_{\ket{\psi}}  + \mathrm{min}_{\phi}P(\phi)_{\ket{\psi}} },
\end{equation}
we find that, in the limit of high losses, ($\eta \rightarrow 0 $) the visibility is:
\begin{equation}
    V = \abs{\alpha}^2.
\label{eq:vis_s}
\end{equation}

This result can be extended to include partial photon distinguishability between the photon modes. This is obtained by considering an input state of the form:
\begin{equation}
    \ket{\psi_{\mathrm{in}}^{x_1, x_2}} = \left(\alpha +\beta \sqrt{x_A} a_1^\x + \beta \sqrt{1-x_A} b_1^\x\right)\left(\alpha+\beta e^{i\phi} \sqrt{x_B} a_2^\x + \beta e^{i \phi} \sqrt{1-x_B} c_2^\x\right) \ket{0},
\end{equation}
where $b_1^\x$ and $c_2^\x$ are two additional fictious modes, mutually orthogonal and orthogonal to $(a_1^\x, a_2^\x)$, that are used to model the effect of partially distinguishability due to the photon internal degrees of freedom. Parameters ($x_A, x_B$) are related to the Hong-Ou-Mandel visibility between the two photons as $V_{\mathrm{HOM}} = x_A x_B$. Moreover, we include the purity of the state generated by the source. 

{Indeed, the unnormalized density matrix that our source generates, when considering the detection efficiency $\eta$, would read:
\begin{equation*}
    \ketbra{0}{0} + \eta \left( \lambda \rho_{pure} + (1-\lambda) \ketbra{1}{1} \right) + \mathcal{O}(\eta^2)
\end{equation*}
where all incoherent vacuum contributions (photon losses) are included in the first term and we neglect the $\mathcal{O}(\eta^2)$ terms, since $\eta \rightarrow 0$. Since, with our detection scheme, we have no way to measure the incoherent vacuum contributions, then the remaining state, conditioned to event detections, reads:
%Indeed, each of the two states impinging on the beam splitter can be modeled as follows:
\begin{equation}
    \rho = \lambda \rho_{\text{pure}} + (1-\lambda) \rho_{\text{mixed}}
\end{equation}
where $\rho_{\text{pure}} = \ket{\psi} \bra{\psi}$ and $\rho_{\text{mixed}} = |\alpha|^2 \ket{0}\bra{0} + |\beta|^2 \ket{1}\bra{1}$. }
{In this sense, we use the term “conditional” purity. This parameter does not depend on the exact amount of losses \cite{loredo2019generation}.}
By repeating the calculation above, and by tracing over the internal degrees of freedom of the photons, the fringe visibility of the state at the output of the interferometer is found to be $V = \lambda^2 \sqrt{x_A x_B} |\alpha|^2$. We finally obtain:
\begin{equation}
V = \lambda^2 \sqrt{V_{\mathrm{HOM}}} |\alpha|^2
\end{equation}

This relationship can be employed to $\lambda$, from the linear fit reported in the inset of Fig.~3a of the main text. Indeed, by considering also the high loss regime, the measured signal rate is proportional to: 
\begin{equation}
S_c \propto \langle P_1(\phi) \rangle_\phi \propto (1 - \alpha^2)
\end{equation}
which means that the visibility $V$ is by itself linearly proportional to $S_c$, with slope dependent from the purity $\lambda$.

{
\section{Characterization of the entangled states via a Mach-Zehnder interferometer}}

\begin{figure}[ht]
    \centering
    \includegraphics[width=0.4\textwidth]{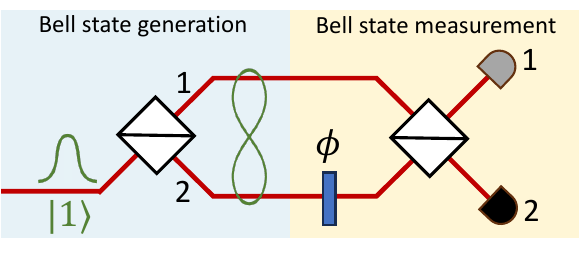}
    \caption{{\textbf{Bell state generation and analyzer setup.} The first beam splitter generates the Bell state while the second one is used to perform a partial Bell state measurement.}
    \label{fig:MZI_bell_state}}
\end{figure} 

{
The quantum resource required for a genuine teleportation is an entangled state shared by Alice and Bob. Here, we demonstrate that Alice and Bob share such quantum resources. Indeed, we measured the fidelity of our entangled state by using the setup shown in Fig.~\ref{fig:MZI_bell_state}. which is a standard Mach-Zehnder interferometer.
The Bell state $\ket{\psi^+}$ is generated by sending one photon in the first unbiased beam splitter while the Bell state measurement (BSM) is performed by the second one. %standard photonic setup for partial Bell state measurements in the path--photon-number degree of freedom. 
More precisely, % one photon enters a balanced Mach-Zehnder interferometer.
 the ideal state after the first symmetric beam splitter is the following:
\begin{equation}
    \ket{\psi^+} = \frac{\ket{10}_{12}+\ket{01}_{12}}{\sqrt{2}}
    \label{eq:bell_state}
\end{equation}
The second beam splitter mirrors implements the inverse transformation of the first one such that the detection probability of one photon at output modes 1,2  oscillates as $P_{1,\text{click}} = \cos^2{(\phi/2)};~ P_{2,\text{click}} = \sin^2{(\phi/2)}$.
%On the other hand, the fidelity of the state above with the Bell state 
%\begin{equation*}
%    \ket{\psi^+} = \frac{\ket{10}_{12}+\ket{01}_{12}}{\sqrt{2}}
%\end{equation*}
The oscillating phase factor $\phi$ is only due to the optical path difference in the Mach-Zehnder interferometer. %, hence, we can suppose to fix its value to zero.
Consequently, the fidelity of our experimental entangled state with the Bell state $\ket{\psi^+}$ is given by the $\max_{\phi}{P^{exp}_{1,\text{click}}(\phi)}$.
At the experimental level, the value of the fidelity with the Bell state can be retrieved from the measurement of the visibility $V$ of the fringes at the output of the MZI, according to the relationship $F=\frac{1+V}{2}$. Such a derivation and argument is largely discussed in Refs. \cite{lombardi2002teleportation, fattal2004quantum}, the two works that reported the first implementations of a vacuum-one-photon teleportation protocol.
Discrepancies from the ideal fidelity are mainly due to a partially asymmetric beam splitter and a non-vanishing $g^2(0)$ value. In our setup the MZI fringes visibility is near unity, thus demonstrating, on one hand, the presence of coherence in the superposition state in \eqref{eq:bell_state}, and on the other, the high fidelity with the Bell state. The high fidelity obtained ensures that Alice and Bob share an entangled resource and, therefore, that the teleportation protocol here reported exploits genuine quantum resources and would not be reproducible with only classical resources.
}

{
\section{Classical bound for the fringe visibility}}

{We now discuss the formal derivation of the visibility one would observe when using the classical strategy reported in \cite{Classical_limit_1995} and known as the \textit{measure-and-prepare} protocol.
From \cite{Classical_limit_1995}, we know that the state {closest to the target} that Alice can prepare without performing a full quantum state tomography of the qubit is the following:
%In this protocol, the most similar state to the target that Alice can prepare has the following form:
\begin{equation}
    \rho_{CT} = F \ketbra{\psi}{\psi} + (1-F) \ketbra{\psi_{\perp}}{\psi_{\perp}}
\label{eq:classical_teleported}
\end{equation}
where the parameter $F$ is the fidelity with the target state $\ket{\psi}$, {which} can {only take values in $[1/3, 2/3]$}, while $\ket{\psi_{\perp}} = \beta^* \ket{0} - \alpha^* \ket{1}$. 
In particular, in the best case, this qubit will have $F=2/3$ with the target state $\ket{\psi}$ and can also be written as $\rho_{CT} = \frac{1}{3} \ketbra{\psi}{\psi}+ \frac{1}{3} \ketbra{0}{0}+\frac{1}{3} \ketbra{1}{1}$. Consequently, in such a case, the fidelity with $\ket{\psi_{\perp}}$ is $F=1/3$.}

{Considering our implementation, the classical bound can be retrieved by evaluating the fringe visibility recorded in Bob's station when a probe qubit $\ket{\psi}_{\phi}$ and the state $\rho_{CT}$ interfere.
The single-photon detection probability in this case reads:
\begin{equation}
    P(\phi)_{\rho_{CT}} = F P(\phi)_{\ket{\psi}} + (1-F)  P(\phi)_{\ket{\psi_{\perp}}}
\end{equation}
The term $P(\phi)_{\ket{\psi}}$ was derived in Supplementary Note~I and is defined in Eq.~\eqref{eq:prob_det_1}.
Instead, the detection probability when the states $\ket{\psi}_{\phi}$ and $\ket{\psi_{\perp}}$ interfere reads:
\begin{equation}
    P(\phi)_{\ket{\psi_{\perp}}} =  \frac{\eta}{2}(1-2\abs{\alpha}^2\abs{\beta}^2\cos{\phi}) - \frac{\abs{\alpha}^2\abs{\beta}^2}{2}\eta^2
\end{equation}
In the high-loss limit ($\eta \longrightarrow 0$), the fringe visibility reads:
\begin{equation}
    V_{\rho_{CT}} = \frac{2 \abs{\alpha}^2\abs{\beta}^2\abs{2F-1}}{1+F(2\abs{\beta}^2 -1)}
\end{equation}
which, using $V = \abs{\alpha}^2$ (derived in Eq.~\eqref{eq:vis_s} of Supplementary Note~I), becomes:
\begin{equation}
    V_{\rho_{CT}} = \frac{2 V (1-V) \abs{2F-1}}{1+F(1-2V)}
    \label{eq:vsvr}
\end{equation}
We analyze $V_{\rho_{CT}}$ as a function of $V$ in Supplementary Fig.~\ref{fig:quantum-classical-limit}, {for different values of the fidelity $F$}. 
We first observe that, for fixed values of $F$, the curve is always monotonously decreasing for $ V \in [0, 1/2]$ and monotonously increasing for $V \in [1/2,1]$. In particular, for $F = 1/2$, the curve $V_{\rho_{CT}}(V)$ is constantly zero.
We also notice that, for {$F \in (1/3, 2/3)$}, all curves {are bounded by zero and the red and the blue plots. As an example, we show the curves for five different values of the fidelity $F = \{1/2,~3/5,~2/5,~3/7,~3/8\} \in (1/3,2/3)$.} {We highlighted this area in purple to represent} the convolution of the two curves for $F=1/3$ and $F=2/3$. 
This means that the whole classically accessible region (highlighted in {purple}) is delimited by the value zero and the {function $\max_{F}\{{V_T(V,F)}\}$}, which define the bound achievable by using the classical strategy {\textit{measure-and-prepare}}. This region is also highlighted in {purple} in Fig.~3a of the main text.\\
\begin{figure}[t]
    \centering
    \includegraphics[width = 0.65\textwidth]{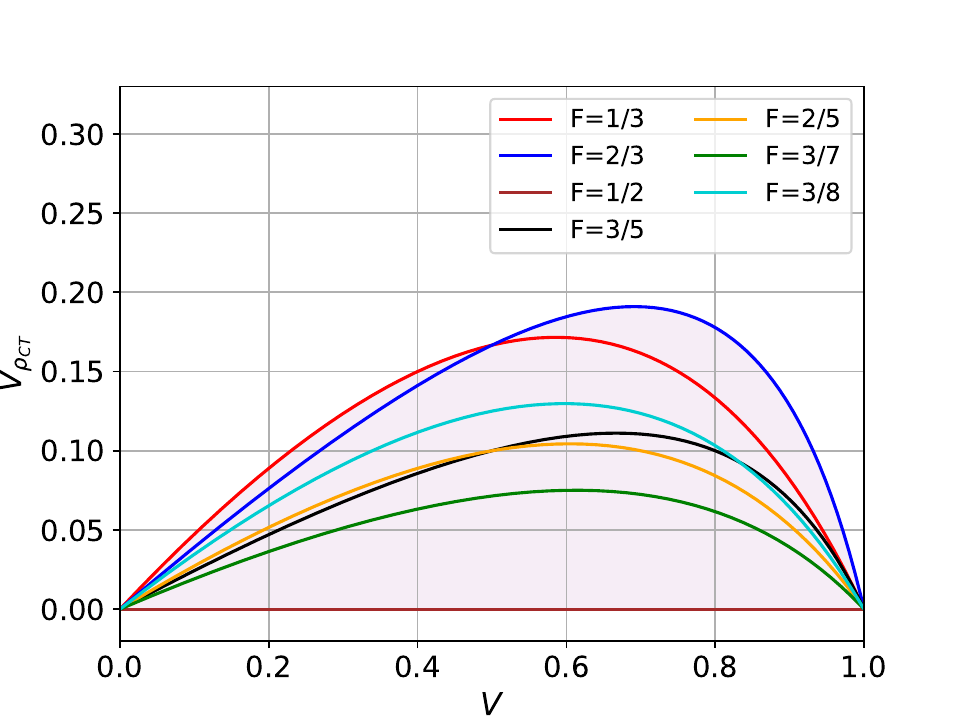}
    \caption{{\textbf{Classical bound on the fringe visibility in the \textit{measure-and-prepare} protocol. a)} The purple {area} defines the classical {region} for the visibility of the state $\rho_{CT}$ ($V_{\rho_{CT}}$) in terms of the visibility of the target state ($V$), using the \textit{measure-and-prepare} strategy. {This region is bounded by the red and blue curves, corresponding respectively to $F=1/3$ and $F=2/3$.}  For $F \in (1/3, 2/3)$, all curves are below the {red and blue} ones, which therefore represents the classical bound on $V_{\rho_{CT}}$.}}
    \label{fig:quantum-classical-limit}
\end{figure}

\section{Full model of the quantum teleportation scheme}

In this section, we develop a model to describe the expected visibility of states teleported with our platform, $V_T$, and discuss its relationship with the visibility of the target state $V$.  We will refer to the spatial modes, to the beam-splitters (BS), and to the delay loops as labeled in Fig.~\ref{fig:teleportation_scheme}.

\begin{figure}[ht!]
    \centering
    \includegraphics[width=0.8\textwidth]{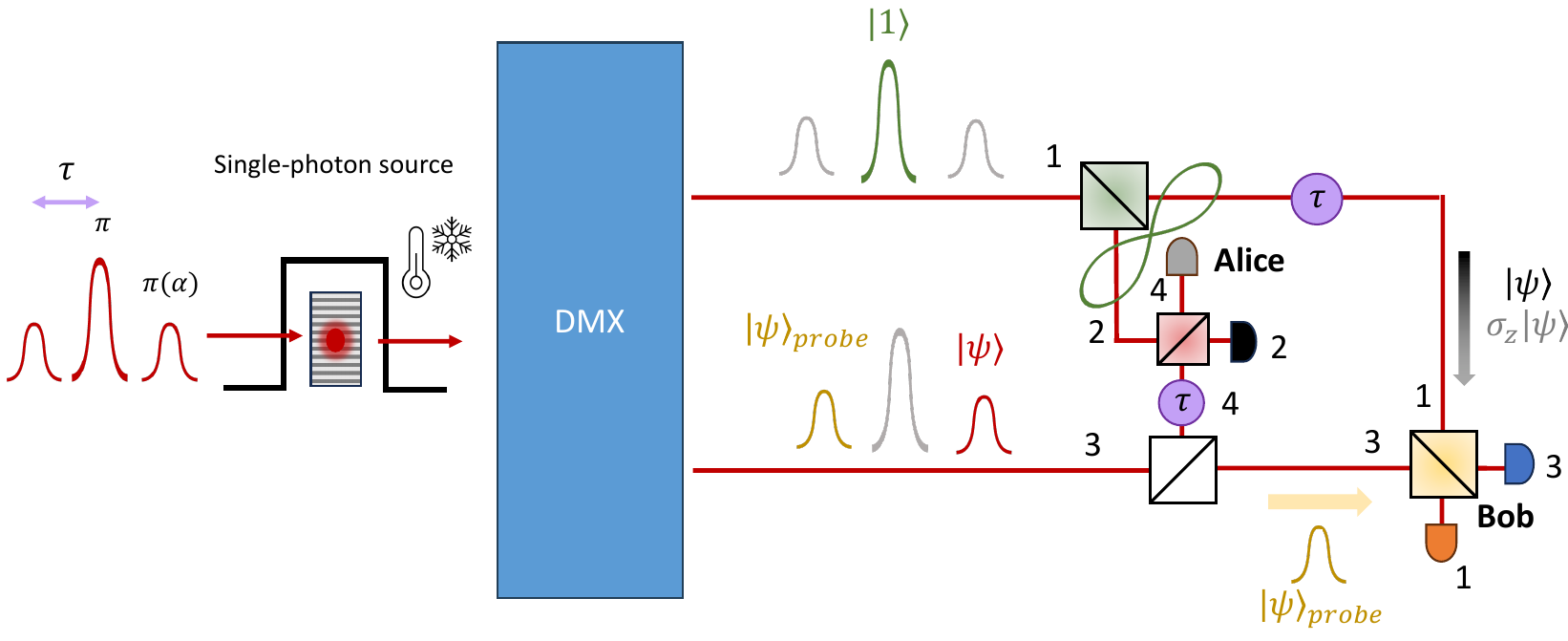}
    \caption{\textbf{Teleportation scheme.} Here we recall the teleportation scheme, already reported in the main text, by adding the modes labels used for the calculations carried out in this section.} 
    \label{fig:teleportation_scheme}
\end{figure}

We first consider an initial state where partial photon indistinguishability is not included, as:
\begin{equation}
    \ket{\psi}_{3, 2\tau} \otimes \ket{1}_{1,\tau} \otimes \ket{\psi}_{3,0} =   \left( \alpha \ket{0} + \beta \ket{1}  \right)_{3, 2\tau} \otimes \ket{1}_{1, \tau}\otimes   \left( \alpha \ket{0} + \gamma \ket{1}  \right)_{3, 0} = \left( \alpha + \beta a^{\x}_{3, 2\tau}  \right)  a^{\x}_{1, \tau}  \left( \alpha + \gamma a^{\x}_{3, 0}  \right) \ket{0},
\end{equation}
{The qubit we want to teleport is $\ket{\psi} =  \alpha\ket{0} + \gamma \ket{1}$}, where $\gamma = |\beta|e^{i\chi}$, $\beta = |\beta|e^{i\phi}$ and $\phi - \chi = \delta$.

After application of $BS_{12}$ the state evolves into:
\begin{equation}
    \left( \alpha + \beta a^{\x}_{3, 2\tau}  \right) \frac{1}{\sqrt{2}} \left( a^{\x}_{1, \tau} +  a^{\x}_{2, \tau}\right) \left( \alpha + \gamma a^{\x}_{3, 0}  \right) \ket{0}.
\end{equation}
{This operation transforms the qubit $\ket{1}_{1,\tau}$ in the following Bell state over modes 1 and 2:
\begin{equation}
    \ket{1}_{1,\tau} \rightarrow \frac{\left( \ket{1}_{1,\tau}^{\mathrm{B}} \ket{0}_{2,\tau}^{\mathrm{A}} + \ket{0}_{1,\tau}^{\mathrm{B}} \ket{1}_{2,\tau}^{\mathrm{A}} \right) }{\sqrt{2}}
\end{equation}
where we highlighted with the superscripts A and B that mode 1 travels to Bob while mode 2 travels to Alice.}
Then, after application of $BS_{34}$ we find:
\begin{equation}
    \left( \alpha + \beta  \frac{a^{\x}_{3, 2\tau} + a^{\x}_{4, 2\tau}}{\sqrt{2}}   \right)  \frac{1}{\sqrt{2}} \left( a^{\x}_{1, \tau} +  a^{\x}_{2, \tau}\right)
   \left( \alpha + \gamma  \frac{a^{\x}_{3, 0} + a^{\x}_{4, 0}}{\sqrt{2}}   \right) \ket{0}.
\end{equation}
After application of delay $\tau$ on mode 4 we obtain:
\begin{equation}
    \left( \alpha + \beta  \frac{a^{\x}_{3, 2\tau} + a^{\x}_{4, 3\tau}}{\sqrt{2}}   \right)  \frac{1}{\sqrt{2}} \left( a^{\x}_{1, \tau} +  a^{\x}_{2, \tau}\right)
   \left( \alpha + \gamma  \frac{a^{\x}_{3, 0} + a^{\x}_{4, \tau}}{\sqrt{2}}   \right) \ket{0}
\end{equation}
{At this point, Alice performs a Bell state measurement between the qubit to be teleported, i.e. $\ket{\psi}$, and the subsystem of the Bell pair that she kept. Indeed, a}fter application of $BS_{24}$, we find:
\begin{equation}
    \left( \alpha + \frac{\beta}{2} ( \sqrt{2}a^{\x}_{3, 2\tau} + a^{\x}_{2, 3\tau} - a^{\x}_{4, 3\tau} )  \right) \frac{1}{2} \left( \sqrt{2}a^{\x}_{1, \tau} +  a^{\x}_{2, \tau} + a^{\x}_{4, \tau}\right)
    \left( \alpha + \frac{\gamma}{2} ( \sqrt{2}a^{\x}_{3, 0} + a^{\x}_{2, \tau} - a^{\x}_{4, \tau} )  \right)  \ket{0}
\end{equation}

{Let us now analyze the state in which the qubit outgoing from mode 1 of $BS_{12}$ and traveling to Bob is left after Alice's Bell state measurement.
According to the teleportation protocol, we expect this qubit to be in a state identical to $\ket{\psi}$, up to a unitary transformation.
We can temporarily neglect the term in the first bracket and consider only the evolution of subsystems $\ket{\psi}_{3,0}$ and $\ket{1}_{1,\tau}$:
\begin{equation}
    \frac{1}{2} \left( \sqrt{2}a^{\x}_{1, \tau} +  a^{\x}_{2, \tau} + a^{\x}_{4, \tau}\right)
    \left( \alpha + \frac{\gamma}{2} ( \sqrt{2}a^{\x}_{3, 0} + a^{\x}_{2, \tau} - a^{\x}_{4, \tau} )  \right)  \ket{0}
\end{equation}
When post-selecting on the occurrence of one detection event in Alice's detector $2$ at time $\tau$, we are left with the following state:
\begin{equation}
    \frac{1}{2} 
    \left( \alpha a^{\x}_{2, \tau}+ \frac{\gamma}{\sqrt{2}} a^{\x}_{2, \tau}a^{\x}_{1, \tau} + \frac{\gamma}{\sqrt{2}} a^{\x}_{3, 0}a^{\x}_{2, \tau} + \frac{\gamma}{2} (a^{\x}_{2, \tau})^2 \right)  \ket{0}
\end{equation}
The term $\frac{\gamma}{\sqrt{2}} a^{\x}_{3, 0} a^{\x}_{2, \tau}$ appears due to the fact that we are employing non-deterministic photon routing to divide $\ket{\psi}$ and $\ket{\psi}_{probe}$.
Moreover, the two-photon contribution $\frac{\gamma}{2} (a^{\x}_{2, \tau})^2$ cannot be distinguished experimentally from a single detection event without using photon-number-resolving detectors.
If we could experimentally exclude such terms (with active photon routing and photon-number-resolving detectors), the teleported state appearing on mode 1 would read:
\begin{equation}
\rho_T^{\mathrm{ideal}} = 
    \begin{pmatrix}
        | \alpha |^2 & \alpha \gamma^* \\
        \alpha^* \gamma & | \gamma |^2
    \end{pmatrix}
    \label{eq:ideal_teleported}
\end{equation}
which is identical to the state $\rho = \ket{\psi} \bra{\psi}$ that we aim to teleport.
In this ideal situation, the fringe visibility $V_T$ at the output of Bob's self-homodyne measurement would be identical to visibility $V$ of the state to be teleported, estimated via a standard MZI-based technique. We show this ideal case in the grey plot in Fig.~3a of the main text.}\\

{However, when taking into account those two terms, the normalized state traveling along mode 1 reads:
\begin{equation}
\rho_T = 
    \frac{1}{|\alpha|^2 + \frac{3}{2} |\gamma|^2}
    \begin{pmatrix}
        | \alpha |^2 + | \gamma |^2 & \frac{\alpha \gamma^*}{\sqrt{2}} \\
        \frac{\alpha^* \gamma}{\sqrt{2}} & \frac{| \gamma |^2}{2}
    \end{pmatrix}
\label{eq:nonideal_tel1}
\end{equation}
Analogous reasoning can be applied to the case in which one detection event occurs in Alice's detector $4$ at time $\tau$, where the final state would read:
\begin{equation}
\rho_T = 
    \frac{1}{|\alpha|^2 + \frac{3}{2} |\gamma|^2}
    \begin{pmatrix}
        | \alpha |^2 + | \gamma |^2 & -\frac{\alpha \gamma^*}{\sqrt{2}} \\
        -\frac{\alpha^* \gamma}{\sqrt{2}} & \frac{| \gamma |^2}{2}
    \end{pmatrix}
\label{eq:nonideal_tel2}
\end{equation}
The difference between the state in Eq.~\eqref{eq:ideal_teleported} and the ones in Eq.~\eqref{eq:nonideal_tel1}-\eqref{eq:nonideal_tel2} causes the observed fringe visibility at Bob's station to be different from the ideal case $V_T = V$.}\\

{We now continue with the calculations to quantify the expected visibility $V_T$ in such a non-ideal case, which is shown in the green plot of Fig.~3a of the main text.}
After application of delay $\tau$ on mode 1 we obtain:
\begin{equation}
    \left( \alpha + \frac{\beta}{2} ( \sqrt{2}a^{\x}_{3, 2\tau} + a^{\x}_{2, 3\tau} - a^{\x}_{4, 3\tau} )  \right)  \frac{1}{2} \left( \sqrt{2}a^{\x}_{1, 2\tau} +  a^{\x}_{2, \tau} + a^{\x}_{4, \tau}\right)
  \left( \alpha + \frac{\gamma}{2} ( \sqrt{2}a^{\x}_{3, 0} + a^{\x}_{2, \tau} - a^{\x}_{4, \tau} )  \right)  \ket{0}
\end{equation}
Finally, after application of $BS_{13}$ the state can be written as:
\begin{equation}
    \left[ \alpha + \frac{\beta}{2} ( a^{\x}_{3, 2\tau} + a^{\x}_{1, 2\tau} + a^{\x}_{2, 3\tau} - a^{\x}_{4, 3\tau} )  \right] \frac{1}{2} \left( a^{\x}_{3, 2\tau} - a^{\x}_{1, 2\tau} +  a^{\x}_{2, \tau} + a^{\x}_{4, \tau}\right)\left[ \alpha + \frac{\gamma}{2} ( a^{\x}_{3, 0} + a^{\x}_{1, 0}  + a^{\x}_{2, \tau} - a^{\x}_{4, \tau} )  \right]  \ket{0}
\end{equation}

The output state at the detection can be also rearranged as:
\begin{gather}
    \begin{split}
        &\frac{\alpha^2}{2} \left[ a^{\x}_{3, 2\tau} - a^{\x}_{1, 2\tau} +  a^{\x}_{2, \tau} - a^{\x}_{4, \tau}  \right] + \frac{\alpha \gamma}{4} \left[  \left(a^{\x}_{3, 2\tau} - a^{\x}_{1, 2\tau} \right) \left(a^{\x}_{3, 0} + a^{\x}_{1, 0} \right) +\left(a^{\x}_{3, 2\tau} - a^{\x}_{1, 2\tau} \right)  \left(a^{\x}_{2, \tau} - a^{\x}_{4, \tau} \right)+ \right.\\
        & +\left. \left(a^{\x}_{2, \tau} + a^{\x}_{4, \tau} \right) \left(a^{\x}_{3, 0} + a^{\x}_{1, 0} \right) +\left(a^{\x}_{2, \tau} + a^{\x}_{4, \tau} \right) \left(a^{\x}_{2, \tau} - a^{\x}_{4, \tau} \right) \right ] + \frac{\alpha \beta}{4} \left[  \left(a^{\x}_{3, 2\tau} + a^{\x}_{1, 2\tau} \right) \left(a^{\x}_{3, 2\tau} - a^{\x}_{1, 2\tau} \right) + \right.\\
        & +\left. \left(a^{\x}_{2, 3\tau} - a^{\x}_{4, 3\tau} \right)  \left(a^{\x}_{2, \tau} + a^{\x}_{4, \tau} \right)+ \left(a^{\x}_{3, 2\tau} + a^{\x}_{1, 2\tau} \right) \left(a^{\x}_{2, \tau} + a^{\x}_{4, \tau} \right) +\left(a^{\x}_{2, 3\tau} - a^{\x}_{4, 3\tau} \right) \left(a^{\x}_{3, 2\tau} - a^{\x}_{1, 2\tau} \right) \right ] + \\ &+\frac{\beta \gamma}{8} \left[ \left( a^{\x}_{3, 2\tau} + a^{\x}_{1, 2\tau} \right) \left( a^{\x}_{3, 2\tau} - a^{\x}_{1, 2\tau} \right) \left(a^{\x}_{3, 0} + a^{\x}_{1,0} \right) + \left( a^{\x}_{3, 2\tau} + a^{\x}_{1, 2\tau} \right) \left(a^{\x}_{3, 2\tau} - a^{\x}_{1, 2\tau} \right) \left(a^{\x}_{2, \tau} - a^{\x}_{4,\tau} \right) \right.+\\
        &+ \left( a^{\x}_{3, 2\tau} + a^{\x}_{1, 2\tau} \right) \left(a^{\x}_{2, \tau} + a^{\x}_{4, \tau} \right) \left(a^{\x}_{3, 0} + a^{\x}_{1,0} \right) + \left( a^{\x}_{3, 2\tau} + a^{\x}_{1, 2\tau} \right) \left(a^{\x}_{2, \tau} + a^{\x}_{4, \tau} \right) \left(a^{\x}_{2, \tau} - a^{\x}_{4,\tau} \right) +\\
        &+ \left( a^{\x}_{2, 3\tau} - a^{\x}_{4, 3\tau} \right) \left( a^{\x}_{3, 2\tau} - a^{\x}_{1, 2\tau} \right) \left(a^{\x}_{3, 0} + a^{\x}_{1,0} \right) + \left( a^{\x}_{2, 3\tau} - a^{\x}_{4, 3\tau} \right) \left(a^{\x}_{3, 2\tau} - a^{\x}_{1, 2\tau} \right) \left(a^{\x}_{2, \tau} - a^{\x}_{4,\tau} \right)  +\\
        &+ \left.\left( a^{\x}_{2, 3\tau} - a^{\x}_{4, 3\tau} \right) \left(a^{\x}_{2, \tau} + a^{\x}_{4, \tau} \right) \left(a^{\x}_{3, 0} + a^{\x}_{1,0} \right) + \left( a^{\x}_{2, 3\tau} - a^{\x}_{4, 3\tau} \right) \left(a^{\x}_{2, \tau} + a^{\x}_{4, \tau} \right) \left(a^{\x}_{2, \tau} - a^{\x}_{4,\tau} \right) \right] \ket{0}
    \end{split}
\end{gather}

From this expression, we can know the probability of having an event where one detector per station clicks. More specifically, we need to consider coincidences event between one detector of the pair (1,3), clicking at time $2 \tau$, and one detector of the pair (2,4) clicking at time $\tau$. The four combinations for all probabilities are found to be:
\begin{gather}
\begin{split}
    P(1,2) &= \frac{ |\beta|^2 }{16} \left(  2 - 2|\alpha|^2 \cos{\delta} \right),\\ 
    P(1,4) &= \frac{ |\beta|^2 }{16} \left(  2 + 2|\alpha|^2 \cos{\delta} \right),\\
    P(3,2) &= \frac{ |\beta|^2 }{16} \left(  2 + 2|\alpha|^2 \cos{\delta} \right),\\
    P(3,4) &= \frac{ |\beta|^2 }{16} \left(  2 - 2|\alpha|^2 \cos{\delta} \right),
\end{split}
\end{gather}
leading to the following visibility:
\begin{equation}
    V_T = |\alpha|^2
\end{equation}

{Until this point, we did not take into account other experimental effects such as losses, purity, and partial indistinguishability. However, a}nalogously to the previous section, we need to consider {also such} different effects.
We begin by including the effect of losses in our setup, analogously to the derivation above. Since the loss operator commutes with the other operators, we can compute the teleported visibility in a lossy platform, by {including all the losses in the detector efficiency. 
With respect to the previous derivation, the probabilities of the interesting events become:
\begin{gather}
\begin{split}
    P(1_{2 \tau},2_{\tau}) &= \frac{ |\beta|^2 \eta_1\eta_2}{32} \left(  6-\eta_1-\eta_2- (2-\eta_1-\eta_2)|\alpha|^2 - 4|\alpha|^2 \cos{\delta} \right),\\ 
    P(1_{2 \tau},4_{\tau}) &= \frac{ |\beta|^2 \eta_1\eta_4}{32} \left(  6-\eta_1-\eta_4- (2-\eta_1-\eta_4)|\alpha|^2 + 4|\alpha|^2 \cos{\delta} \right),\\ 
    P(3_{2 \tau},2_{\tau}) &= \frac{ |\beta|^2 \eta_3\eta_2}{32} \left(  6-\eta_3-\eta_2- (2-\eta_3-\eta_2)|\alpha|^2 + 4|\alpha|^2 \cos{\delta} \right),\\ 
    P(3_{2 \tau},4_{\tau}) &= \frac{ |\beta|^2 \eta_3\eta_4}{32} \left(  6-\eta_3-\eta_4- (2-\eta_3-\eta_4)|\alpha|^2 - 4|\alpha|^2 \cos{\delta} \right),
\end{split}
\end{gather}
where $\eta_1 \dots \eta_4$ are the detection probabilities at the four outputs. In the limit $\eta_{i} \rightarrow 0$, all probabilities have the same visibility which is found to be:
\begin{equation}
    V_T = \frac{2 \abs{\alpha}^2}{3-\abs{\alpha}^2}
\label{eq:v_eff_zero}
\end{equation}
Finally, with analogous procedure to the one discussed in the previous section, we now take into account also the state non-ideal purity $\lambda < 1$ and HOM visibilities. 
%By considering the notation $\alpha = \alpha_T$ for the teleported state, we find that
Hence, the formula in Eq.~\eqref{eq:v_eff_zero} becomes:
\begin{equation}
    V_T = \frac{2 \lambda^2 \sqrt{V^{\text{Alice}}_{\text{HOM}}V^{\text{Bob}}_{\text{HOM}}} |\alpha|^2}{3-|\alpha|^2}.
\end{equation}
In this way, by using Eq.~\eqref{eq:vis_s}, the relationship between the target state visibility $V$ and the visibility $V_T$ of the teleported state in our platform can be written as:
\begin{equation}
    V_T = \frac{2 \lambda^2 \sqrt{V^{\text{Alice}}_{\text{HOM}}V^{\text{Bob}}_{\text{HOM}}} V}{3\lambda^2 \sqrt{V^{\text{Alice}}_{\text{HOM}}} - V}.
\end{equation}
This expression has been used as the green line in Fig.~3a of the main text to identify the expected visibility from the collected experimental data. The experimental values of the target state visibilities and vacuum populations, and teleported visibilities, are reported in Table \ref{tab:results_dmx}.

\begin{table}[h!]
    \centering
        \begin{tabular}{c|c|c}
        $\abs{\alpha}^2$ & $V$ & $V_T$   \\
        \hline
        \hline
        $0.211 \pm 0.001$ & $0.197 \pm 0.001$  & $0.13 \pm 0.02$      \\
        $0.323 \pm 0.005$ & $0.303 \pm 0.005$  & $0.21 \pm 0.02$      \\
        $0.425 \pm 0.002$ & $0.398 \pm 0.002$  & $0.26 \pm 0.02$      \\
        $0.545 \pm 0.006$ & $0.510 \pm 0.006$  & $0.36 \pm 0.03$      \\
        $0.632 \pm 0.004$ & $0.591 \pm 0.004$  & $0.41 \pm 0.04$      \\
        $0.769 \pm 0.005$ & $0.720 \pm 0.005$  & $0.52 \pm 0.05$ 
        \end{tabular}
    \caption{Comparison between the visibility of the teleported state $V_T$ with the visibility of the target state $V$ with vacuum population $|\alpha|^2$, estimated independently via the self-homodyning procedure in a MZI.  %{ analisi ner regime di alte perdite}
    }
    \label{tab:results_dmx}
\end{table}

\begin{comment}

\section{Quantum state teleportation calculations (without DMX)}
    \input{Calcoli}

\end{comment}

\section{Full model of the entanglement swapping experiment}

In this section, we derive the model employed to describe the expected visibility of the output state after the entanglement swapping protocol. In this derivation, we first consider the case where the two interfering states are two single-photon states, and the BS transmittance is $T=0.5$. Then, we will consider a noise model where the states at the input are allowed to have a nonzero vacuum component. We refer to Fig. \ref{fig:enter-label} for mode labelling and for the sequence of operations considered in the model.

\begin{figure}[ht!]
    \centering
    \includegraphics[width = 0.7\textwidth]{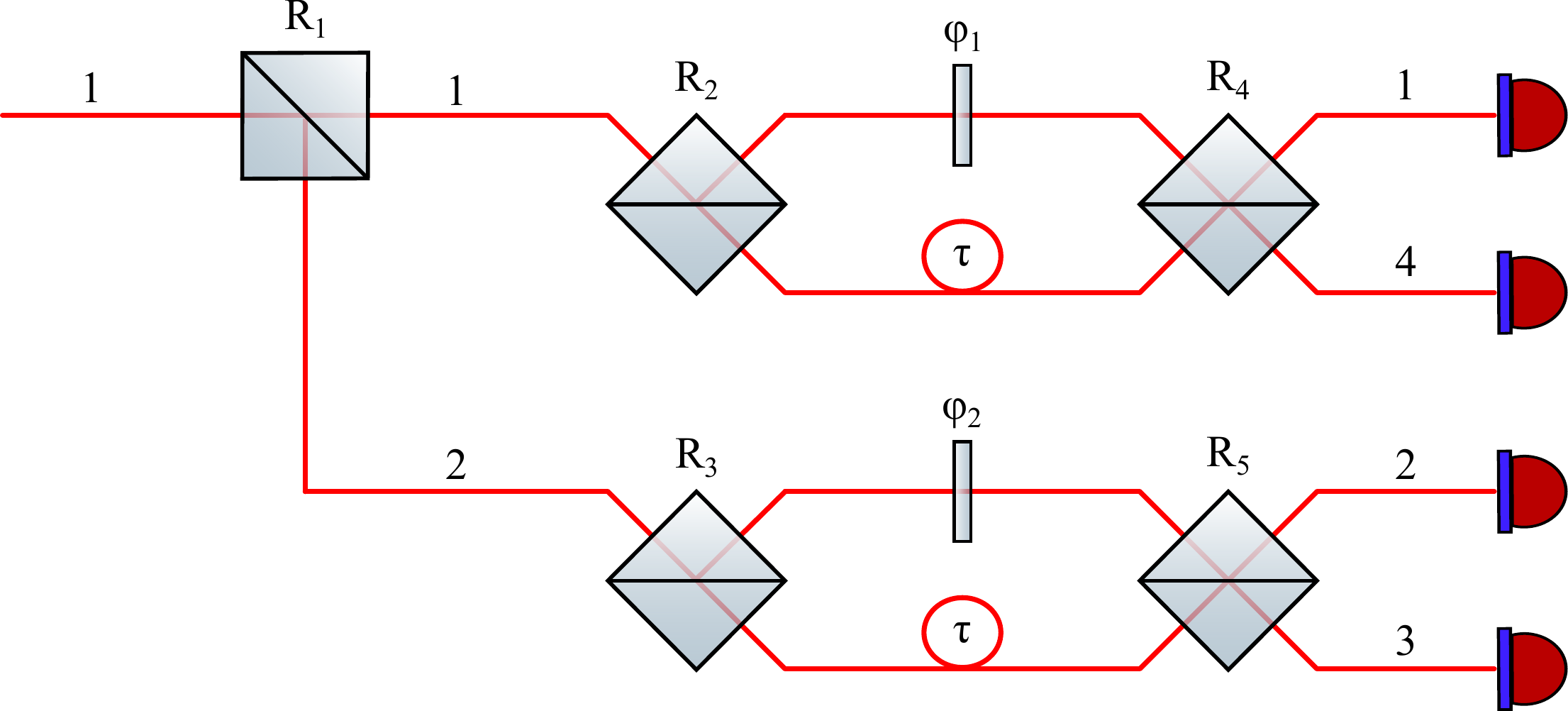}
    \caption{\textbf{Entanglement swapping apparatus.} Scheme of the optical modes and beam-splitters reflectivies ($R_i = 1 - T_i$)to perform the entanglement swapping.}
    \label{fig:enter-label}
\end{figure}

\subsection{Single-photon input case}
We consider here the ideal case where all BS in Fig. \ref{fig:enter-label} have transmittance $t_i^2 = T_i = 0.5$, and the input state is a set of two single photons:
\begin{equation} 
\ket{1}_\tau\ket{1}_0.
\end{equation}
Such a state can be written through creation operators as:
\begin{equation}
\left[a^\x_{1,\tau}\right]\left[a^\x_{1,0}\right]\ket{0}.
\label{eq: swap_state_init}
\end{equation}
After propagation through the first BS, that acts as $a^\x_{1}\rightarrow\left(a^\x_{1}+a^\x_{2}\right)/\sqrt{2}$, the state becomes:
\begin{equation}
    \frac{1}{2}\left[a^\x_{1,\tau}+a^\x_{2,\tau}\right]\left[a^\x_{1,0}+a^\x_{2,0}\right]\ket{0}.
    \label{eq: swap_state_after_I_BS}
\end{equation}
Then, after the first layer of two BS evolutions, the state will evolve into:
\begin{equation}
    \frac{1}{4}\left[a^\x_{1,\tau}+a^\x_{4,\tau}+a^\x_{3,\tau}+a^\x_{2,\tau}\right]\left[a^\x_{1,0}+a^\x_{4,0}+a^\x_{3,0}+a^\x_{2,0}\right]\ket{0}.
\end{equation}
Before the last layer of two BSs, we observe that:
\begin{itemize}
    \item mode 1 and 2 will acquire a time delay equal to $\tau$;
    \item mode 4 and mode 3 will acquire a relative phase shift $e^{i\phi}$ and $e^{i\phi'}$ respectively, due to the propagation in the Mach-Zehnder interferometer.
\end{itemize}  
The state before the final layer of BSs becomes:
\begin{equation}
    \frac{1}{4}\left[a^\x_{1,2\tau}+e^{i\phi}a^\x_{4,\tau}+e^{i\phi'}a^\x_{3,\tau}+a^\x_{2,2\tau}\right]\left[a^\x_{1,\tau}+e^{i\phi}a^\x_{4,0}+e^{i\phi'}a^\x_{3,0}+a^\x_{2,\tau}\right]\ket{0}.
\end{equation}
By looking at a specific time bin $t=\tau$, we can neglect the other terms and keep only the following part of the state:
\begin{equation}
    \frac{1}{4}\left[e^{i\phi}a^\x_{4,\tau}+e^{i\phi'}a^\x_{3,\tau}\right]\left[a^\x_{1,\tau}+a^\x_{2,\tau}\right]\ket{0}.
\end{equation}
By applying the evolution of the final layer of two BSs, we obtain that the relevant part of the final state reads:
\begin{equation}
    \frac{1}{8}\left[e^{i\phi}a^\x_{4,\tau}+e^{i\phi}a^\x_{1,\tau}+e^{i\phi'}a^\x_{2,\tau}-e^{i\phi'}a^\x_{3,\tau}\right]\left[a^\x_{4,\tau}-a^\x_{1,\tau}+a^\x_{2,\tau}+a^\x_{3,\tau}\right]\ket{0}.
\end{equation}
The probabilities $\text{P}_{4,2}$, $\text{P}_{4,3}$, $\text{P}_{1,2}$, $\text{P}_{1,3}$ for the relevant two-fold coincidence events $\text{CC}_{4,2}$, $\text{CC}_{4,3}$, $\text{CC}_{1,2}$, $\text{CC}_{1,3}$, can be then obtained as:
\begin{equation}
    P_{1,2}^{\text{ID}} = \frac{1}{32}\biggl(1-cos \xi \biggl); \qquad P_{1,3}^{\text{ID}} = \frac{1}{32}\biggl(1+cos \xi \biggl); \qquad P_{4,2}^{\text{ID}} = \frac{1}{32}\biggl(1+cos \xi \biggl); \qquad P_{4,3}^{\text{ID}} = \frac{1}{32}\biggl(1-cos \xi \biggl).
\end{equation}
where $\xi = \phi-\phi'$. In this ideal scenario, all visibilities are then equal to 1.

\subsection{Model of the experiment}
We now consider the scenario where the two states contain a small superposition with a vacuum component. The input state is then written as:
\begin{equation}
\left[\alpha\ket{0}+\beta\ket{1}\right]_\tau\left[\alpha\ket{0}+\gamma\ket{1}\right]_0,
\end{equation}
where $\gamma$ is related to $\beta$ through a phase shift $\gamma=e^{i\delta}\beta$. Expressing the one-photon state through the creation operators, the initial state is written as:
\begin{equation}
    \left[\alpha+\beta a^\x_{1,\tau}\right]\left[\alpha+\gamma a^\x_{1,0}\right]\ket{0}.
\end{equation}
We then follow the same assumptions of the ideal case, except that the transmittance of the BSs can be different than $0.5$. Therefore, the first BS would transform the creation operators as $a^\x_{1}\rightarrow\left(t_1 a^\x_{1}+r_1 a^\x_{2}\right)$, and the state becomes:
\begin{equation}
       \left[\alpha+\beta\left(t_1 a^\x_{1, \tau}+r_1 a^\x_{2, \tau}\right)\right]\left[\alpha+\gamma \left(t_1 a^\x_{1, 0}+r_1 a^\x_{2, 0}\right)\right]\ket{0}. \\
\end{equation}
Then, by considering the evolution induced by BS$_2$ and BS$_3$, we obtain the following state:
\begin{equation}
\left[\alpha+\beta t_1 \left( t_2 a^\x_{1, \tau}+ r_2 a^\x_{4, \tau}\right)+\beta r_1 \left(r_3 a^\x_{3, \tau} + t_3 a^\x_{2, \tau}\right)\right]\left[\alpha+\gamma t_1 \left( t_2 a^\x_{1, 0}+ r_2 a^\x_{4, 0}\right)+\gamma r_1 \left(r_3 a^\x_{3, 0} + t_3 a^\x_{2, 0}\right)\right]\ket{0},
\end{equation}
that can be also expanded:
\begin{equation}
       \left[\alpha+\beta t_1 t_2 a^\x_{1, \tau}+\beta t_1 r_2 a^\x_{4, \tau}+\beta r_1 r_3  a^\x_{3, \tau} + \beta r_1 t_3 a^\x_{2, \tau}\right]\left[\alpha+\gamma t_1 t_2 a^\x_{1, 0}+\gamma t_1 r_2 a^\x_{4, 0}+\gamma r_1 r_3  a^\x_{3, 0} + \gamma r_1 t_3 a^\x_{2, 0}\right]\ket{0}.
\end{equation}
As for the previous calculations, we have that: 
\begin{itemize}
    \item mode 1 and 2 will acquire a time delay equal to $\tau$;
    \item mode 4 and mode 3 will acquire a relative phase shift $e^{i\phi}$ and $e^{i\phi'}$ respectively, due to the propagation in the Mach-Zehnder interferometer.
\end{itemize}  
Hence, the state evolves into:
\begin{gather}
    \begin{split}
       &\left[\alpha+\beta t_1 t_2 a^\x_{1, 2\tau}+e^{i\phi}\beta t_1 r_2 a^\x_{4, \tau}+e^{i\phi'}\beta r_1 r_3   a^\x_{3, \tau}+\beta r_1 t_3 a^\x_{2, 2\tau}\right]\times\\
       &\times\left[\alpha+\gamma t_1 t_2 a^\x_{1, \tau}+e^{i\phi}\gamma t_1r_2 a^\x_{4, 0}+e^{i\phi'}\gamma r_1 r_3   a^\x_{3, 0}+\gamma r_1 t_3 a^\x_{2, \tau}\right]\ket{0}.
    \end{split}
\end{gather}
By keeping only terms related to time bin $t=\tau$, and by neglecting the vacuum components that will not lead to a useful two-fold coincidence events, we can write the relevant portion of the state as:
\begin{equation}
       \left[e^{i\phi}\beta t_1 r_2 a^\x_{4, \tau}+e^{i\phi'}\beta r_1 r_3   a^\x_{3, \tau}\right]\left[\gamma t_1 t_2 a^\x_{1, \tau}+\gamma r_1 t_3 a^\x_{2, \tau}\right]\ket{0}.
\end{equation}
After the evolution of the final layer of two BSs, the revelant part of the final state reads:
\begin{gather}
   \begin{split}
       &\left[e^{i\phi}\beta t_1r_2r_4 a^\x_{1, \tau}+e^{i\phi}\beta t_1r_2t_4 a^\x_{4, \tau}+e^{i\phi'}\beta r_1r_3r_5   a^\x_{2, \tau} -e^{i\phi'}\beta r_1r_3t_5 a^\x_{3, \tau}\right]\times \\
       &\times\left[\gamma t_1t_2r_4 a^\x_{4, \tau}-\gamma t_1t_2t_4 a^\x_{1, \tau} +\gamma r_1t_3t_5   a^\x_{2, \tau}+\gamma r_1t_3r_5 a^\x_{3, \tau}\right]\ket{0}.
    \end{split}
\end{gather}
We can now obtain the probabilities $\text{P}_{4,2}$, $\text{P}_{4,3}$, $\text{P}_{1,2}$, $\text{P}_{1,3}$ for the relevant two-fold coincidence events $\text{CC}_{4,2}$, $\text{CC}_{4,3}$, $\text{CC}_{1,2}$, $\text{CC}_{1,3}$.
\begin{itemize}
    \item For Channels 1 and 2 we obtain:
    \begin{align}
        P_{1,2} &= \abs{\beta}^2 R_1T_1 \biggl(R_2T_3R_4T_5+T_2R_3T_4R_5 + 2 \sqrt{R_2T_2R_3T_3R_4T_4R_5T_5} \cos \xi \biggr), \\
        V_{1,2} &= \frac{2 \sqrt{R_2T_2R_3T_3R_4T_4R_5T_5}}{R_2T_3R_4T_5+T_2R_3T_4R_5}.
    \end{align}

 \item For Channels 1 and 3 we obtain:
   \begin{align}
        P_{1,3} &= \abs{\beta}^2 R_1T_1 \biggl(R_2T_3R_4R_5+T_2R_3T_4T_5 + 2 \sqrt{R_2T_2R_3T_3R_4T_4R_5T_5} \cos \xi \biggr), \\
        V_{1,3} &= \frac{2 \sqrt{R_2T_2R_3T_3R_4T_4R_5T_5}}{R_2T_3R_4R_5+T_2R_3T_4T_5}.
    \end{align}
    
\item For Channels 4 and 2 we obtain:
      \begin{align}
        P_{4,2} &= \abs{\beta}^2 R_1T_1 \biggl(R_2T_3T_4T_5+T_2R_3R_4R_5 + 2 \sqrt{R_2T_2R_3T_3R_4T_4R_5T_5} \cos \xi \biggr), \\
        V_{4,2} &= \frac{2 \sqrt{R_2T_2R_3T_3R_4T_4R_5T_5}}{R_2T_3T_4T_5+T_2R_3R_4R_5}.
    \end{align}
    
\item For Channels 4 and 3 we obtain
   \begin{align}
        P_{4,3} &= \abs{\beta}^2 R_1T_1 \biggl(R_2T_3T_4R_5+T_2R_3R_4T_5 + 2 \sqrt{R_2T_2R_3T_3R_4T_4R_5T_5} \cos \xi \biggr), \\
        V_{4,3} &= \frac{2 \sqrt{R_2T_2R_3T_3R_4T_4R_5T_5}}{R_2T_3T_4R_5+T_2R_3R_4T_5}.
    \end{align}
    
\end{itemize}

As an additional source of noise, we also take into account partially distinguishable photons with HOM visibility $V_{HOM}$. The result is that all four visibility are scaled of the same factor $m = V_{HOM}$.

Let us now define the following four parameters:
\begin{equation}
    x := \frac{T_2R_3}{R_2T_3}; \qquad y := \frac{T_4}{R_4}; \qquad z := \frac{T_5}{R_5}; \qquad w := 2 m \sqrt{xyz}.
\end{equation}
In this way, we can rewrite the four visibilities as:
\begin{equation}
    \begin{cases}
        V_{1,2} = \frac{w}{xy+z},\\
        V_{1,3} = \frac{w}{xyz+1},\\
        V_{4,2} = \frac{w}{x+yz},\\
        V_{4,3} = \frac{w}{y+xz}.
    \end{cases}
\end{equation}
Through this parametrization, we are able to retrieve some relations between the visibilities, the reflectivity of the BSs and the photons indistinguishability parameters. Indeed, after some calculations, we retrieve:
\begin{equation}
    \begin{cases}
        z = \frac{t_2+t_3-t_1^2-1\pm \sqrt{(t_2+t_3-t_1^2-1)^2-4(t_2t_3-t_1)^2}}{2(t_2t_3-t_1)},\\
        x = \frac{t_2-t_3z}{t_1-z},\\
        y = \frac{t_3-zt_2}{t_1-z},\\
        w = \frac{V_{1,2}(1-z^2)}{t_1-z},
    \end{cases}
    \qquad
    \begin{cases}
        t_1 := \frac{V_{1,2}}{V_{1,3}},\\
        t_2 := \frac{V_{1,2}}{V_{4,2}},\\
        t_3 := \frac{V_{1,2}}{V_{4,3}}.\\
    \end{cases}
    \label{eq:inv_eq}
\end{equation}
where the sign is chosen such that $z\geq 0$. We will use such relations to validate our experimental results, as described in the following subsection.

\subsection{Entanglement swapping results analysis}
We first report the measured visibilities in our apparatus:
\begin{equation}
    \begin{cases}
        V_{A_1,C} = 0.942 \pm 0.002\\
        V_{A_1,B} = 0.862 \pm 0.002\\
        V_{A_2,C} = 0.879 \pm 0.002\\
        V_{A_2,B} = 0.903 \pm 0.002\\
    \end{cases}
    V_{\text{ave}} = 0.896 \pm 0.001.
\end{equation}
obtained at the four output combinations of the entanglement swapping protocol. To validate the experiment, we compare these results with the model derived above. We first assume that the BS reflectivities have the ideal values:
\begin{equation}
    R_2=R_3=R_4=R_5 = 0.5,
\end{equation}
while $V_{HOM} = 0.902$ is the Hong-Ou-Mandel visibility averaged over the two MZI interferometers. In this case, we find that the expected visibility for all possible two-fold events amounts to $V^{\text{theo}} = 0.902$.
%\begin{equation}
%    \begin{cases}
%        V_{1,2}^{\text{theo}} = 0.902 \\
%        V_{1,3}^{\text{theo}} = 0.902 \\
%        V_{4,2}^{\text{theo}} = 0.902 \\
%        V_{4,3}^{\text{theo}} = 0.902 
%    \end{cases}
%\end{equation}
The difference between the theoretical prediction and the experimental one can be attributed to unbalanced splitting ratios of the various BSs and to different coupling and detection efficiencies in the apparatus. All these effects can be summarized by considering non-ideal reflectivities of the BSs. 

If we follow the inverse approach, using the equations~\ref{eq:inv_eq} and deriving the reflectivities of the BSs from the experimental visibilities we obtain:
\begin{equation}
    x = 1.16 \pm 0.11; \qquad R_4 = 0.44 \pm 0.04; \qquad R_5 = 0.38 \pm 0.08; \qquad V_{HOM} = 0.92 \pm 0.04.
\end{equation}
These value are compatible with the parameters of the experimental apparatus.

\section{Characterization of the single-photon source}

We characterize the single-photon purity of our source by measuring the second-order auto-correlation function in a Hanbury-Brown-Twiss interferometer at both Alice's and Bob's stations. Here, we show the normalized coincidences as a function of the delay in one of the two arms of the interferometer for both stations in Fig.~\ref{fig:caratterizzazione}a-b. 
\begin{figure}[ht!]
    \centering 
    \includegraphics[width = 0.99\textwidth]{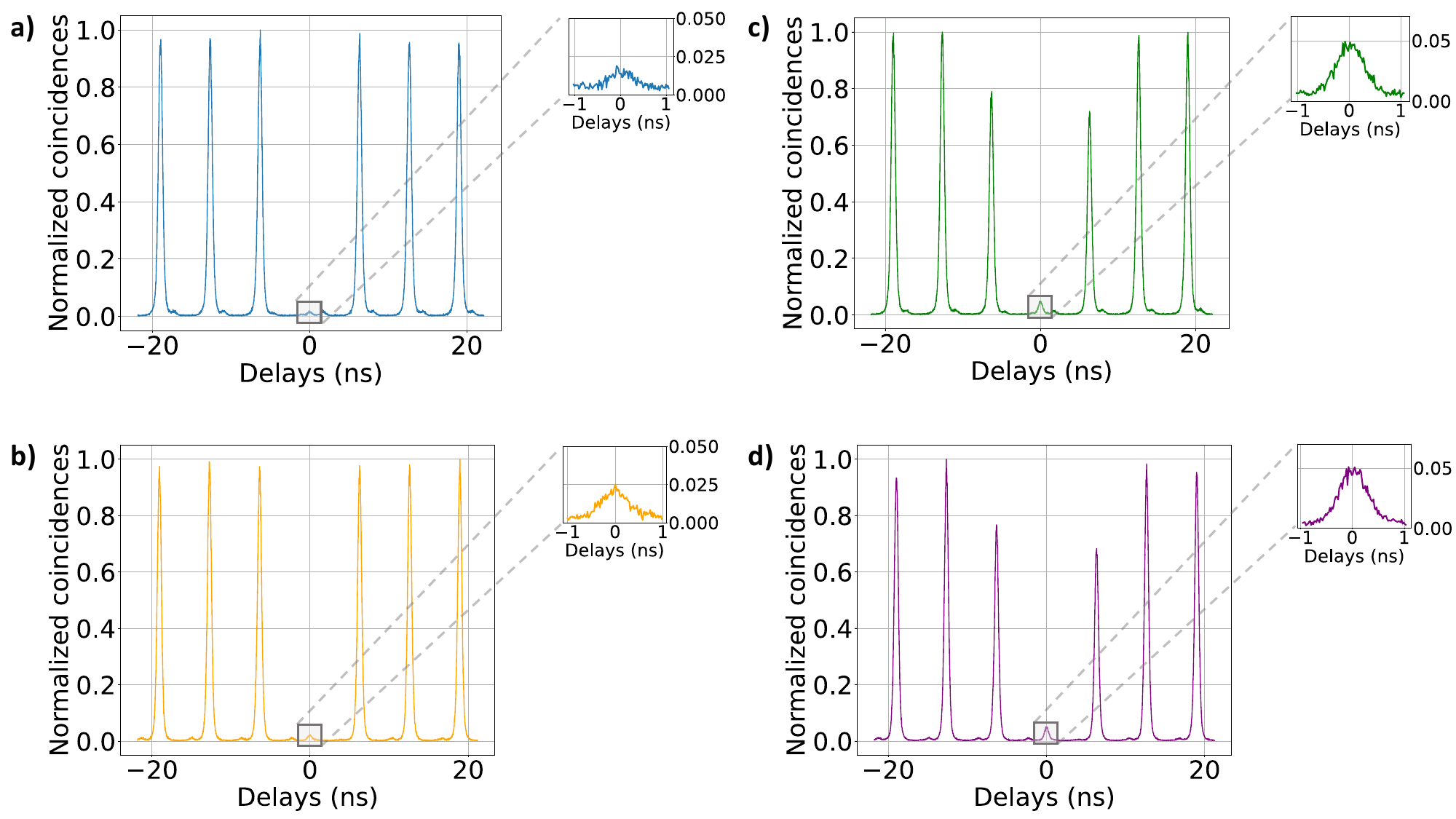}
    \caption{\textbf{Characteristics of our source. a)-b)} Plot of the normalized coincidences against time delay in a Hanbury-Brown-Twiss interferometer for the measurement of the second-order autocorrelation $g_2(0)$ of respectively Alice's and Bob's stations. \textbf{c)-d)} Plot of the normalized coincidences against time delay between in a Mach-Zehnder interferometer for the measurement of the HOM visibility $V_{HOM}$ of respectively Alice's and Bob's stations.}
    \label{fig:caratterizzazione}
\end{figure}
The single-photon indistinguishability is characterized by measuring the HOM visibility of two consecutive photons in a MZI interferometer as the one shown in Fig.~2b of the main text, for both Alice's and Bob's stations. Here, we show the normalized coincidences as a function of the delay in one of the two arms of the interferometer for both stations in Fig.~\ref{fig:caratterizzazione}c-d.

Finally, we report typical values of such quantities throughout the whole experiment:
\begin{gather}
\begin{split}
    g_2^{\text{Alice}}(0) &= 0.0146 \pm 0.0006 \\
    g_2^{\text{Bob}}(0) &=0.0192 \pm 0.0007 \\
    V_{\text{HOM}}^{\text{Alice}} &= 0.9055 \pm 0.0015 \\
    V_{\text{HOM}}^{\text{Bob}} &= 0.8987 \pm 0.0012
\end{split}
\end{gather}

{\section{Estimation of the visibility from the single-photon counts}}

In this section, we report the derivation of a reliable procedure that we employed to estimate the fringe visibility. As discussed here, such a method allows an accurate estimation of the visibility of a single-count trace even in the presence of low statistics, whereas standard methods would significantly overestimate it, as depicted in Fig.~\ref{fig:visibility_estim}.
A single-count time trace can be expressed as:
\begin{equation}
    n(t) = \mathcal{P}\left( N\frac{1+V\cos(\phi(t))}{2} \right)
    \label{eq:generaltimetrace}
\end{equation}
where $\mathcal{P}(\cdot)$ is a sample from a Poissonian distribution with mean value equal to the quantity in the parenthesis, $\phi(t)$ is a stochastic variable that varies slowly compared to signal sampling times, $N$ is the average number of photons and $V$ is the fringes visibility of the signal that is the parameter to be estimated.

\begin{figure}[ht!]
    \centering
    \includegraphics[width = \textwidth]{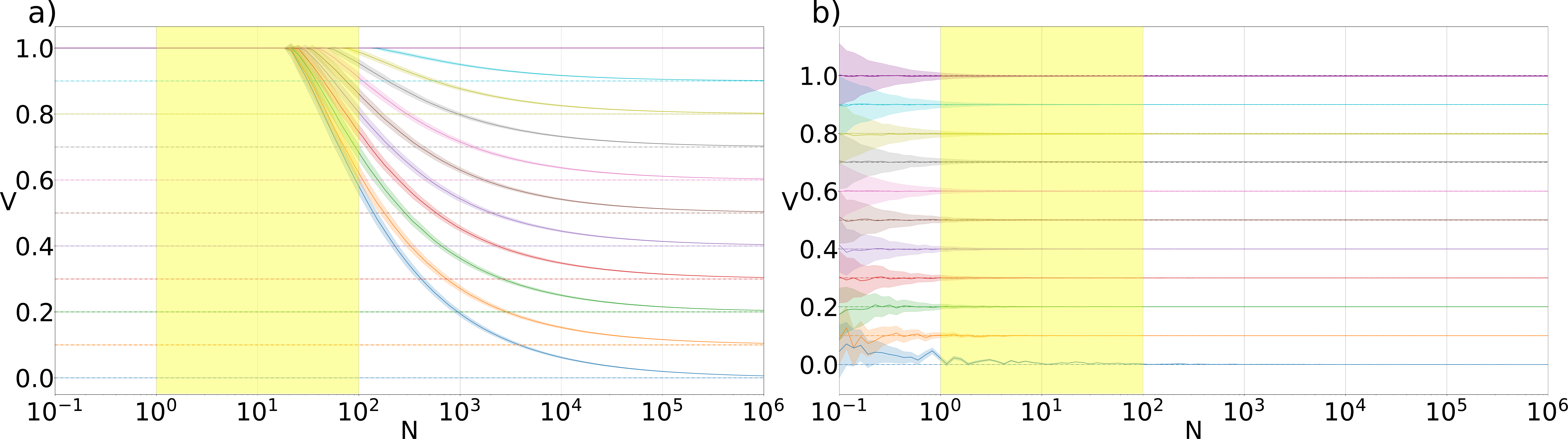}
    \caption{\textbf{Estimation of fringes visibility.} Numerical simulation of the visibility measurement for a typical time trace of the experiment that covers $10^5$ time bins of $50$ms, as a function of the average number of photons $N$ in each time bin. The dashed line represents the true visibility of the fringes. The solid line is the mean value reconstructed by averaging on $1000$ instances of the simulation. The shaded regions represent the variance of such estimations. In a) the visibility is estimated according to Eq. (\ref{eq:vminmax}); b) represents the reconstruction based on the measure of the trace variance. The region highlighted in yellow is the range of the $N$ in our teleportation experiment.}
    \label{fig:visibility_estim}
\end{figure}

%To correctly analyze the data, we need a reliable procedure to estimate the visibility.
The standard procedure to measure the visibility of a single-count trace, as the one in Eq.~\eqref{eq:generaltimetrace}, consists of applying the following formula:
\begin{equation}
    V = \frac{\max(n(t))-\min(n(t))}{\max(n(t))+\min(n(t))}
    \label{eq:vminmax}
\end{equation}
but, as shown in Fig.~\ref{fig:visibility_estim}a, if the number of photons $N$ is small, then a significant bias is obtained in the estimation of the visibility, especially in the region of the teleportation and entanglement swapping experiments (that is, $N = 10^0 \divisionsymbol 10^2$, highlighted in yellow in the figure).

To overcome this issue, we employed a different approach based on the measurement of the signal variance. Under the reasonable assumption that $\braket{\cos(\phi(t))}_t = 0$ and $\braket{\cos(2\phi(t))}_t = 0$, and that the time trace is sufficiently long and hence statistically significant, then we can compute the time average of the signal and of the square of the signal as:
\begin{align}
    \braket{n}_t &= \left\langle N\frac{1+V\cos(\phi(t))}{2} \right \rangle = \frac{N}{2}\\
    \braket{n^2}_t &= \left\langle\left(N\frac{1+V\cos(\phi(t))}{2}\right)^2 \right\rangle + \left\langle N\frac{1+V\cos(\phi(t))}{2} \right\rangle = \frac{N^2}{4}\left(1+\frac{V^2}{2}\right) + \frac{N}{2}
\end{align}
By combining previous equations we obtain:
\begin{equation}
    V^2 = 2 \frac{\braket{n^2}_t - \braket{n}_t^2 - \braket{n}_t } {\braket{n}_t^2}
\end{equation}
As we can observe, in Fig.~\ref{fig:visibility_estim}b this approach is more reliable compared to the previous one, due to the fast convergence of the estimated value of the visibility in the regime of the low values of $N$, i.e. of low statistics.